\documentclass[aps,prb, reprint,superscriptaddress]{revtex4-2}

\usepackage{graphicx}
\usepackage{dcolumn}
\usepackage{subfigure, caption}
\usepackage{xcolor, amsfonts, amsmath, amssymb}
\usepackage{svg, comment}
\usepackage{soul}
\usepackage[colorlinks=true, citecolor=blue, linkcolor=blue, urlcolor=blue]{hyperref}

\newcommand{\cartoonfigsize}{0.45}
\newcommand{\fullfigsize}{0.4}
\newcommand{\fullfigsizecb}{0.36}
\newcommand{\halffigsize}{0.30}
\newcommand{\largehalffigsize}{0.55}

\newcommand{\spin}{\alpha}
\newcommand{\pauli}{\sigma}
\newcommand{\relaxrate}{\Gamma}
\newcommand{\neighbor}{\boldsymbol{\delta}}
\newcommand{\gap}{\Delta}
\newcommand{\vac}{\Omega}
\newcommand{\bogoliubon}{D}
\newcommand{\bogangle}{\varphi}
\newcommand{\insteq}{\text{(relax)}}
\newcommand{\field}{\mathcal{E}}

\newcommand{\tunnelrate}{\gamma}
\newcommand{\production}{n}
\newcommand{\dos}{\rho_{\text{dh}}}
\newcommand{\mphindex}{\mu}
\newcommand{\depth}{l}
\newcommand{\wsenergy}{E}

\newcommand{\ex}[1]{\left\langle {#1} \right\rangle}
\newcommand{\ket}[1]{\left| {#1} \right\rangle}

\newcommand{\mel}[3]{\left\langle {#1} | {#2} | {#3} \right\rangle}

\newcommand{\op}{\left(}
\renewcommand{\cp}{\right)}
\newcommand{\os}{\left[}
\newcommand{\cs}{\right]}
\newcommand{\oa}{\big{(}}
\newcommand{\ca}{\big{)}}
\newcommand{\ob}{\Big{(}}
\newcommand{\cb}{\Big{)}}
\renewcommand{\oc}{\bigg{(}}
\newcommand{\cc}{\bigg{)}}

\newcommand{\e}{\epsilon}

\newcommand{\dd}{\text{d}}

\newcommand{\dbar}{d\hspace*{-0.08em}\bar{}\hspace*{0.1em}} 

\newcommand{\N}{\mathbb N}
\newcommand{\OO}{\mathcal O}

\newcommand{\trace}{\operatorname{Tr}}

\begin{document}

\title{\textbf{Subgap pumping of antiferromagnetic Mott insulators:\\ photoexcitation mechanisms and applications} 
}

\author{Radu~Andrei}
\affiliation{Institute for Theoretical Physics, ETH Z\"urich, 8093 Z\"urich, Switzerland}

\author{Mingyao~Guo}
\affiliation{Department of Physics, California Institute of Technology, Pasadena, CA 91125, USA}
\affiliation{Institute for Quantum Information and Matter, California Institute of Technology, Pasadena, CA 91125, USA}

\author{Mustafa~Ali}
\affiliation{Department of Physics, University of California San Diego, La Jolla, CA 92093, USA}

\author{Hoon~Kim}
\affiliation{Department of Physics, California Institute of Technology, Pasadena, CA 91125, USA}
\affiliation{Institute for Quantum Information and Matter, California Institute of Technology, Pasadena, CA 91125, USA}

\author{Richard~D.~Averitt}
\affiliation{Department of Physics, University of California San Diego, La Jolla, CA 92093, USA}

\author{David~Hsieh}
\affiliation{Department of Physics, California Institute of Technology, Pasadena, CA 91125, USA}
\affiliation{Institute for Quantum Information and Matter, California Institute of Technology, Pasadena, CA 91125, USA}

\author{Eugene~Demler}
\affiliation{Institute for Theoretical Physics, ETH Z\"urich, 8093 Z\"urich, Switzerland}

\date{\today}

\begin{abstract}
    We study the behavior of the 2D repulsive Hubbard model on a square lattice at half filling, under strong driving with ac electric fields, by employing a time-dependent Gaussian variational approach. Within the same theoretical framework, we analytically obtain the conventional Keldysh crossover between multiphoton and tunneling photoexcitation mechanisms, as well as two new regimes beyond the Keldysh paradigm. We discuss how dynamical renormalization of the Mott-Hubbard gap feeds back into the photoexcitation process, modulating the carrier generation rate in real time. The momentum distribution of quasiparticle excitations immediately after the drive is calculated, and shown to contain valuable information about the generation mechanism. Finally, we discuss experimental probing of the pump-induced nonequilibrium electronic state.
\end{abstract}


\maketitle

\section{Introduction} \label{sec:introduction}

Developments in optical technology have enabled the application of strong driving to condensed matter systems, across wide ranges of the electromagnetic spectrum \cite{basov_towards_2017, giannetti_ultrafast_2016}. This, in turn, has opened the door towards new regimes of experimental probing, in which responses of many-body systems can no longer be understood from the perspective of expansion in powers of the incident light's electric field. Usually, the analysis of perturbative probes can be performed by considering only the dynamics of a few elementary excitations; however, understanding ultrafast experiments with correlated materials requires unraveling non-equilibrium dynamics of many photogenerated excitations, often at different energy scales, and with strong interactions among them. Ultrafast techniques have been recently used to investigate interactions between charge carriers \cite{renormalizing_u_lco_ultrafast, singla_thz_frequency_2015, nio_renormalization, beaulieu_ultrafast_2021, gray_ultrafast_2018, kaiser_optical_2014}, to study the coupling of fermionic quasiparticles to collective bosonic degrees of freedom \cite{kaiser_optical_2014, dressed_quasiparticle_formation, wang_ultrafast_2022, hortensius_coherent_2021, radovskaia_photoengineering_2025}, and to analyze mechanisms of energy dissipation and thermalization \cite{perfetti_time_2006, gillmeister_ultrafast_2020, dean_ultrafast_2016, choi_light_induced_2024}. These experiments have also made it possible to access metastable states and ‘hidden phases’ \cite{stojchevska_ultrafast_2014, janod_resistive_2015}, thus paving a way for the study of exotic properties unattainable in thermal equilibrium \cite{de_la_torre_colloquium_2021, photoinduced_nonequilibrium_review}.\\

Materials that have attracted considerable attention in the field of ultrafast optical probes are Mott insulators (MI). In these systems, noninteracting band theory predicts metallic behavior, but strong interactions localize electrons and give rise to an insulating state instead. Magnetic ordering, often accompanying the MI state, points to the strong interplay between charge and spin degrees of freedom. When chemically doped, these materials exhibit some of the most puzzling phenomena in electronic systems, including high-temperature superconductivity and the pseudogap phase. A key feature of parent (undoped) Mott insulators is the presence of an energy gap for fermionic quasiparticle excitations; thus, when subject to a strong optical drive, one could expect these systems to respond similarly to semiconductors. However, the dynamics in MI is richer, because photoexcited quasiparticles suppress magnetic order and the Mott gap, thus changing the nature of quasiparticles themselves. This should be contrasted to the case of semiconductors, in which the bandgap is minimally affected by the photoexcited quasiparticles.\\

Another aspect that makes strong optical driving of MI nontrivial is the interplay of the real- and momentum-space character of quasiparticle generation. Following the work of Keldysh on ionization of atomic gases under strong driving \cite{keldysh_original_ionization}, the conventional picture of doublon/hole pair photoexcitation in MI consists of two distinct regimes: a multiphoton one at high frequencies of the drive, and respectively Landau-Zener tunneling in strong fields. The former corresponds to photoexciting fermionic quasiparticles from the lower Hubbard band to the upper one, while conserving their quasi-momenta. The latter can be visualized as electrons tunneling in real space between the two bands, due a strong energy gradient created by the driving field. The two competing mechanisms are separated by a Keldysh crossover, occurring when the driving field $\field$ and frequency $\omega_d$ are comparable, $e \field \xi \sim \hbar \omega_d$; here, $\xi$ is the doublon-hole correlation length \footnote{In the limit of strong interactions $U / \tau \gg 1$ (see Section \ref{sec:theoretical_approach} for definitions of the interaction $U$ and electron hopping $\tau$), the correlation length $\xi$ is on the order of the lattice constant $a$. For the detailed dependence of $\xi / a$ on $U/\tau$ in the 1D Hubbard model, see also Ref. \cite{millis_correlation_length} and Figure 3a of Ref. \cite{oka_nonlinear_doublons}.}. Photoexcitation rates corresponding to these regimes have been calculated both analytically and numerically for the 1D Hubbard model in \cite{oka_nonlinear_doublons}, under a rigid-band approximation. Since then, there has been considerable interest in the effects of strong dc fields on MI \cite{dmft_dc_dielectric_breakdown, dmft_dc_usa_dielectric_breakdown, dmft_doublon_production_heating, dmft_two_dielectric_breakdowns, resistive_switching_hysteresis, gap_collapse_and_switching, iTEBD_study}, as well as nonequilibrium steady states under ac driving \cite{floquet_dmft_introduction, floquet_prethermalization_resonant_hubbard, driven_steady_state, photoinduced_nonequilibrium_review}. Recently, evidence for the Keldysh crossover has also been observed experimentally \cite{dave_experimental_crossover}.\\

Suppression of the Mott gap following strong optical driving has been observed experimentally \cite{dressed_quasiparticle_formation, wang_ultrafast_2022, renormalizing_u_lco_ultrafast, nio_renormalization} and analyzed from a theoretical perspective \cite{dynamics_screening_photodoped, multiband_photodoped_screening, golez_theory_2025}. However, little emphasis has been placed so far on the feedback of dynamical gap renormalization into the photoexcitation process itself, and the influence this in turn has on the carrier production rate. When driving is realized by strong-field, sub-picosecond laser pulses, such a mechanism will become relevant. An accurate description of the system dynamics will then need to fully track the real-time evolution of the electronic state, as opposed to approaches commonly employed in previous literature: computing instantaneous production rates at time $t = 0$ from Fermi’s golden rule \cite{oka_nonlinear_doublons} or, in the opposite limit, finding steady states under ac driving \cite{driven_steady_state}. Developing a theoretical model for nonequilibrium dynamics of strongly driven Mott insulators that captures this feedback is one of the primary goals of this paper. We also discuss several types of experimental probes that can elucidate the nature of transient states arising during strong driving of MI, and reveal the nontrivial feedback between photoexcitation of quasiparticles and dynamics of the magnetic order and the Mott gap.\\

Variational methods are a powerful tool in the analysis of strongly correlated systems, both in terms of describing ground state properties and for understanding real-time evolution \cite{geometry_variational_methods, variational_non_gaussian}. In particular, within the context of superconductivity, time-dependent extensions of BCS theory have been successfully employed to treat dynamics under external driving, or upon quenching certain system parameters \cite{transport_equations_superconductors, transport_equations_clean_superconductors, bcs_formation_solitons, solution_bcs_central_spin, nonequilibrium_cooper_nonadiabatic, quenching_bcs_no_eom, nonequilibrium_hubbard_sc_controversial}. The central role played in these cases by a dynamical gap, set by electronic interactions and self-consistently evolved in time, forms a strong analogy with Mott systems. In this paper, we apply a similar time-dependent, self-consistent Gaussian variational method, based on the SDW approach to the Hubbard model \cite{sdw_bag_superconductivity}, to the problem of carrier photoexcitation under subgap pumping in AF Mott insulators.\\

\begin{figure}
	\centering	
    \scalebox{\fullfigsize}{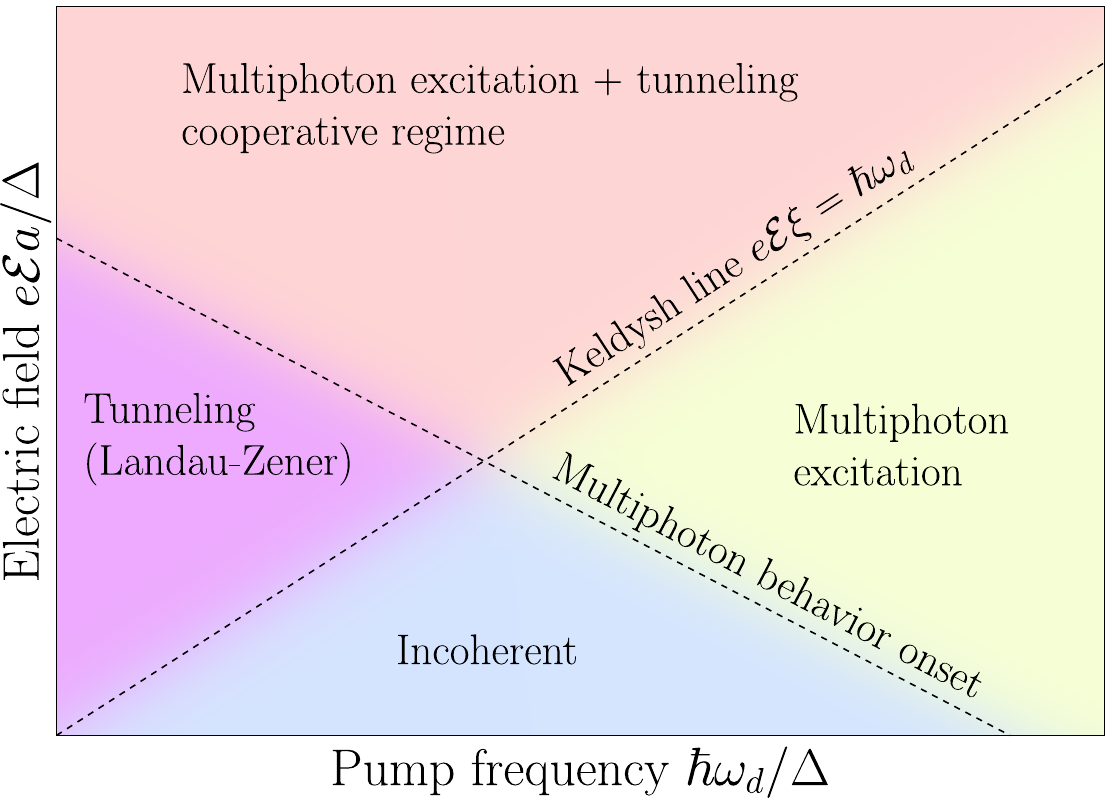}
	\caption{Schematic depiction of photoexcitation regimes in a subgap-driven Mott insulator, as a function of the pump frequency $\omega_d$ and electric field $\mathcal{E}$. Taking the dc limit recovers Landau-Zener tunneling for a wide range of electric fields, while the high-frequency region is dominated by multiphoton behavior. Rather than plain competition between these regimes, and their separation by a narrow Keldysh crossover, we instead find a broad cooperative region at strong drive. Finally, the low-field area is dominated by weak, incoherent photoexcitation, arising from inevitable carrier decoherence in any real solid-state system. Also see Figure \ref{fig:qp_momentum_variance}, which provides quantitative distinction between these regimes, based on the momentum distribution of excited quasiparticles.}
	\label{fig:regime_sketch_2d}
\end{figure}

Since direct measurements of the distribution and total density of photogenerated carriers are not always readily available, an important question arises regarding the probing of such systems after the photoexcitation process. One possibility is provided by electron-phonon coupling, an important feature of all solid state systems, which yields additional richness to the properties of correlated states both in and out of equilibrium. Within the context of pump-probe experiments on MI, a commonly observed feature \cite{reflected_phonons_thin_film, long_lived_oscillations_thin_film} is the generation of coherent acoustic phonons, which manifests itself as oscillations in the transient reflectivity, on timescales significantly longer than those relevant for electronic dynamics. Theoretical understanding of this process has been mostly phenomenological \cite{phonon_mechanism_review_macroscopic}, based on a strain wave induced in the material due to the local heating effect of the pump. Here, in order to provide a microscopic counterpart to the aforementioned picture, we employ the Su-Schrieffer-Heger (SSH) model \cite{ssh_original, phonons_in_cuprates} of electron-phonon coupling; we note however that our analysis is general and should apply to a broad class of models for electron-phonon interactions. By connecting the amplitude of such acoustic phonons with the density of photogenerated carriers, we highlight a novel option for probing the quasiparticle excitations, which is complementary to the usual techniques that rely on direct changes to the optical conductivity on picosecond timescales.\\

This paper is organized as follows: in Section \ref{sec:theoretical_approach}, we describe the theoretical assumptions and variational method used to model the photoexcitation process. In the limit of negligible gap renormalization, we show that an approximate analytic solution allows us to recover the four regimes depicted in Figure \ref{fig:regime_sketch_2d}. Throughout the following sections, we provide numerical calculations with experimental parameters typical for transition metal oxides, such as the cuprate parent compounds. Section \ref{sec:results_1_post_pump} focuses on the system state immediately after the pump pulse, analyzing how the photocarrier density and momentum distribution depend on the pump frequency $\omega_d$ and electric field $\field$. In Section \ref{sec:results_2_real_time_dynamics}, we discuss how dynamical renormalization of the Mott-Hubbard gap feeds back into the photoexcitation process, modifying the doublon-hole production rate in real time, and altering the multiphoton frequency conditions. The coupling of quasiparticles to acoustic phonons, enabling the employment of the latter to probe the former, is investigated in Section \ref{sec:acoustic_phonons}. Finally, Section \ref{sec:conclusions} presents conclusions as well as possible avenues for future research.\\

\section{Theoretical approach} \label{sec:theoretical_approach}

\subsection{Evolution equations}

As a prototype of interacting electron systems, we study a single-band Hubbard model, at half filling, on a square lattice in two dimensions. Hopping is taken to be nearest-neighbor only. Coupling to the external electric field of the drive is implemented via a Peierls phase, making the Hamiltonian time-dependent:
\begin{equation} \label{eq:hubbard_hamiltonian}
    H_\text{sys}(t) = - \tau \sum_{\mathbf{j}, \neighbor, \spin} e^{i e \neighbor \cdot \mathbf{A}(t)} \; c^\dag_{\mathbf{j} + \neighbor, \spin} c_{\mathbf{j}, \spin} + U \sum_\mathbf{j} n_{\mathbf{j} \uparrow} n_{\mathbf{j} \downarrow}.
\end{equation}
Here, the sums run over lattice sites $\mathbf j$, while $\neighbor \in \{\pm \hat x, \pm \hat y\}$ points to nearest-neighbor sites, and $\spin = \pm 1$ denotes electronic spin. We focus on the strongly interacting limit $U/\tau \gg 1$; typical one-band models describing cuprates \cite{cuprate_hubbard_models_review} employ $U/\tau \sim 10$.\\

We investigate evolution under the drive using a time-dependent Gaussian variational approach, which relies on the system's tendency towards antiferromagnetic spin ordering, at wavevector $\mathbf Q \equiv (\pi, \pi)$. For simplicity, we pick the $\hat z$ axis to be the direction of long-range N\'eel ordering \footnote{Note that only the short-range part of AFM correlations is essential here, so the results will also be qualitatively valid above $T_N$; see Appendix \ref{ssec:sm_intro_hamiltonian_drive} for further discussion.}. The variational ground state is taken to be the mean-field wavefunction for an antiferromagnetic state:
\begin{equation} \label{eq:variational_gs_wavefunction}
\ket{\vac} \equiv \prod_{\substack{\mathbf k \in \text{BZ'} \\ \spin}} \oa u_{\mathbf k} c_{\mathbf k, \spin}^\dag + v_{\mathbf k} c_{\mathbf k + \mathbf Q, \spin}^\dag \ca \ket{0}.
\end{equation}
Here, $\ket{0}$ is the fermionic vacuum, and the product runs over momenta $\mathbf k$ in the magnetic Brillouin zone BZ', whose size is smaller by a factor of 2 compared to the structural one, due to symmetry breaking in the AF phase. Further symmetry arguments show that the only non-zero two-point correlators are of the form $\ex{c_{\mathbf k, \spin}^\dag c_{\mathbf k, \spin}}$ and $\ex{c_{\mathbf k + \mathbf Q, \spin}^\dag c_{\mathbf k, \spin}}$. 
The Neel order parameter $S \equiv \mel{\vac}{S^z_\mathbf{Q}}{\vac}$ is given by the operator
\begin{equation}
    S^z_{\mathbf Q} \equiv {1 \over N} \sum_{\mathbf k, \spin} \spin \; c^\dag_{\mathbf k + \mathbf Q, \spin} c_{\mathbf k, \spin}.
\end{equation}
Minimizing the energy with respect to the parameters $u_{\mathbf k}$ and $v_{\mathbf k}$ yields the self-consistency equation for $S$, which is again of the BCS type:
\begin{equation}
    {1 \over N} \sum_{\mathbf k \in \text{BZ'}} {1 \over \sqrt{(\e_{\mathbf k}^0 / U)^2 + (S / 2)^2}} = 1.
\end{equation}
In the above, $\e_{\mathbf k}^0 \equiv -2 \tau (\cos k_x + \cos k_y)$ is the kinetic energy on a square lattice, with nearest-neighbor hopping. Coupling to the drive makes it time-dependent, via the substitution $\mathbf k \to \mathbf k - e \mathbf A(t)$. Assuming a spatially uniform vector potential, the translation invariance of the Hubbard Hamiltonian \eqref{eq:hubbard_hamiltonian} is preserved. We will then model time evolution by letting the correlators $\ex{c_{\mathbf k, \spin}^\dag (t) \; c_{\mathbf k, \spin} (t)}$ and $\ex{c_{\mathbf k + \mathbf Q, \spin}^\dag (t) \; c_{\mathbf k, \spin} (t)}$, as well as the order parameter $S(t)$, evolve self-consistently. Switching to a doublon/hole basis via a Bogoliubov transformation, we employ a variational density matrix of the form
\begin{equation}
    \rho_\text{var}(t) =\bigotimes_{\substack{\mathbf k \in \text{BZ'} \\ \spin}} \begin{pmatrix} {1 \over 2} \os {1 - \rho_{\mathbf k}^z (t)} \cs & \spin \rho_{\mathbf k}^+ (t) \\ \spin \rho_{\mathbf k}^- (t) & {1 \over 2} \os {1 + \rho_{\mathbf k}^z (t)} \cs \end{pmatrix},
\end{equation}
where for each momentum $\mathbf k \in \text{BZ'}$ we restrict our attention to the subspace spanned by the states $\ket{h_{- \mathbf k}, d_{\mathbf k}}$, representing a doublon-hole pair with zero total momentum, and respectively $\ket{\vac_{\mathbf k}}$, with no excitations. The ground state \eqref{eq:variational_gs_wavefunction} corresponds to choosing the initial conditions $\rho_{\mathbf k}^+ = \rho_{\mathbf k}^- = 0$ and $\rho_{\mathbf k}^z = 1$, for all $\mathbf k$. More details are presented in Appendices \ref{sec:sm_sdw}~\textendash~\ref{sec:sm_ansatz}.\\

Interaction of doublons and holes with other degrees of freedom, such as magnons or optical phonons, will give rise to scattering between different $\mathbf{k}$ sectors, as well as decay of the off-diagonal terms $\rho_{\mathbf k}^\pm$. Working in the limit of pump duration $T_\text{pump}$ much shorter than the characteristic timescale of momentum relaxation $\tau_\text{intraband rel.}$, the coupling between different momenta will have a weak effect on the photoexcitation process; on the other hand, even small decay rates of $\rho_{\mathbf k}^\pm$ are relevant for off-resonant pumping, so this effect should be taken into consideration. While such decoherence rates will in principle depend on $\mathbf{k}$, we employ the approximation that a single effective dephasing rate $\relaxrate$ acts in all momentum sectors. The resulting time-evolution equations are
\begin{subequations} \label{eq:maintext_evolution_equations}
\begin{align}
	\partial_t \rho_\mathbf{k}^+ &= {\partial_t \bogangle_\mathbf{k} \over 2} \; \rho_\mathbf{k}^z + 2 \os i E_\mathbf{k} - \relaxrate \cs  \rho_\mathbf{k}^+\\ 
	\partial_t \rho_\mathbf{k}^z &= - (\partial_t \bogangle_\mathbf{k}) \; (\rho_\mathbf{k}^+ + \rho_\mathbf{k}^-)\\ 
    \partial_t S &= -{4 \over N_\text{BZ'}} \sum_{\mathbf k \in \text{BZ'}} \e_{\mathbf k} \os \rho_{\mathbf k}^{y} +  {\relaxrate \over 2 E_{\mathbf k}} \rho_{\mathbf k}^x \cs, 
\end{align}
\end{subequations}
where $\bogangle_\mathbf{k}(t) \equiv \arctan [2 \e_{\mathbf k}(t) / U S(t)]$ is the Bogoliubov angle, and the quasiparticle dispersion is given by
\begin{equation} \label{eq:maintext_quasiparticle_dispersion}
    E_\mathbf{k} = \sqrt{\e_{\mathbf k}^2(t) + \op {US(t) \over 2} \cp^2},
\end{equation}
with the corresponding density of states shown in Fig. \ref{fig:density_of_states} of Appendix \ref{sec:sm_sdw}. The complexity of the system \eqref{eq:maintext_evolution_equations} resides mainly in the coupling of different $\mathbf k$ sectors, due the implicit appearance of the time-dependent order parameter $S(t)$ through the factors of $\bogangle_\mathbf{k}$ and $E_\mathbf{k}$. In this way, our variational approach describes how the dynamics of the collective gap $\gap(t) \equiv U S(t)$ feeds back into the photoexcitation process, giving rise to several new phenomena under strong driving, to be discussed in Section \ref{sec:results_2_real_time_dynamics} below. On the other hand, when the external pump is weak enough, the order parameter will suffer negligible variation, and to leading order can be approximated by its equilibrium value $S(t) \approx S_0$. This recovers the rigid-band, semiconductor-like model, and allows for approximate analytic calculation of photoexcitation rates; we describe the main steps and results of this procedure in the following subsection.\\

\subsection{Analytic photoexcitation rates in the rigid-band approximation} \label{ssec:results_1_analytical_approximations}

When the dynamical nature of the Mott-Hubbard gap is ignored via the approximation $S(t) \approx S_0$, different $\mathbf{k}$ sectors are decoupled in the evolution equations \eqref{eq:maintext_evolution_equations}, and in consequence each one may be treated separately. The remaining time-dependence on the RHS, arising from $\e_{\mathbf k}(t)$, still precludes a complete analytic solution of $\rho_\mathbf{k}^z(t)$ or $\rho_\mathbf{k}^\pm(t)$. Instead, we employ a perturbative approach to find $\rho_\mathbf{k}^z(t \to \infty)$, from which we recover the excited quasiparticle density after pumping, via $n_{\text{qp}, \mathbf k} = (1 - \rho_\mathbf{k}^z) / 2$. For simplicity, we assume the driving to be realized by a monochromatic electric field of amplitude $\field_0$ and frequency $\omega_d$. With this approach we recover the standard Keldysh crossover in the field dependence of $n_{\text{qp}}$, as well as nontrivial behavior against the driving frequency. Here, we discuss the main steps of the calculation, as well as its conclusions; detailed derivations are presented in Appendix \ref{sec:analytic_approximate_solutions}. The intuition gained by examining these regimes will also prove useful in the case of strong driving, discussed in Sections \ref{ssec:results_1_2d_diagram} and onward.\\

Importantly, the expansion parameter we employ is not the electric field strength; indeed, the tunneling rate \eqref{eq:maintext_tunneling_production_rate} is exponential in $1/\field_0$, which is a nonperturbative result. Instead, the form of eqs. \eqref{eq:maintext_evolution_equations} is naturally amenable to expanding in $\partial_t \bogangle_\mathbf{k}$, which in the rigid-band approximation can be written as
\begin{equation} \label{eq:maintext_approximate_driving_term}
	\partial_t \bogangle_{\mathbf k}(t) \approx - {US / 2 \over \os U S / 2 \cs^2 + \os \e_{\mathbf k}(t) \cs ^2} \; \boldsymbol{\field}(t) \cdot \nabla_{\mathbf A} \e_{\mathbf k} (t).
\end{equation}
Although this is apparently proportional to the electric field $\field$, the implicit appearance of the drive in $\e_{\mathbf k}(t) \equiv \e_{\mathbf k - A(t)}$ means that $\partial_t \bogangle_{\mathbf k}$ in fact encodes contributions to infinite order in $\field$. Furthermore, we note that $\partial_t \bogangle_{\mathbf k}$ is smaller in magnitude compared to $\field$, by a factor on the order of $\tau / U$.\\

The relevant propagator for eqs. \eqref{eq:maintext_evolution_equations} depends on two time variables, owing to the modulation of $E_\mathbf{k}(t)$ through the drive; its Fourier transform will correspondingly be a function of two frequencies:
\begin{subequations}
    \begin{align}
    H_{\mathbf k}(x, \delta) &= \cos \op 2 \int_{x - \delta/2}^{x + \delta/2} dv \; E_\mathbf{k}(v) \cp \\
    H_{\mathbf k}(\Omega, \omega) &= \int dx \; d\delta \; H_{\mathbf k}(x, \delta) \; e^{-i x \Omega} \; e^{-i \delta \omega}.
\end{align}
\end{subequations}
In the limit $\relaxrate \to 0$, photoexcitation is described by a generalized resonance condition (see Appendix \ref{ssec:perturbative_approach} for a detailed derivation):
\begin{widetext}
    \begin{equation} \label{eq:maintext_no_decoherence_resonance}
        n_{\text{qp}, \mathbf k} (t \to \infty) = {1 \over 4} \int \dbar \Omega \; \dbar \omega \; (\partial \bogangle_{\mathbf k})_{\omega - \Omega / 2} \; H_{\mathbf k}(\Omega, \omega) \; (\partial \bogangle_{\mathbf k})_{- \omega - \Omega / 2}.
    \end{equation}
\end{widetext}
Here, $(\partial \bogangle_{\mathbf k})_\omega$ is defined \footnote{Note that Fourier transforming eq. \eqref{eq:maintext_approximate_driving_term} directly is much more convenient than computing $\bogangle_{\mathbf k}(\omega)$ and multiplying by $i \omega$; hence the unusual notation. The resonance condition \eqref{eq:maintext_no_decoherence_resonance} is also more transparent in this form.} as the Fourier transform of $\partial_t \bogangle_{\mathbf k}(t)$, i.e. $(\partial \bogangle_{\mathbf k})_\omega \equiv i \omega \bogangle_{\mathbf k}(\omega)$. Calculating the production rates in various driving regimes, then, reduces to matching poles of $H_{\mathbf k}$ with peaks of $(\partial \bogangle_{\mathbf k})$. Both rely heavily on $\e_{\mathbf k}(t)$ for their time dependence, so we turn to analyzing its spectrum.\\

Since the vector potential for our monochromatic drive will have amplitude $\field_0 / \omega_d$, the kinetic energy schematically looks like $\e_\mathbf{k} \sim \cos \os k + (\field_0 / \omega_d) \sin (\omega_d t) \cs$. In the weak-field regime $e \field_0 a \ll \hbar \omega_d$, one may directly expand in $\field_0 / \omega_d$ to find frequency components which are harmonics of $\omega_d$. To match the main pole of $H_{\mathbf k}$, located at $2 E_\mathbf{k}$, we must go up to the order $2 E_\mathbf{k} / \omega_d$; from the two factors of $(\partial \bogangle_{\mathbf k})$ in eq. \eqref{eq:maintext_no_decoherence_resonance}, this yields
\begin{equation} \label{eq:maintext_mph_1}
    n_{\text{qp}, \mathbf k} (t \to \infty) \propto \op {e \field_0 a \over \hbar \omega_d} \cp^{4 E_\mathbf{k} / \omega_d} \quad [e \field_0 a \ll \hbar \omega_d].
\end{equation}
Recalling that $2 E_\mathbf{k} / \omega_d$ is the number of pump photons required to match the energy of a doublon-hole pair, we conclude that eq. \eqref{eq:maintext_mph_1} represents the usual electric-field dependence of a multiphoton process. Moreover, since the density of states for our excitations is peaked at the bottom of the Hubbard band (Fig. \ref{fig:density_of_states}), let us consider for now $\gap / \omega_d \in \mathbb{Z}$; we integrate over $\mathbf k$ while approximating $2 E_\mathbf{k} \sim \gap$, to find the conventional form \cite{oka_nonlinear_doublons} of total production rate versus electric field: 
\begin{equation} \label{eq:maintext_multiphoton_production_rate}
    \production_\text{mph} \propto \op {e \field_0 a \over \hbar \omega_d} \cp^{2 \gap / \omega_d}.
\end{equation}
Note however that $2 E_\mathbf{k} / \omega_d$ must be an integer for the argument leading towards \eqref{eq:maintext_mph_1} to work; since the quasiparticle excitations are still dispersive, it follows that the multiphoton mechanism for general $\omega_d$ will be momentum-selective towards a resonant contour of the Hubbard bands (see e.g. panels b, e in Fig. \ref{fig:main_text_momentum_distributions}). In this sense, eq. \eqref{eq:maintext_multiphoton_production_rate} should be understood as a special case of the more accurate description in eq. \eqref{eq:maintext_mph_1}. It follows that the production rate will also be sensitive to the density of states for doublon-hole pairs at the resonant energy, an aspect which we explore in eq. \eqref{eq:maintext_mph_freq_dep} below, when analyzing the frequency dependence of photoexcitation rates.\\

Consider now the opposite regime $e \field_0 a \gg \hbar \omega_d$: the amplitude $\field_0 / \omega_d$ of the vector potential is large, so the shifted momentum $\mathbf k - e \mathbf A (t)$ explores the entire Brillouin zone width several times within a given pump cycle. The time dependence of $\e_\mathbf{k}$ will be mainly set by the field strength: for example, in the dc limit $\omega_d \to 0$, the vector potential increases linearly with time $\mathbf A(t) = - \boldsymbol{\field} t$, and in consequence $\e_\mathbf{k}(t)$ oscillates at the Bloch frequency $\omega_\text{Bloch} = e \field_0 a / \hbar$. This is to be contrasted with the multiphoton case, where $\omega_d$ and its harmonics were setting the time dependence of $\e_\mathbf{k}(t)$. We now apply the same generalized resonance procedure; the resulting production rate will contain an exponent which looks like $2 \gap / \omega_\text{Bloch}$, therefore giving rise to the exponential behavior in $1/\field_0$, which is characteristic to Landau-Zener tunneling. Note that such broad excursions in momentum space will produce excitations across the entire BZ, which agrees with the intuition that tunneling is a process well-localized in real space. Going back to finite $\omega_d$, we employ a Jacobi-Anger expansion of $\e_\mathbf{k}$ and $H_{\mathbf k}$, finally arriving at the generalized tunneling expression:
\begin{equation} \label{eq:maintext_tunneling_production_rate}
    \production_\text{tunnel} \propto \field_0^2 \; \exp \os - \tunnelrate(\omega_d) \; {\gap \over e \field_0 a} \cs,
\end{equation}
where $\gamma(\omega_d)$ is a frequency-dependent tunneling rate. Crossover between the regimes \eqref{eq:maintext_multiphoton_production_rate} and \eqref{eq:maintext_tunneling_production_rate} is given, as expected, by the Keldysh condition \footnote{As discussed in the introduction, the length scale characterizing this crossover in the literature is the doublon-hole correlation length $\xi$, rather than the lattice constant $a$. However, for our strongly-interacting regime $U/\tau \sim 10$, we have $\xi \approx a$, and for simplicity we use $a$ as our typical length scale.} $e \field_0 a \sim \hbar \omega_d$. We may understand this as the point where the amplitude of the vector potential $\mathbf{A}$ becomes comparable with the BZ size, and the important frequency scale in $\e_\mathbf{k}(t)$ shifts from $\omega_d$ to $\omega_\text{Bloch}$. The Keldysh line is one of the two relevant regime boundaries sketched in Figure \ref{fig:regime_sketch_2d}.\\

Having discussed the photoexcitation rate dependence on the driving field $\field_0$, we turn to investigating its behavior versus driving frequency $\omega_d$, an aspect which has received little attention in previous analytical treatments. We will find that the distinction between multiphoton and tunneling processes becomes difficult to articulate clearly: a considerable region of parameter space actually gives rise to collaboration between these two mechanisms, shaping a new regime which lies beyond the Keldysh crossover paradigm.\\

In the low-field, multiphoton regime $e \field_0 a \ll \hbar \omega_d$, we have seen that the resonance condition $n_\text{ph} = 2 E_\mathbf{k} / \omega_d \in \mathbb{Z}$ gives rise to momentum selectivity. With regard to frequency dependence, this will yield a total production rate proportional to the doublon-hole density of states $\dos(n_\text{ph} \; \omega_d)$ evaluated at the corresponding multiple of the driving frequency, i.e.
\begin{equation} \label{eq:maintext_mph_freq_dep}
    \production_\text{mph} \propto \op {e \field_0 a \over \hbar \omega_d} \cp^{2 \gap / \omega_d} \; \dos(n_\text{ph} \; \omega_d),
\end{equation}
where again we approximated $2 E_\mathbf{k} \approx \gap$ in the exponent. In particular, whenever the driving frequency crosses the $n-$photon threshold $\omega_d \ge \gap / n$, there will be a considerable increase of the photoexcitation rate. Such threshold behavior can be seen in the numerical calculations of Fig. \ref{fig:quasiparticles_and_multiphoton}a. The same conclusion remains valid at strong driving, once the gap suppression is taken into account, moving the thresholds of $\gap / n$ towards lower frequency.\\

In the regime $e \field_0 a \gg \hbar \omega_d$, where tunneling is expected, the conventional approach treats low-frequency fields as practically dc; for field strengths below the Schwinger tunneling threshold, this yields a weak and monotonous dependence of the photoexcitation rate on $\omega_d$ (see Figure 7a of Ref. \cite{oka_nonlinear_doublons}). In contrast, the tunneling rate $\tunnelrate(\omega_d)$ given by our approach, which appears in eq. \eqref{eq:maintext_tunneling_production_rate} and is discussed in detail within Section \ref{ssec:low_frequency_analytic}, points to richer behavior:
\begin{itemize}
    \item In the dc limit $\omega_d \to 0$, it reduces to a constant $\tunnelrate(0) \propto \ln (U / \tau)$, in agreement with Ref. \cite{oka_nonlinear_doublons};
    \item However, at finite $\omega_d > 0$, we find significant enhancements of $\tunnelrate(\omega_d)$ whenever the condition $\gap / \omega_d \in \mathbb{Z}$ is satisfied. Technically, this arises because the spectrum of $\e_\mathbf{k}(t)$ consists of peaks at multiples of $\omega_d$, with $\field_0$ giving a cutoff beyond which the amplitude of these peaks quickly decays (Fig. \ref{fig:ac_tunnel_spectra} in Appendix \ref{ssec:perturbative_approach_finite_frequency}). Therefore, the generalized resonance condition will be simultaneously sensitive to both $\field_0$ and $\omega_d$.
\end{itemize}
Enhancement of photoexcitation rates around frequencies $\omega_d = \gap / n$ is reminiscent of the multiphoton behavior discussed previously. Indeed, in Fig. \ref{fig:quasiparticles_and_multiphoton}a the frequency thresholds do not only appear below the Keldysh line, where the multiphoton regime is expected to reside, but rather extend far above it as well. There will be, in consequence, a broad region in parameter space, where photoexcitation rates display the tunneling dependence \eqref{eq:maintext_tunneling_production_rate} on the electric field, but multiphoton thresholds in the frequency direction. Simultaneous presence of multiphoton and tunneling signatures in the same region of parameter space points to cooperation between these two mechanisms, rather than their plain competition which underlies the Keldysh crossover picture.\\

Finally, we discuss the question of finite $\relaxrate$, which describes the influence onto photoexcitation of the scattering and dephasing mechanisms, that inevitably exist in solid-state systems at finite temperature. The generalized resonance condition \eqref{eq:maintext_no_decoherence_resonance} is relaxed to:
\begin{widetext}
   \begin{equation} \label{eq:maintext_with_decoherence_resonance}
	   n_{\text{qp}, \mathbf k} (t \to \infty) = {1 \over 2} \int \dbar \omega_1 \; \dbar \Omega \; \dbar \omega \; {2 \relaxrate \over (2 \relaxrate)^2 + (\omega - \omega_1)^2} \; \os (\partial \bogangle)_{\omega_1 - \Omega / 2} \; H(\Omega, \omega) \; (\partial \bogangle)_{- \omega_1 - \Omega / 2} \cs.
    \end{equation}
\end{widetext}
One effect thereof will be broadening of the multiphoton thresholds described above, but in the limit $\relaxrate \ll \omega_d$ this will not be a relevant qualitative change. More consequentially, the broadening of the doublon/hole spectral function allows for linear absorption from a subgap pump. Such a pathway is highly inefficient, especially if $\relaxrate$ is the smallest energy scale in the system, as it comes with a prefactor of $\relaxrate / \gap$. However, when both the pump frequency and field are low enough, such that multiphoton and tunneling mechanisms are strongly suppressed, the incoherent pathway will dominate, making this regime relevant in certain experimental contexts. To leading order in $\relaxrate$, for a pulse of fluence $F$ and frequency $\omega_d \ll \gap$, we obtain the excitation density 
\begin{equation} \label{eq:maintext_analytic_decoherence_qp}
n_\text{incoh} = {\eta_0 \relaxrate F \over 4} \op {U S \over 2} \cp^2 \int_{BZ'} {d^2 \mathbf k \over 2 \pi^2} \; {[\hat{\field} \cdot (\nabla \e_{\mathbf k})_0]^2 \over E_{\mathbf k, 0}^6},
\end{equation}
where $\eta_0$ is the free-space impedance, and $(\nabla \e_{\mathbf k})_0$ denotes the gradient of kinetic energy in momentum space. The integrand of \eqref{eq:maintext_analytic_decoherence_qp} also yields a prediction for the momentum distribution of the photogenerated carriers, which is plotted in Fig. \ref{fig:analytic_incoherent_momentum_distribution}, and compares very well to the distribution found from the numerical solution of the full evolution equations, in the incoherent regime (Fig. \ref{fig:main_text_momentum_distributions}a,d).\\

The four regimes derived in this part comprise the photoexcitation landscape sketched in Figure \ref{fig:regime_sketch_2d}. In the following section, we compare these analytic results to the numerical solution, including gap renormalization effects; and we highlight the momentum distribution of photogenerated carriers as a powerful tool to distinguish excitation mechanisms (Fig. \ref{fig:qp_momentum_variance}).\\

\section{Electronic state after the drive} \label{sec:results_1_post_pump}

In the case of strong driving, the rigid-band approximation fails, and we turn to numerically solving the full system \eqref{eq:maintext_evolution_equations} of real-time evolution equations. For illustration, we take $U = 1.93$ eV, $\tau = 189$ meV, $\relaxrate = 1.64$ meV, and $T_\text{pump} = 100$ fs, which yield a static order parameter $S_0 \approx 0.93$ and the corresponding equilibrium gap $\Delta_0 = 1.8$ eV.

\subsection{Photoexcited carrier density for arbitrary pump parameters} \label{ssec:results_1_2d_diagram}

\begin{figure*}
	\centering	
    \scalebox{\fullfigsize}{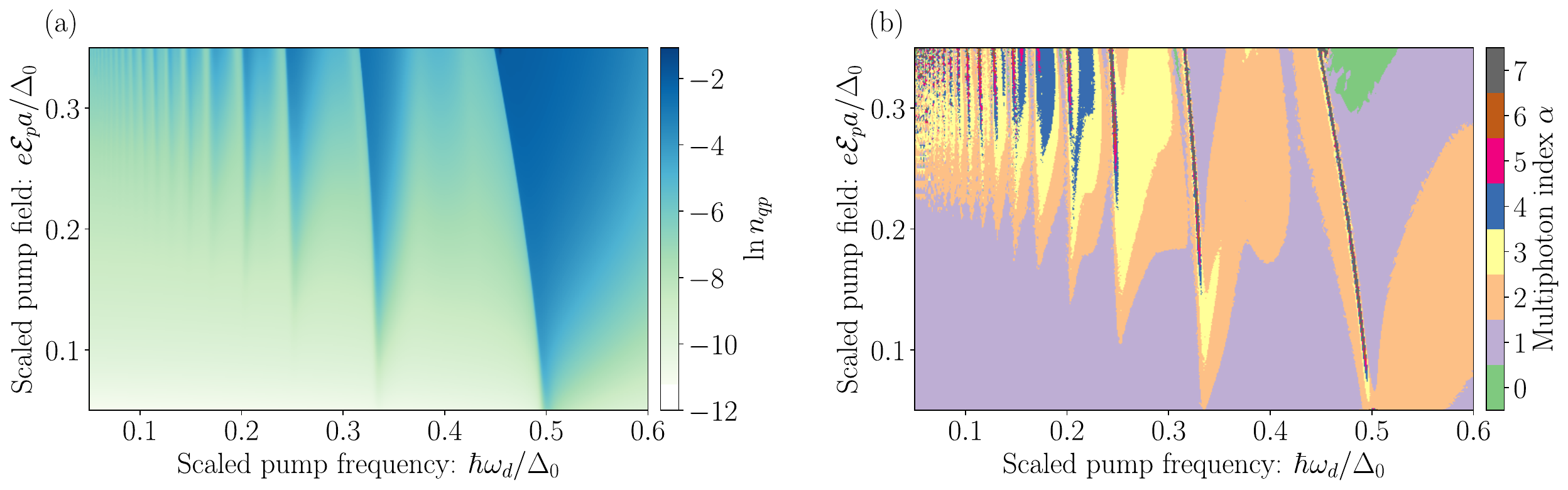}
	\caption{\textbf{(a)} Photoexcited carrier density obtained from numerically solving evolution equations for a single-band 2D Hubbard model, for pump duration $T_\text{pump} = 100$ fs, hopping amplitude $\tau = 189$ meV, interaction $U = 1.93$ eV, and dephasing rate $\relaxrate = 1.64$ meV. The color scale is logarithmic. We vary the pump frequency $\omega_d$ and peak electric field $\field_p$ in order to capture all four regimes sketched in Figure \ref{fig:regime_sketch_2d}. \textbf{(b)} Extracted multiphoton index $\mphindex \equiv (1/2) \; \dd(\ln n)/ \dd (\ln \field_p)$, as a function of the same pump parameters.}
	\label{fig:quasiparticles_and_multiphoton}
\end{figure*}

Start by extracting the total density of doublon-hole pairs, at times immediately following the pump, as a function of the driving frequency $\omega_d$ and peak electric field $\field_p$. Results are shown, using a logarithmic color scale, in panel (a) of Figure \ref{fig:quasiparticles_and_multiphoton}, for driving frequencies $\omega_d$ ranging from under $\Delta_0/ 10$ to above $\Delta_0/2$. To further investigate the differences between photoexcitation regimes, in panel (b) we extract across our parameter space the \textbf{multiphoton index} $\mphindex$, defined by
\begin{equation} \label{eq:multiphoton_index_definition}
    \mphindex \equiv {1 \over 2} \frac{\mathrm{d} (\ln n_\text{qp})}{\mathrm{d} (\ln \field_p)}.
\end{equation}
This is expected to be 1 in the incoherent regime, cf. eq. \eqref{eq:maintext_analytic_decoherence_qp}, while in the multiphoton case it should give the corresponding order $n = \Delta / \omega_d$, as evidenced by eq. \eqref{eq:maintext_multiphoton_production_rate}. On the other hand, when the tunneling mechanism \eqref{eq:maintext_tunneling_production_rate} dominates, $\mphindex$ will not have a simple expression, as the production rate is not a power-law dependence on the electric field.\\

Prominently featuring in panel (a) are sharp frequency thresholds, across which the photoexcited carrier density increases abruptly. As argued in Section \ref{ssec:results_1_analytical_approximations}, these thresholds are located at $\omega_d = \Delta_0 / n$ when the peak pump field $\field_p$ is weak, and therefore we interpret them as reflecting an $n-$photon resonance condition. Upon stronger driving, the thresholds bend towards lower frequency, which can be understood as arising from renormalization of the collective gap to a lower value $\Delta_r(t) < \Delta_0$, due to the presence of doublons and holes; further discussion of this effect is presented in section \ref{ssec:results2_threshold_lowering}. Nontrivial frequency dependence of $n_\text{qp}$ continues far into the region of parameter space above the Keldysh line, $e \field_p a > \hbar \omega_d$. As mentioned previously, this is in stark contrast to previous analytical work on photoexcitation rates \cite{oka_nonlinear_doublons}, which finds a smooth frequency dependence \footnote{We remark that resonant features have been numerically found in a different context, when analyzing nonequilibrium steady states of continuously driven Mott insulators in infinite dimensions \cite{driven_steady_state}. However, we expect the extremely high field strengths considered in that case to either produce significant gap suppression, which would in turn shift down the resonance conditions, or break down the insulating state altogether. These consequences of dynamical gap renormalization seem to be absent from steady-state double occupancy plots in that study.}.\\

The Keldysh line does not show any change in the qualitative behavior of total photocarrier density, since multiphoton thresholds are present on both sides thereof. In panel (b), we focus further on electric field dependence, via the multiphoton index $\mphindex$ defined in \eqref{eq:multiphoton_index_definition}. The expectation that $\mphindex = n$ for frequencies $\omega_d \gtrsim \Delta / n$ is true above the two-photon threshold at low fields, as well as in a narrow region near the three-photon frequency, at slightly stronger drives \footnote{At first, it may seem surprising that the regionsAt first, it may seem surprising that the regions in Fig. \ref{fig:quasiparticles_and_multiphoton}b where $\mu$ matches the multiphoton order $n$ are very narrow, compared to the area marked `Multiphoton excitation' in Figure \ref{fig:regime_sketch_2d}. Part of the cause is the strongly-peaked density of states $\dos$ in our model, whereas in a real material it would be broader. On the other hand, gap renormalization effects also alter electric field dependence, and therefore reduce its utility for identifying the multiphoton mechanism. We will find that momentum distribution of photocarriers offers a better tool for this purpose.}. However, from the $4-$ and $5-$ photon frequencies downwards, the absence of a well-defined region with the correct multiphoton $\mphindex$ hints to a different functional form of the field dependence; as expected from the Keldysh crossover picture, and discussed in Section \ref{ssec:results_1_low_frequency}, a tunneling expression is appropriate for this region. Also note that, for strong fields and low multiphoton orders (high frequencies), the index $\mphindex$ is found \emph{below} the value expected from a weak-field analysis \eqref{eq:maintext_multiphoton_production_rate}. Such behavior indicates the onset of a non-perturbative regime, characterized by gap renormalization, and real-time saturation of photocarrier density (Section \ref{ssec:results_gap_suppresison}).\\

Notably, multiphoton thresholds in Figure \ref{fig:quasiparticles_and_multiphoton}a do not extend all the way down to $\field_p = 0$, but instead require stronger driving in order to be observed at low frequencies; the same tendency is visible in Figure \ref{fig:quasiparticles_and_multiphoton}b. This is schematically represented in Figure \ref{fig:regime_sketch_2d} by a `multiphoton behavior onset' line of negative slope. Such behavior can be understood as follows: below the Keldysh line $e \field_p a < \hbar \omega_d$, the highly-nonlinear multiphoton processes must compete with incoherent, off-resonant absorption. The $\field^2$ scaling of the latter mechanism offers an advantage under low pump fields; moreover, multiphoton transitions are constrained by strict resonance conditions, and will only be sensitive to the density of states close to energy $n \omega_d$. If that does not match the DOS peak at $\Delta$, the number of carriers produced by multiphoton excitation is further reduced; this explains why the incoherent region in our calculations (as estimated by simultaneous $\mphindex \approx 1$, $e \field_p a \lesssim \hbar \omega_d$, and low $n_\text{qp}$) extends to higher fields in-between the various multiphoton resonances. Note that more sophisticated modelling, such as including hopping terms beyond nearest-neighbor in the Hamiltonian \eqref{eq:hubbard_hamiltonian}, will spread out the doublon-hole DOS $\dos$. In turn, the multiphoton mechanism will occupy a larger portion of parameter space, at the expense of the incoherent regime.\\

Finally, we remark that the region in parameter space which lies below the multiphoton onset boundary (bottom-left corner in panels of Figure \ref{fig:quasiparticles_and_multiphoton}) is still bisected by the Keldysh line, but the multiphoton index $\mphindex$ offers no clarification regarding the competition of excitation pathways. This is to be expected, since $\mphindex$ was not constructed a relevant indicator of tunneling; in the next subsection we consider electric-field fitting forms more appropriate for that case.

\subsection{Low-frequency behavior, and multiphoton-tunneling cooperation} \label{ssec:results_1_low_frequency}

\begin{figure*}
	\centering	
    \scalebox{\fullfigsize}{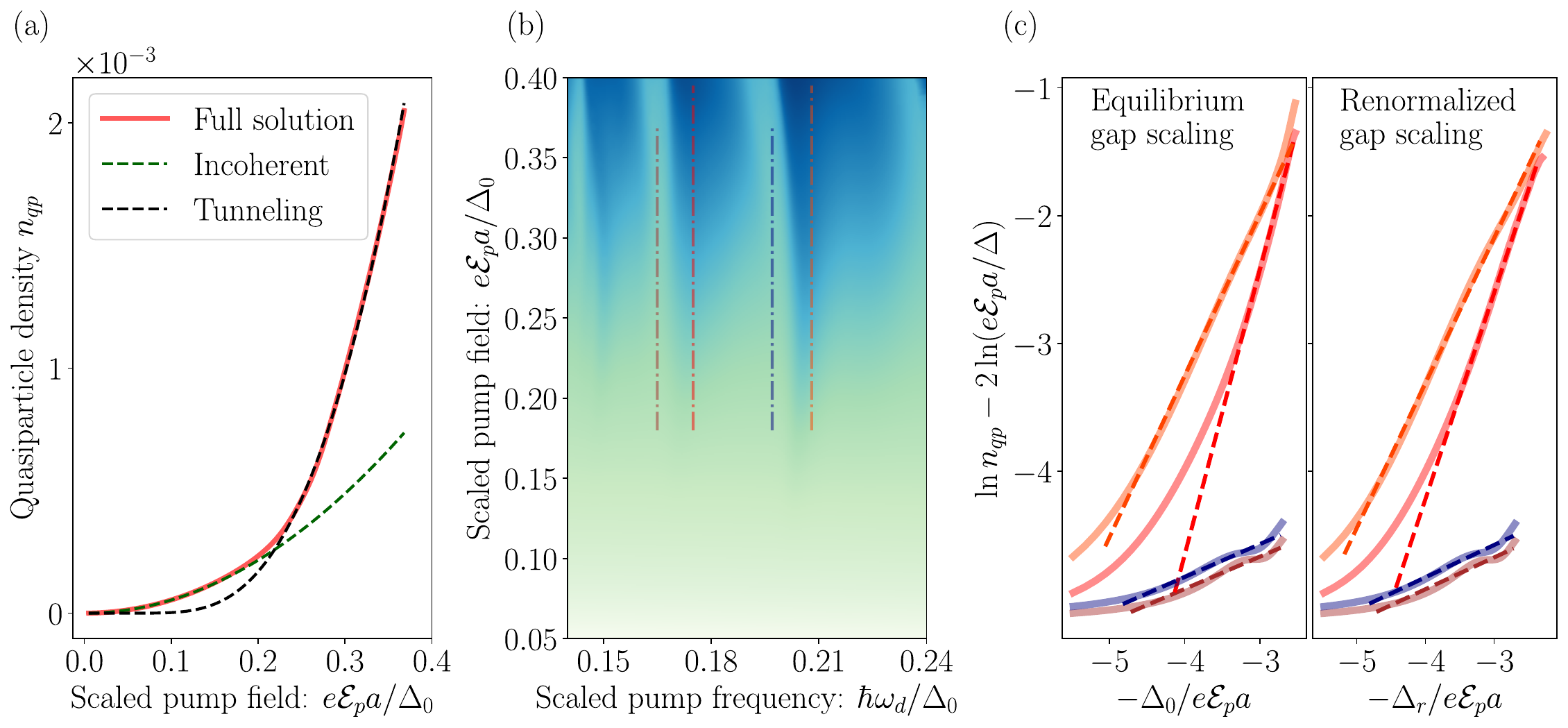}
	\caption{\textbf{(a)} Field dependence of photoexcited carrier density $n_\text{qp}$, at frequecy $\omega_d = \Delta_0 / 10$, together with incoherent and tunneling fits. There is a narrow crossover region between them, and no multiphoton dependence is visible. \textbf{(b)} Linecuts for investigating electric field dependence, below and above the $5-$ and $6-$photon resonances. \textbf{(c)} Comparison between tunneling fits, scaling by the equilibrium gap, or respectively the renormalized one. The latter option gives improvement above the resonances, where the excited quasiparticle density is large and gap renormalization is significant.}
	\label{fig:low_frequency_composite_figure}
\end{figure*}

When the driving frequency is much smaller than the gap, and the pump field is weak, the highly nonlinear multiphoton production rate \eqref{eq:maintext_multiphoton_production_rate} will be dominated by the incoherent mechanism \eqref{eq:maintext_analytic_decoherence_qp}. If the electric field $\field_p$ is increased, we may cross the Keldysh line before reaching strong enough driving for \eqref{eq:maintext_multiphoton_production_rate} to become visible. In that case, the relevant crossover will take place between incoherent and tunneling regimes, with the multiphoton one completely absent, as shown in Figure \ref{fig:low_frequency_composite_figure}a for a pulse with $\omega_d = \Delta_0 / 10$. The dashed line marked `incoherent' is a polynomial fit to the numerical solution at low fields; the resulting exponent versus $\field_p$ is indeed 2, corresponding to the index $\mphindex = 1$.\\

Increasing the pump frequency gives rise to more efficient photoexcitation, in turn allowing for a detailed study of the multiphoton-tunneling cooperation process. Focusing on a region of parameter space situated above the Keldysh line, but which nevertheless displays frequency thresholds for $n_\text{qp}$, we investigate the field-dependence of photocarrier density. As indicated in Figure \ref{fig:low_frequency_composite_figure}b, we take constant-frequency linecuts, immediately above and below multiphoton resonances, for pump fields situated around $\field_p \sim \hbar \omega_d / e a$ and above. From eqn. \eqref{eq:maintext_tunneling_production_rate} it follows that there should be a linear dependence between $\ln n_{qp} - 2 \ln (\field_0 / \Delta)$ and $- \Delta / e \field_0 a$, with slope given by $\gamma(\omega_d)$. Since we are now dealing with pulsed rather than continuous driving, we use the peak electric field of the pump $\field_p$ in place of the amplitude $\field_0$; the tunneling rate strongly increases with applied field, so the latter's maximum value should dominate the photoexcitation process.\\

Under strong driving, significant gap renormalization due to the photoexcited carriers will contradict the rigid-band assumption on which the derivation of \eqref{eq:maintext_tunneling_production_rate} is based. Therefore, it is not a priori clear that such exponential field dependence should still hold. Furthermore, even if it approximately does, the value of the gap $\Delta$ used in the scaling must be an effective one, situated in between the equilibrium and final values. Panel \ref{fig:low_frequency_composite_figure}c compares fits which use the static ($\Delta_0$), and respectively post-pump renormalized ($\Delta_r$), gaps for the scaling. Below-resonance results (brown, blue) do not visibly differ between the two normalization choices, as the gap is only weakly altered in those regions. On the other hand, at above-threshold frequencies (red, orange), there is improved agreement with the tunneling functional form, if scaling is performed using the final gap value. This not only confirms the importance of taking gap dynamics into consideration, but also suggests that most of the photoexcitation takes place after the gap has been lowered. As tunneling is controlled by exponential suppression in $\Delta / \field_0$, it is indeed expected that lowering $\Delta$ would increase the production rate.\\

Note that the sub-resonance tunneling fits in Figure \ref{fig:low_frequency_composite_figure}c both have the same slope $\gamma(\omega_d)$. As discussed in Section \ref{ssec:results_1_analytical_approximations}, such lack of frequency dependence matches the expectation derived from the conventional picture of tunneling. On the other hand, the tunneling rate $\gamma(\omega_d)$ is significantly increased just above threshold frequencies, in agreement with the calculation detailed in Section \ref{ssec:low_frequency_analytic}. Therefore, once eventual gap renormalization is taken into account, the conclusions about low-frequency behavior drawn in Section \ref{ssec:results_1_analytical_approximations} qualitatively hold true even beyond the region of validity for the rigid-band approximation.

\subsection{Momentum distribution of quasiparticles} \label{ssec:results_1_momentum_distribution}

\begin{figure*}
	\centering	
    \scalebox{\fullfigsize}{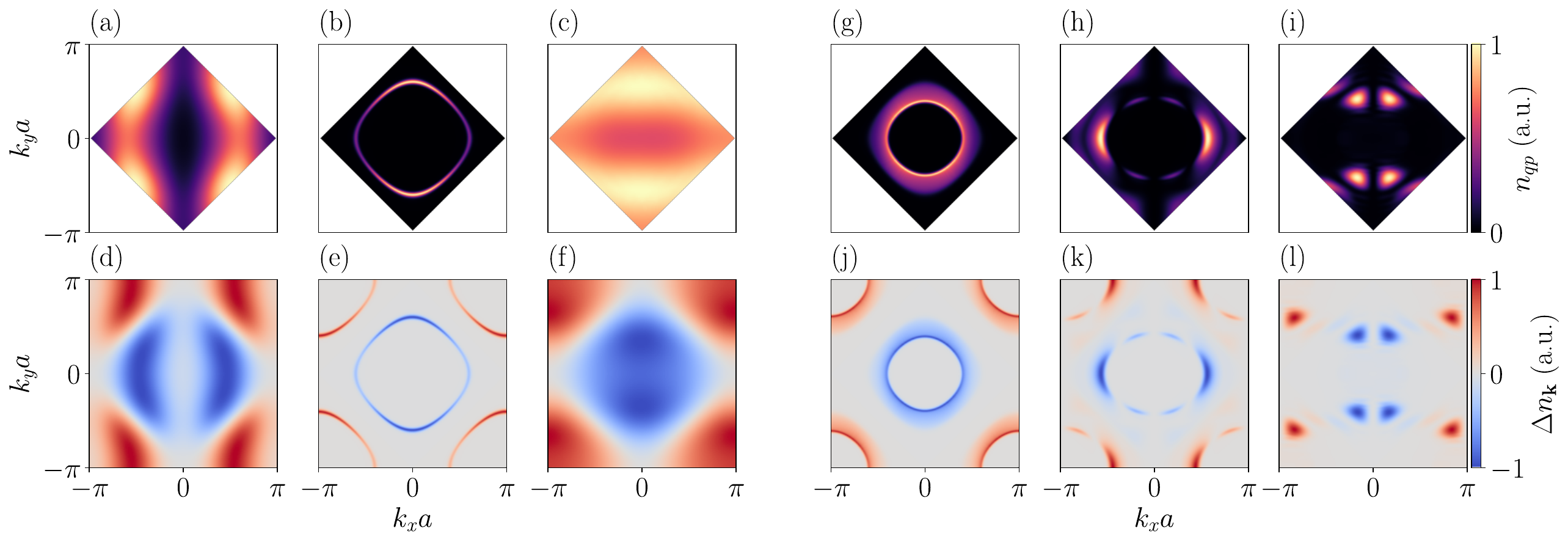}
	\caption{Momentum distribution of photoexcited quasiparticles $n_{\text{qp}, \mathbf{k}}$ (top row) and corresponding changes $\Delta n_{\mathbf k}$ in electronic occupation (bottom row), in the absence (a - f) as well as presence (g - l) of gap renormalization. The pump polarization is chosen along the $\hat x$ axis. Left panels (a - f) illustrate, under pumping with dimensionless parameters ($\hbar \omega_d / \Delta_0$, $e \field_p a / \Delta_0$), the following regimes accessible at weak driving: \textbf{(a, d)} Incoherent linear absorption - (0.075, 0.075); \textbf{(b, e)} Resonant 2-photon excitation - (0.52, 0.075); \textbf{(c, f)} Landau-Zener tunneling - (0.05, 0.17). The distributions match expectations arising from a rigid-band treatment. Right panels (g - l) explore the case of strong driving: \textbf{(g, j)} Multiphoton regime with gap suppression - (0.52, 0.25); \textbf{(h, k)} Near the Keldysh line, tunneling signatures start to appear - (0.33, 0.33); \textbf{(i, l)} Far inside the cooperative region - (0.145, 0.4).}
	\label{fig:main_text_momentum_distributions}
\end{figure*}

Having investigated the behavior of total photocarrier density versus pump parameters, we now consider their distribution in the Brillouin zone, an aspect which so far has received relatively little theoretical attention in 2D Mott insulators. It has been argued in Ref. \cite{oka_nonlinear_doublons} that for a 1D model, the momentum distribution would differ significantly between the multiphoton and tunneling pathways. In the weak-field limit, our approach yields a similar qualitative conclusion. Indeed, the multiphoton mechanism relies on a resonance condition which, when the effective Hubbard bands disperse, will only be satisfied by a particular subset of momentum points; on the other hand, tunneling can be understood as a local process in real-space, so it should give rise to a broad distribution of relative momenta for doublon-hole pairs. Experimentally, ultrafast broadband optical reflectivity measurements can offer indirect information about the momentum distribution of photocarriers; this approach has been used, in conjunction with electric-field dependence, to confirm the crossing of the Keldysh line in Ca$_2$RuO$_4$ by varying the pump fluence \cite{dave_experimental_crossover}. Other experimental methods, such as time- and angle-resolved photoemission spectroscopy (trARPES), promise to offer more detailed insight into the momentum configuration of photoexcited carriers \cite{band_build_up_original, band_build_up_nw}, as well as the dominant relaxation mechanisms they experience.\\

We begin by discussing the full momentum distribution of photocarriers, at various points in the two-dimensional $\omega_d - \field_p$ parameter space. The analytic approach presented in Section \ref{ssec:results_1_analytical_approximations}, and detailed in Appendix \ref{sec:analytic_approximate_solutions}, also gives us information about momentum-resolved occupations in the rigid-band limit. As a preliminary, recalling the quasiparticle dispersion \eqref{eq:maintext_quasiparticle_dispersion}, we see that the direct gap between the upper and lower Hubbard bands is situated on the diamond $|k_x| + |k_y| = \pi$ that defines the edge of the magnetic Brillouin zone, while the maximum energy difference is reached at $\mathbf k = 0$. Figure \ref{fig:qp_energy_contours} shows the shape of energy contours for doublon-hole pairs at zero total momentum. Under weak pumping conditions, we find the following behaviors:
\begin{itemize}
    \item In the incoherent regime, carriers are excited across most of the band, although in a nonuniform manner. Their distribution is dictated by the integrand of \eqref{eq:maintext_analytic_decoherence_qp}, where the $\hat{\field} \cdot (\nabla \e_{\mathbf k})_0$ selects only certain `stripes' in momentum space, while the denominator $E_{\mathbf k, 0}^6$ penalizes higher-energy regions close to the BZ center. The result is a concentration of carriers near momenta $(\pm \pi/2, \pm \pi/2)$, with lower density between these points. Numerical results of the full evolution are presented in Fig. \ref{fig:main_text_momentum_distributions}a,d; for comparison, the integrand of \eqref{eq:maintext_analytic_decoherence_qp} is shown in Figure \ref{fig:analytic_incoherent_momentum_distribution}.
    \item For the multiphoton mechanism, the dominant excitation region is around the energy contour on-resonance with $n$ photons. This follows from considering the resonance condition $2 E_{\mathbf k} / \omega_d \in \N$ implicit in \eqref{eq:maintext_mph_1}. An example for the 2-photon regime is shown in Fig. \ref{fig:main_text_momentum_distributions}b,e and further illustration of how tuning the pump frequency can change the selected energy contour is given in Fig. \ref{fig:low_field_frequency_tuning}.
    \item In the tunneling case, effective momentum shifts of $e \mathbf A(t)$, arising from coupling to the driving field, explore the entire width of the Brillouin zone. Momentum dependence enters eqs. \eqref{eq:tunneling_pre_result} and \eqref{eq:tunneling_result} only via the energy $\tilde{E}_{\mathbf k}$ averaged over such trajectories. The averaging procedure ensures that $\tilde{E}_{\mathbf k}$ only has weak momentum dependence, yielding a comparatively uniform photocarrier density throughout the magnetic BZ (Fig. \ref{fig:main_text_momentum_distributions}c,f).
\end{itemize}
Going beyond the weak-driving limit, two important qualitative changes occur: dynamical gap renormalization, and the emergence of multiphoton-tunneling cooperation. The right panels of Figure \ref{fig:main_text_momentum_distributions} illustrate the effects they have on the momentum distribution of carriers:
\begin{itemize}
    \item In the multiphoton region of parameter space, strong pumping will start by exciting carriers along an energy contour, in the same manner as before. However, as gap suppression starts to take effect but the driving frequency is unchanged, the resonant contour shifts higher within the band, i.e. inward to $\mathbf k = 0$. The resulting momentum distribution of photocarriers traces out the contour's trajectory, as illustrated in panels (g, j). 
    \item Near the Keldysh line but above the multiphoton onset one, momentum distributions such as the one shown in panels (h, k) combine features of both conventional excitation pathways. A slightly distorted energy contour is still visible, and the carrier density is negligible at energies above it. However, at lower energies (exterior of the contour in panel h), there is a broader carrier spread in momentum, as would be expected from a tunneling process.
    \item Within the cooperative region, the distribution displays sharp features in momentum space, as exemplified in panels (i, l). Given the low driving frequencies $\omega_d$ needed to reach this regime, several multiphoton resonance conditions can be simultaneously satisfied due to the finite quasiparticle bandwidth. At the same time, large values of the effective momentum shift $e \mathbf A(t)$ give rise to a complex pattern of interference between tunneling events. The result is that photoexcitation becomes concentrated in narrower regions of the BZ, strongly deviating from the weak-field tunneling results of panels (c, f).
\end{itemize}
By focusing on other pump parameters, analysis of momentum distributions can yield further insight regarding the strength of gap renormalization, or competition between photoexcitation mechanisms near the boundaries that separate them in parameter space. Further discussion is presented in Appendix \ref{sec:appendix_momentum_distribution}.\\
\begin{figure}
	\centering	
    \scalebox{\fullfigsizecb}{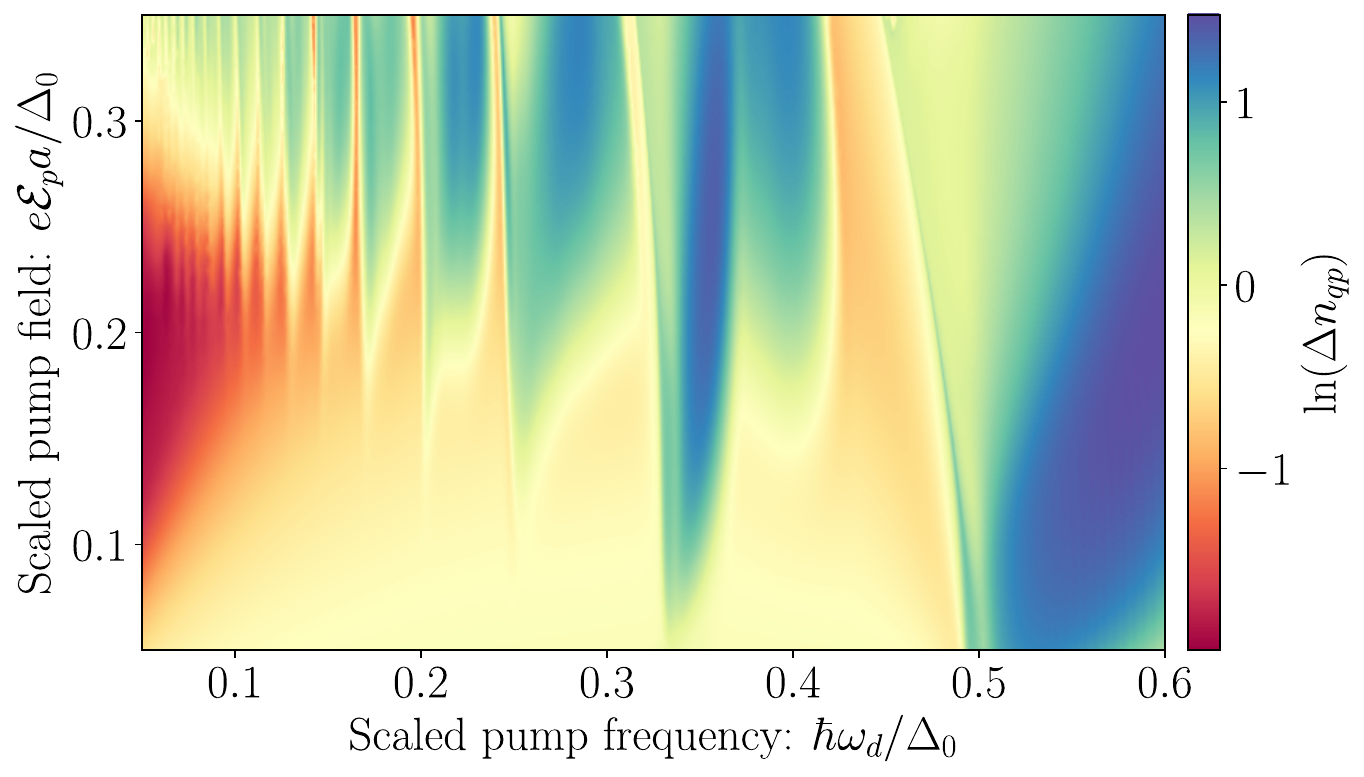}
	\caption{Normalized deviation in photocarrier density across the Brillouin zone, as a function of pump parameters; color scale is logarithmic. All four photoexcitation regimes depicted in Figure \ref{fig:regime_sketch_2d}, as well as their approximate boundaries, can be observed: on the left side, a broad momentum distribution (red) points to a Landau-Zener tunneling mechanism. To the right, sharp features in reciprocal space (blue), situated at $\omega_d = \Delta / n$ and above, signal a multiphoton pathway. Intermediate $\mathbf k-$spread at low field strength (yellow) marks the incoherent area. Finally, the presence of multiphoton thresholds above the Keldysh line, accompanied by increasingly wider momentum distributions (green, top left), reveals the cooperative region.}
	\label{fig:qp_momentum_variance}
\end{figure}
To emphasize the great potential of momentum-resolved carrier density for distinguishing photoexcitation regimes, we show in Figure \ref{fig:qp_momentum_variance} the normalized deviation $\Delta n_\text{qp}$ in doublon-hole pair occupation, across the magnetic Brillouin zone BZ', as a function of the pump parameters $\omega_d$ and $\field_p$:
\begin{subequations}
    \begin{align}
    \Delta n_\text{qp} & \equiv {1 \over \overline{n}_\text{qp}} \sqrt{{1 \over N_\text{BZ'}} \sum_{\substack{\mathbf k \in \text{BZ'} }} \oa n_\text{qp}(\mathbf k) - \overline{n}_\text{qp} \ca^2}\\
    \overline{n}_\text{qp} & \equiv {1 \over N_\text{BZ'}} \sum_{\substack{\mathbf k \in \text{BZ'} }} n_\text{qp}(\mathbf k).
\end{align}
\end{subequations}
In this way, quantitative distinction can be made between the four photoexcitation regimes schematically depicted in Figure \ref{fig:regime_sketch_2d}. The onset of multiphoton behavior can be estimated by abrupt increases in $\Delta n_\text{qp}$, and the relevant boundary in $\field_p - \omega_d$ space is seen to have a negative slope. Notably, the incoherent/tunneling regime boundary, which was invisible when considering total photocarrier density (Figure \ref{fig:quasiparticles_and_multiphoton}), is clearly observed in Figure \ref{fig:qp_momentum_variance}. The visibility of regime boundaries, as well as of sharp frequency thresholds corresponding to multiphoton resonance conditions, makes $\Delta n_\text{qp}$ an extremely valuable metric for understanding and characterizing subgap photoexcitation in MI.\\

\section{Quasiparticle dynamics, and feedback effects of gap suppression} \label{sec:results_2_real_time_dynamics}

\subsection{Real-time carrier density and micromotion} \label{ssec:real_time_stroboscopic_micromotion}

In the case of weak and off-resonant driving, few or no photocarriers will be present after the pump ends. This is to be expected from the results of Section \ref{ssec:results_1_analytical_approximations}, and can also be formulated within the Floquet viewpoint, as follows: sidebands will appear while the drive is active, but in the absence of a multiphoton resonance they are far detuned. Weak Floquet dressing of the upper and lower Hubbard bands will occur, but the two are still separated by a considerable gap. Therefore, after the driving is over, we expect the system to return close to its ground state: quantities such as $\ex{d_{\mathbf k}^\dag d_{\mathbf k}}$ and $\ex{h_{-\mathbf k}^\dag h_{-\mathbf k}}$ will be negligible at late times. However, this does not directly imply that these expectations will vanish at every moment throughout the evolution, as there may be nontrivial micromotion.\\

Following the discussion in Section \ref{ssec:static_solution_variational}, the definitions of doublon and hole operators $d_{\mathbf k}, h_{-\mathbf k}$ change with time, since they are momentum-dependent, and $\mathbf k$ effectively acquires a time-dependent shift through the Peierls substitution. Therefore, some transient occupation of the doublon-hole pair states will be generated even for off-resonant pumping. After the drive stops, operators describing elementary excitations return to their original definitions; if the density matrix in the original electronic basis has not changed significantly, we are left with no photocarriers. The micromotion is physically relevant, as a probe will couple to the transient carriers, affecting observables such as the optical conductivity; in experimental measurements such as ultrafast reflectivity changes, this may be observed as a coherent pump-probe artifact.\\

\begin{figure}
	\centering	
    \scalebox{\fullfigsize}{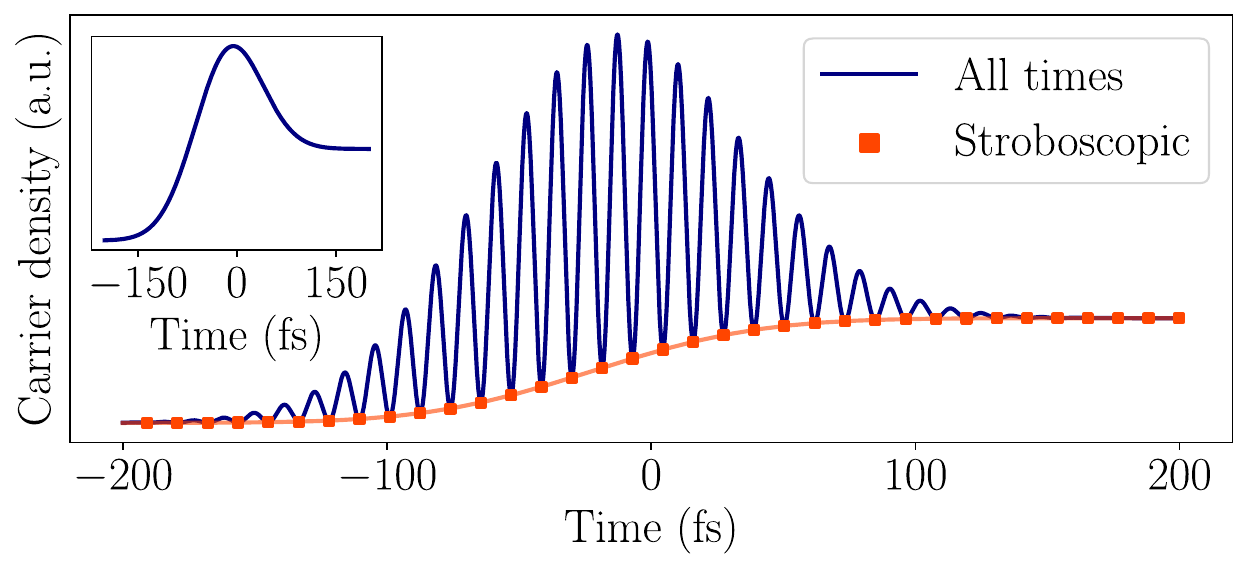}
	\caption{Typical time dependence of total carrier density $n_\text{qp}(t)$ under a weak, low frequency drive described by ($\hbar \omega_d / \Delta_0$, $e \field_p a / \Delta_0$) = (0.1, 0.05). Values at stroboscopic times $t_n$, defined via $\boldsymbol{\field}(t_n) = 0$, are denoted by orange squares; the smooth line connecting them serves as a guide to the eye. Time zero is taken to be the center of the pump pulse. \textbf{Inset:} the same carrier density $n_\text{qp}(t)$, convolved with a probing profile of duration $T_\text{pump}$, which is too wide to resolve subcycle oscillations. The coherent peak during the presence of the pump can be attributed to micromotion.}
	\label{fig_realtime_illustration}
\end{figure}

In order to understand the evolution in real-time, it is natural to separate the stroboscopic behavior, by evaluating the system state when the external driving field crosses zero. This will contain the slower dynamics, such as `permanent' carrier excitation throughout the pump duration $T_\text{pump}$, while ignoring details on subcycle timescales. On the other hand, the remaining micromotion will offer information about pump-probe coherent artifacts, high-harmonic generation, etc. The typical weak-driving behavior of quasiparticle density $n_\text{qp}(t)$, as exemplified in Figure \ref{fig_realtime_illustration}, consists of a smooth variation at stroboscopic times, upon which an oscillatory component is superimposed, with frequency $2 \omega_d$ and envelope following that of the pump. The driving parameters chosen for this example place it in the incoherent linear absorption regime, and the stroboscopic carrier density $n_\text{qp}(t_n)$ indeed follows the integral of the pump envelope, as would also follow from the arguments of Section \ref{ssec:results_1_analytical_approximations}. Eliminating this excitation mechanism, by letting $\relaxrate \to 0$, would take all $n_\text{qp}(t_n) \to 0$ without considerably affecting the micromotion. Therefore, by varying material and/or pump parameters, the relative amplitudes of stroboscopic- versus micro-motion can be significantly modified.\\

Ultrafast optical experiments, such as measurements of transient reflectivity, typically employ probe pulses with duration similar to that of the pump. As this is significantly longer than $\omega_d^{-1}$, there is no way to directly observe subcycle dynamics in such setups. However, as shown in the inset of Figure \ref{fig_realtime_illustration}, the nonzero average value of the micromotion component in $n_\text{qp}(t)$ gives rise to a coherent peak in the signal. The relative strength of this feature, compared to the measured value after the end of the pump, can therefore offer information about the relevance of micromotion for the system evolution. Throughout the rest of this section, we analyze the signatures that gap renormalization is expected to produce in real-time, when pumping in the vicinity of a multiphoton resonance.

\subsection{Effects of gap suppression on high-frequency driving} \label{ssec:results_gap_suppresison}

The effects of moderately strong driving at low frequency have been discussed in Section \ref{ssec:results_1_low_frequency}. As shown in Figure \ref{fig:low_frequency_composite_figure}, the main consequence of gap suppression is an enhancement of the tunneling rate under high electric fields $\field_p$. In principle, renormalization of the gap should also affect multiphoton resonance conditions $\Delta = n \omega_d$, and indeed weak bending of frequency thresholds is visible in panel \ref{fig:low_frequency_composite_figure}b. However, this is generally a weak effect for small $\omega_d$, because of two reasons: first, photoexcitation efficiency in this region of parameter space is low, so the absolute change of the gap $\Delta(t) - \Delta_0$ is correspondingly small. Secondly, the frequencies that need to be compared for a multiphoton condition are $\omega_d$ and $\Delta / n$; at larger orders $n$, any given renormalization of the gap $\Delta(t) - \Delta_0$ will produce a smaller effect towards altering the frequency thresholds.\\

On the other hand, when moving towards higher frequencies, i.e. lower multiphoton orders, the effects of gap renormalization on resonance conditions are expected to be more dramatic. We will analyze two such situations: starting with the pump at resonance and being moved away from it, which will give rise to saturation of the excitation density; and the opposite case of being taken into resonance by the gap renormalization, which will manifest itself as a lowering of the multiphoton frequency threshold.

\begin{figure*}
	\centering	
    \scalebox{\cartoonfigsize}{\includegraphics{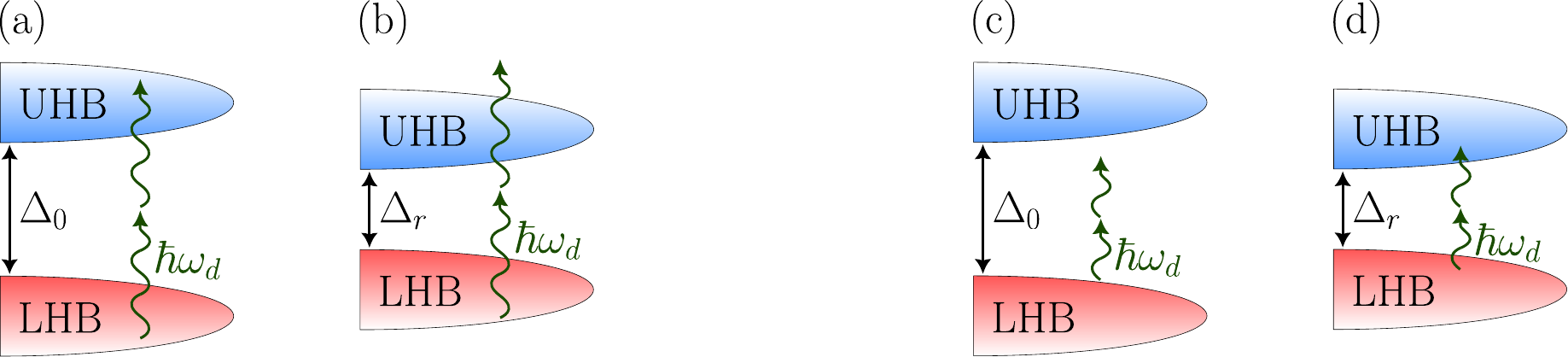}}
	\caption{Effects of gap renormalization on multiphoton frequency conditions. Left: pumping the system on-resonance with a multiphoton transition \textbf{(a)} will photoexcite carriers, which in turn suppress the gap. This way, the pump can be taken out of resonance \textbf{(b)}, inhibiting further photogeneration of carriers. Right: driving just below a multiphoton threshold \textbf{(c)} may still give rise to nonzero doublon-hole pair occupation, due to micromotion. If the gap suppression this produces is enough to overcome the detuning \textbf{(d)}, the pump is brought into resonance and carrier production accelerates.}
	\label{fig:gap_suppression_cartoons}
\end{figure*}

\subsubsection{Moving away from resonance: photocarrier saturation}

Since gap renormalization is particularly strong when a considerable density of photocarriers is generated, we begin by considering the case of driving on resonance with a multiphoton transition. A typical situation is sketched in the left panels of Figure \ref{fig:gap_suppression_cartoons}: under a strong enough applied field, the multiphoton mechanism will initially give a moderate generation rate for doublon-hole pairs. Afterwards, the presence of these carriers will reduce the gap to some renormalized value $\Delta_r(t)$; the upper Hubbard band eventually moves down, exiting resonance with the initial transition. The corresponding production rate will significantly decrease, and therefore the density of photoexcited carriers saturates in real time. Note that such a mechanism fully works only when the sum of bandwidths for the upper and lower Hubbard bands is less than $\omega_d$, i.e. driving is realized at high enough frequency; otherwise, we would also successively enable lower-order multiphoton transitions upon suppressing the gap. However, if the density of states is strongly nonuniform within each band, it will suffice to bring the high-DOS regions out of resonance, and carrier production will still decrease dramatically.\\

\begin{figure*}
	\centering	
    \scalebox{\fullfigsize}{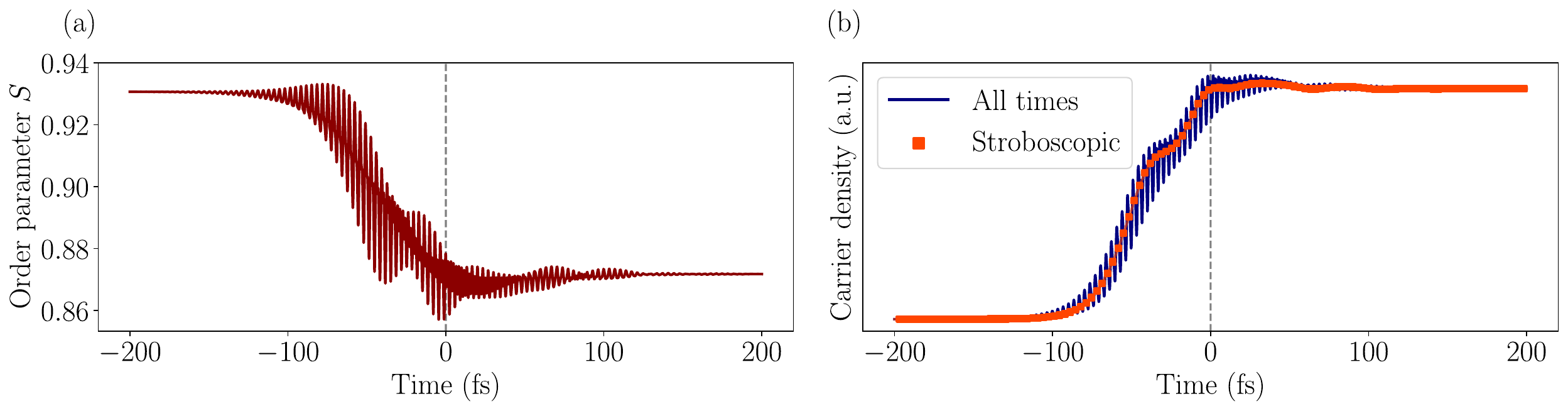}
	\caption{Real-time illustration of photocarrier saturation, due to exiting a resonance region. Driving parameters are ($\hbar \omega_d / \Delta_0$, $e \field_p a / \Delta_0$) = (0.34, 0.3), starting on-resonance with a 3-photon transition. \textbf{(a)} Suppression of the gap order parameter $S = \Delta / U$. \textbf{(b)} Carrier density $n_\text{qp}(t)$, with stroboscopic values highlighted. Notably, at time $t = 0$ (dashed gray lines), the driving field is strongest, but changes in both $S(t)$ and $n_\text{qp}(t)$ have mostly stopped.}
	\label{fig:resonance_exiting_illustration}
\end{figure*}

We illustrate the resonance-exiting process in Figure \ref{fig:resonance_exiting_illustration}, by investigating the real-time behavior of the order parameter $S(t)$, as well as the total carrier density $n_\text{qp}(t)$, under strong driving near a multiphoton threshold. The pump frequency $\omega_d$ is selected such that the DOS peaks in the Hubbard bands are initially on resonance with a 3-photon transition, and the electric field $\field_p$ is taken at the higher end of the range plotted in Figure \ref{fig:quasiparticles_and_multiphoton}. As the gap is continuously reduced throughout the first half of the pump pulse, the photoexcitation rate slows down even though the electric field is increasing. When driving is strongest, at $t=0$, the DOS peaks have moved so far out of resonance that carrier generation practically stops, and saturation is achieved. It follows that a single, field-dependent production rate such as \eqref{eq:maintext_multiphoton_production_rate} fails to describe the evolution, and full consideration of real-time dynamics is instead necessary.

\subsubsection{Moving towards resonance: threshold lowering} \label{ssec:results2_threshold_lowering}
We now turn to investigating the complementary process, namely entering a multiphoton resonance through decreasing the gap. As sketched in the right panels of Figure \ref{fig:gap_suppression_cartoons}, we consider pumping at a frequency $\omega_d$ whose value is just below $\Delta_0 / n$. Gap renormalization down to $\Delta_r(t)$ will reduce the detuning, and under strong enough fields can eliminate it completely. The final result is efficient photoexcitation even for pump frequencies slightly below the static threshold $\Delta_0 / n$, which effectively manifests as a redshift of the multiphoton edge for strong drives, as seen in Figure \ref{fig:quasiparticles_and_multiphoton}a.\\ 

\begin{figure*}
	\centering	
    \scalebox{\fullfigsize}{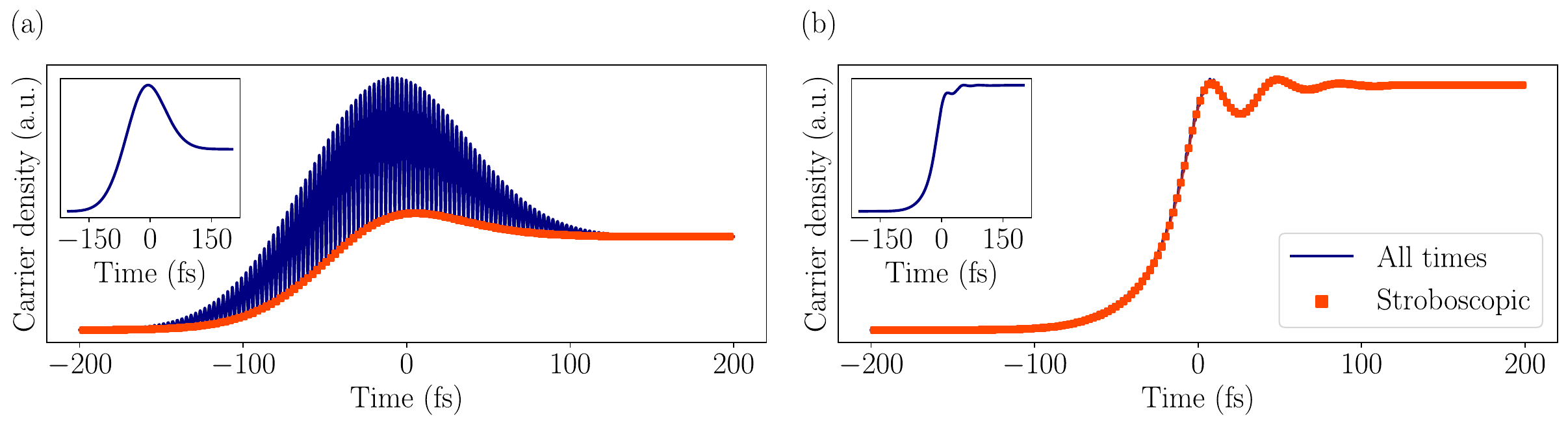}
	\caption{Effective lowering of a multiphoton threshold, due to gap renormalization. Real-time photocarrier density $n_\text{qp}(t)$ is shown, under driving at a moderate field $e \field_p a / \Delta_0 = 0.15$, for two frequencies below $\Delta_0 / 2$, which sets the static 2-photon resonance: \textbf{(a)} $\hbar \omega_d / \Delta_0 = 0.41$ is also below the \emph{reduced} threshold visible in Figure \ref{fig:quasiparticles_and_multiphoton}a, while \textbf{(b)} $\hbar \omega_d / \Delta_0 = 0.49$ lies above it. Strong micromotion is visible in (a), as also indicated by the coherent peak of the probe-convolved signal shown in the inset. In contrast, (b) is dominated by permanent photocarrier generation, and optical probes are expected to only see an abrupt change, without coherent features.}
	\label{fig:resonance_entering_illustration}
\end{figure*}

It may be counterintuitive that by starting with a below-resonance pump, whose photoexcitation rate is expected to be very low, there should be any initial gap reduction at all. However, the transient occupation of the doublon-hole states (micromotion) discussed in Section \ref{ssec:real_time_stroboscopic_micromotion}, which dominates during off-resonant pumping, plays an important role in this case. As seen in Figure \ref{fig_realtime_illustration}, as well as panel \ref{fig:resonance_entering_illustration}a below, carrier density at intermediate times can be significantly higher than the stroboscopic value. This imposes a transient reduction of the gap via self-consistency, and if the multiphoton detuning is small enough, resonance can be achieved through this mechanism. Then, the usual multiphoton processes take over, sharply increasing the carrier density beyond what can be achieved with micromotion only (panel \ref{fig:resonance_entering_illustration}b). From the perspective of total $n_\text{qp}$ produced, the process looks resonant, even if driving was realized at frequency slightly below $\Delta_0 / n$; an effective lowering of the multiphoton threshold has been achieved.\\

By comparing the insets in panels (a) and (b) of Figure \ref{fig:resonance_entering_illustration}, we observe that micromotion tends to play a more important role under far-detuned driving, as opposed to the resonant case. This does not necessarily imply a smaller absolute amplitude for subcycle dynamics of $n_\text{qp}(t)$ in the latter case, but rather reflects the significantly greater density of `permanent' photocarriers produced. We also remark that, once the order parameter is suppressed enough for the pump to go into resonance, carrier production accelerates; this should yield even further gap reduction, and consequently leave the resonance regime which had just been achieved. Indeed, Figure \ref{fig:resonance_entering_illustration}b shows practical carrier saturation shortly before the peak in driving field at around $t = 0$, although further oscillations are visible afterwards due to the strongly non-equilibrium nature of the resulting carrier distribution.\\

Because renormalization of the collective gap can dramatically affect photoexcitation rates over the course of driving, it follows that a full real-time calculation of the dynamics is necessary to understand the system's behavior under general pumping parameters. This way, one may recover within a single framework the case of almost-constant production rate (as calculated analytically in Section \ref{ssec:results_1_analytical_approximations} for weak drive), various quasi-steady states (such as the one reached after exiting a multiphoton resonance condition, Fig. \ref{fig:resonance_exiting_illustration}), as well as intermediate regimes between these two limits. From an experimental point of view, the pump pulse shape and duration will control the crossover between these regimes, and are therefore valuable tuning knobs for exploring the landscape of photoexcited states.\\

\section{Acoustic phonons as a probe of electron kinetic energy} \label{sec:acoustic_phonons}

In this part, we consider the generation of coherent longitudinal acoustic phonons as a consequence of carrier photoexcitation, on timescales much longer than the pump duration or electronic dynamics. Although this phenomenon has been experimentally observed some time ago \cite{reflected_phonons_thin_film, long_lived_oscillations_thin_film}, microscopic connection to the photocarrier density produced by the drive has been lacking. Here, we consider the SSH form of electron-phonon coupling, relevant for the cuprate parent compounds since the coupling strength depends nontrivially on the momentum transferred between electron and phonon degrees of freedom \cite{phonons_in_cuprates, rosch_electron_phonon_2004, devereaux_neutron_1999}. We show that, within this model, the acoustic phonon amplitude probes the additional kinetic energy of electrons in the metastable state reached after pumping. When the total photocarrier density $n_\text{qp}$ is low, the change in kinetic energy is directly proportional to it, and it follows that acoustic phonons can be used as an indirect probe of $n_\text{qp}$. On the other hand, dispersion details matter at strong photoexcitation, and the dependence of phonon amplitude on carrier density deviates from linearity. Nevertheless, we argue that phonons provide a valuable way of probing $n_\text{qp}$, complementary to usual methods such as ultrafast reflectivity measurements.\\

Electronic dynamics after the pump will be dominated by relaxation of the nonequilibrium photocarrier distribution. For large-gap insulators, in the absence of an appropriate mechanism to carry away energies on the scale of $\gap$, recombination of doublon-hole pairs will be strongly suppressed \cite{long_lived_doublons_exp, long_lived_doublons_th, phonon_enhanced_doublon_recombination}. If other low-energy excitations are present, such as magnons and phonons, the doublon-hole pairs will undergo intraband relaxation and reach a nonthermal metastable state, which will persist until the much slower recombination processes take effect (also see Fig. 2c of Ref. \cite{photoinduced_nonequilibrium_review}). We will on the one hand assume the metastable state is reached quickly compared to phonon timescales, and on the other completely ignore recombination, which amounts to working with the following separation of timescales between pump duration, intraband relaxation processes, acoustic phonons, and doublon-hole pair recombination respectively: 
\begin{equation} \label{eq:timescale_separation}
        T_\text{pump} \ll \tau_\text{intraband rel.} \ll \tau_\text{ac. ph.} \ll \tau_\text{recomb.}.
    \end{equation}
To investigate the generation of coherent acoustic phonons, we aim to approximate the electronic state on the intermediate timescale $\tau_\text{acoustic phonons}$. Start by calculating the total QP density $n_\text{qp} = \int_\text{BZ'} (1 - \rho_{\mathbf k}^z) \; d^2 \mathbf k / 2 \pi^2$ immediately after pumping. We then take the total number of photocarriers to remain constant, but their distribution within the respective bands to be set by the lattice temperature (i.e. thermal distributions, with distinct chemical potentials, for the two effective bands). Note that the carrier distribution and the associated order parameter need to be computed self-consistently, as they are interdependent: $S$ does not only depend on the total quasiparticle number, but on their arrangement as well, as evidenced in eqn. \eqref{eq:order_parameter_bog_basis}. Any further electronic dynamics in this intermediate-time regime are ignored: we replace the electronic degrees of freedom by their expectation value, to extract an effective Hamiltonian for the acoustic phonons.\\

The SSH Hamiltonian \footnote{Electron-phonon coupling has not been explicitly considered in previous sections because acoustic phonon energies are too low to significantly alter the high-energy photoexcitaton process. Optical phonons may contribute to $\relaxrate$, an argument which is sketched in Section \ref{ssec:phonon_scattering}.}, which describes electron-phonon coupling, will separate into $\hat x$ and $\hat y$ components for a 2D system: 
\begin{widetext}
    \begin{equation} \label{eq:ssh_hamiltonian}
        H_\text{SSH} = - g \sum_{\mathbf j, \sigma} \os \oa c_{\mathbf j + \hat{\mathbf x}, \sigma}^\dag c_{\mathbf j, \sigma} + \text{h.c.} \ca \oa \mathbf x_{\mathbf j + \hat{\mathbf x}} - \mathbf x_{\mathbf j} \ca + \oa c_{\mathbf j + \hat{\mathbf y}, \sigma}^\dag c_{\mathbf j, \sigma} + \text{h.c.} \ca \oa \mathbf y_{\mathbf j + \hat{\mathbf y}} - \mathbf y_{\mathbf j} \ca \cs,
    \end{equation}
where $\mathbf x_{\mathbf j}, \mathbf y_{\mathbf j}$ are the phonon coordinates (displacements) at site $\mathbf j$, and $g$ is a coupling constant. Focus only on the $\mathbf x$ degrees of freedom, for simplicity. Replacing the electronic degrees of freedom by their expectation value gives a displacive contribution in the effective phonon Hamiltonian:
\begin{equation} \label{eq:main_intro_phonon_effective_hamiltonian}
        H_\text{ph,\textbf{x}}^\text{eff} = \sum_{\mathbf j} \os {\mathbf{p}_{\mathbf j}^2 \over 2M} + \rho \oa \mathbf x_{\mathbf j + \hat{\mathbf x}} - \mathbf x_{\mathbf j} \ca^2 \cs - g \sum_{\mathbf j, \sigma} \ex {c_{\mathbf j + \hat{\mathbf x}, \sigma}^\dag c_{\mathbf j, \sigma} + \text{h.c.}} \oa \mathbf x_{\mathbf j + \hat{\mathbf x}} - \mathbf x_{\mathbf j} \ca,
    \end{equation}
\end{widetext}
where the first two terms comprise the free phonon Hamiltonian. The separation of timescales \eqref{eq:timescale_separation} implies that electronic expectations in the last term change fast enough, relative to acoustic phonon scales, to be approximated as a quench. Spatial nonuniformity of the displacive term is crucial, since the SSH coupling vanishes at zero momentum. Optical absorption in the material will ensure that the effective drive strength decreases with depth inside the sample, naturally giving rise to such a nonuniformity. We employ a local-density approximation to calculate the expectation of kinetic energy in regions of the sample subject to different drive strengths, which yields in each region
\begin{equation} \label{eq:kinetic_energy_expectation}
    \ex{E_\text{kin}} = - \int_\text{BZ'} {\e_{\mathbf k}^2 \over E_{\mathbf k}}  \ex{\rho_{\mathbf k}^z}  {d^2 \mathbf k \over 2 \pi^2}.
\end{equation}
Here, $\ex{\rho_{\mathbf k}^z}$ reflects the occupation of doublon-hole pairs at momenta $(\mathbf k, - \mathbf k)$, which can be calculated for any driving strength, as discussed previously. Assuming a spatial profile of the driving field which decays exponentially with depth inside the sample, we also obtain a nonuniform profile of the displacement in \eqref{eq:main_intro_phonon_effective_hamiltonian}. This will produce a phonon wavepacket, containing a range of momenta, which will propagate inwards through the material.\\

\begin{figure*}
	\centering	
    \scalebox{\fullfigsize}{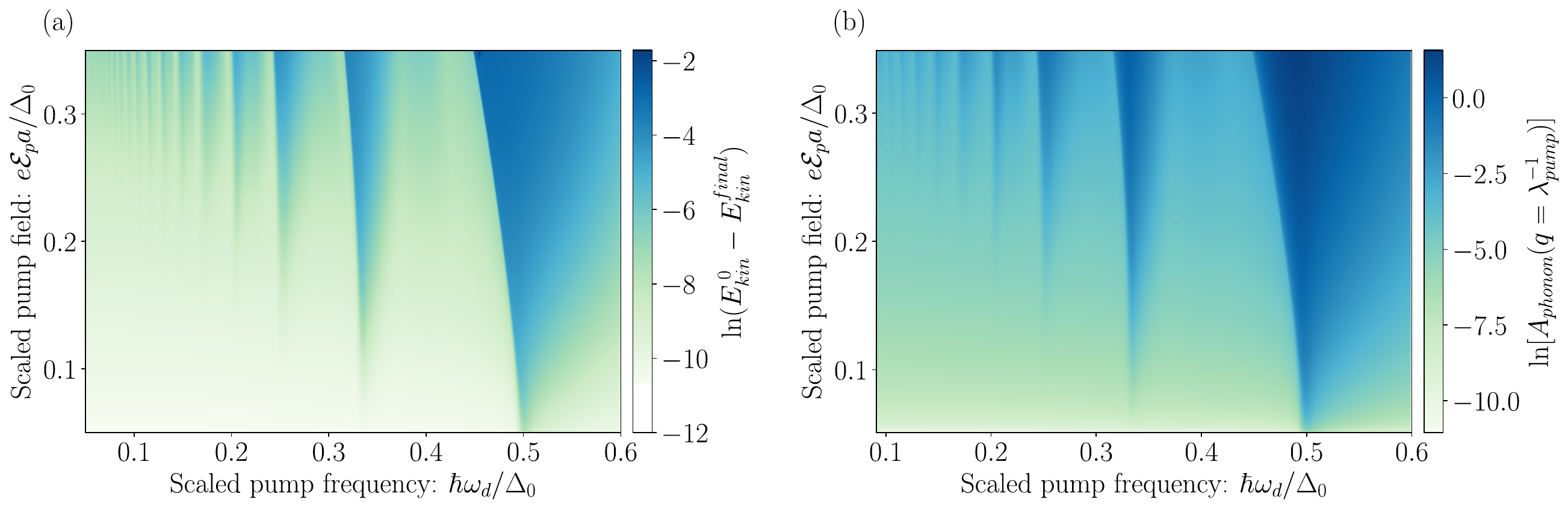}
	\caption{\textbf{(a)} Reduction in local kinetic energy density upon pumping. \textbf{(b)} Amplitude of induced phonon displacement at wavevector $q = \lambda_\text{pump}^{-1}$. Color scales in both panels are logarithmic, and the results closely resemble those for total photocarrier density in Figure \ref{fig:quasiparticles_and_multiphoton}a.}
	\label{fig:kinetic_energy_and_phonon}
\end{figure*}

When few photocarriers are produced, they will mostly settle at the bottom (top) of the upper (lower) Hubbard band; changes in $\ex{E_\text{kin}}$ mainly arise from the $\ex{\rho_{\mathbf k}^z}$ factor, i.e. are set by $n_\text{qp}$ itself. On the other hand, if strong driving produces a high density of doublon-hole pairs, the appearance of different $\e_{\mathbf k}$ and $E_{\mathbf k}$ in \eqref{eq:kinetic_energy_expectation} will modify this simple relationship. Numerical results are plotted in Figure \ref{fig:kinetic_energy_and_phonon}a, which has a qualitatively similar aspect to Fig. \ref{fig:quasiparticles_and_multiphoton}a; therefore, the effects arising from details of the band structure only make a secondary contribution, throughout most of the considered parameter space. Note that doublon-hole production will lead to a decrease in kinetic energy density $E_\text{kin}^\text{final} - E_\text{kin}^0 < 0$. This is in accordance with the intuition that the ground state mostly consists of localized electrons at every site, while doublons and holes will tend to delocalize once produced. Such tendency will also frustrate antiferromagnetism, which is reflected in the suppression of the N\'eel order parameter $S$.\\

Having obtained the kinetic energy density for electrons, as a function of pump parameters, we can turn our focus to the effect it has on acoustic phonons. Rewrite \eqref{eq:main_intro_phonon_effective_hamiltonian} as
\begin{widetext}
  \begin{align} 
    H_\text{ph,\textbf{x}}^\text{eff} &= \sum_{\mathbf j} \os {\mathbf{p}_{\mathbf j}^2 \over 2M} + \rho \oa \mathbf x_{\mathbf j + \hat{\mathbf x}} - \mathbf x_{\mathbf j} \ca^2 \cs - g \sum_{\mathbf j, \sigma} \ex {c_{\mathbf j + \hat{\mathbf x}, \sigma}^\dag c_{\mathbf j, \sigma} + \text{h.c.}} \oa \mathbf x_{\mathbf j + \hat{\mathbf x}} - \mathbf x_{\mathbf j} \ca \nonumber\\
    &= \sum_{\mathbf q} \os \omega_\text{ph} (\mathbf q) \; b_{\hat{\mathbf x}, \; \mathbf q}^\dag b_{\hat{\mathbf x}, \; \mathbf q} - g_1  \op b_{\hat{\mathbf x}, \; \mathbf q}^\dag \; \sin {q_x \over 2} \; \os \sum_{\mathbf j,  \spin}  \ex{ c_{\mathbf j + \hat{\mathbf x}}^\dag c_{\mathbf j} + c_{\mathbf j}^\dag c_{\mathbf j + \hat{\mathbf x}} } e^{-i \mathbf q (\mathbf j + \hat{\mathbf x} / 2)} \cs + \text{h.c.} \cp \cs, \label{eq:main_ph_res_phonon_effective_hamiltonian}
\end{align}  
\end{widetext}
where we only considered the $\hat {\mathbf x}$ polarized phonons for simplicity, and $\omega_\text{ph} (\mathbf q) = c q$ for acoustic phonons. Coupling to the electronic degrees of freedom acts as a displacement of the phonon equilibrium at every momentum, with amplitude characterized by the Fourier transform of the kinetic energy:
\begin{equation} \label{eq:phonon_displacement_amplitude}
    A_\text{phonon}(q) \propto \int d \depth  \ex{E_\text{kin}^\text{final}}_{\depth} e^{-i q \depth}.
\end{equation}
This displacement corresponds to the \emph{deformation potential} mechanism for coherent acoustic phonon generation, discussed phenomenologically in Ref. \cite{phonon_mechanism_review_macroscopic}: modifications of the electronic state alter the equilibrium distance between neighboring ions, thereby leading to the emission of acoustic phonons as the lattice spacing tries to adjust. The equilibrium shift will appear on the timescale $T_\text{pump}$ of carrier generation, and then will be slightly readjusted over $\tau_\text{intraband rel.}$, as doublons and holes redistribute within the Hubbard bands. As mentioned previously, since both of these processes are assumed much shorter than the characteristic timescale for acoustic phonons $\omega_\text{ph}^{-1}$, we can approximate the process as a quench of the effective phonon Hamiltonian. This will set up a wave packet propagating inwards through the sample, which can be detected in pump-probe measurements, e.g. via transient reflectivity oscillations.\\

For any given pulse frequency and amplitude, we use a driving profile which exponentially decays away from the edge, with penetration depth $\lambda_\text{pump} = 100 \; a$. From the results shown in Figure \ref{fig:kinetic_energy_and_phonon}a, we extract the intermediate-time value of kinetic energy density versus depth in the sample. As argued in Section \ref{ssec:coherent_phonons_analytic_introduction}, the main effect on the phonons will appear at wavevectors set by $\lambda_\text{pump}^{-1}$. We plot this contribution $A_\text{phonon}(\lambda_\text{pump}^{-1})$ in Figure \ref{fig:kinetic_energy_and_phonon}b, for a similar range of pump parameters to those investigated earlier. Observe that the shapes and locations of multiphoton thresholds are the same as in Figure \ref{fig:quasiparticles_and_multiphoton}a: despite minor differences, total photocarrier density and acoustic phonon amplitude are closely related. This is not only in general agreement with phenomenological approaches such as \cite{phonon_mechanism_review_macroscopic} but, as advertised, also suggests a complementary way of probing the quasiparticle density in a pump-probe setting, beyond the usual short-time reflectivity peak (which requires sharp time resolution).\\
\section{Conclusions and outlook} \label{sec:conclusions}

We have applied a time-dependent Gaussian variational approach to the problem of strongly driven antiferromagnetic Mott insulators, in order to investigate the doublon-hole pair photoexcitation process under subgap pumping. In contrast to previous rigid-band approaches, we allow the energy gap to vary in time and self-consistently calculate it at every point throughout the evolution. By considering carrier scattering, and including the effects thereof which are relevant on short timescales, we arrive at a compact system of evolution equations given in \eqref{eq:maintext_evolution_equations}.\\

In the case of weak driving, our approach naturally recovers the rigid-band limit, where the system's dynamics are amenable to an analytic treatment. In the coherent case $\relaxrate \to 0$, we discuss how the conventional multiphoton and dc tunneling regimes can be recovered on either side of the Keldysh crossover $\hbar \omega_d \sim e \field a$. The resulting production rates \eqref{eq:maintext_multiphoton_production_rate} and \eqref{eq:maintext_tunneling_production_rate} match the functional forms calculated in Ref. \cite{oka_nonlinear_doublons} for the 1D Hubbard model. Upon investigating the low-frequency ac regime more closely, we find enhanced carrier production when the gap equals an integer multiple of the driving frequency, which points to a cooperation between multiphoton and tunneling processes. Such a mechanism is absent from the analytic treatment of \cite{oka_nonlinear_doublons}, and can be contrasted with the usual picture of multiphoton-tunneling competition giving rise to the Keldysh crossover. Furthermore, we argue that including finite $\relaxrate > 0$ will give rise to an incoherent production pathway, which becomes dominant under weak fields. In the case of low driving frequencies, we highlight the experimental relevance of this mechanism, by showing that it partly overtakes the region in parameter space usually reserved for the multiphoton regime. Although dissipation has been discussed in the context of steady states for both dc \cite{dmft_dc_usa_dielectric_breakdown, resistive_switching_hysteresis} and ac \cite{driven_steady_state, photoinduced_nonequilibrium_review} driving, its role in upsetting the conventional Keldysh crossover for low $\omega_d$ has, to our knowledge, not yet been considered.\\

The analytic arguments give rise to a general picture for subgap photoexcitation in Mott insulators, which is sketched in Figure \ref{fig:regime_sketch_2d}: beyond the usual multiphoton and tunneling regimes, there is a region in parameter space where the two strongly cooperate, as well as a low-field zone where incoherent linear absorption becomes dominant. Instead of the Keldysh line being the only qualitative regime boundary in the $\omega_d - \field_p$ plane, we emphasize the existence of a multiphoton onset line; it marks significant qualitative changes in the photoexcitation mechanism, as evidenced by the appearance of frequency thresholds in carrier production (Fig. \ref{fig:quasiparticles_and_multiphoton}) as well as modifications in the momentum distribution of these excitations (Fig. \ref{fig:qp_momentum_variance}).\\

Using a numerical approach to complement the analytics, we access the strong driving regime, where the commonly used rigid-band assumption ceases to hold, and a richer set of phenomena starts to appear. As the presence of doublons and holes renormalizes the gap, the highly nonlinear production rates calculated previously will in turn be modified. Although effects of a finite photocarrier density on the Mott gap have been theoretically discussed in \cite{dynamics_screening_photodoped, multiband_photodoped_screening}, our self-consistent inclusion of this renormalization's feedback on the photoexcitation process gives rise to new behavior. In the low-frequency case we argue that, for field strengths not yet approaching the dielectric breakdown threshold, the main effect will be a renormalization of the tunneling rate. At high frequencies, on the other hand, the multiphoton condition can be affected strongly enough such that thresholds are crossed without altering the pump frequency. We show how, depending on the starting parameters, this can manifest itself as either saturation of carrier density in real time or, on the other hand, lowering of the multiphoton threshold together with a sharp acceleration of doublon / hole pair production. The failure of interpretations relying on a single production rate (calculated based on driving frequency and field) also points to the experimental relevance of pump shape and duration as control parameters. By tuning these, one should be able to access in both theory and experiment a wide range of behaviors, from short-time rate processes, all the way to saturation and steady states on long timescales.\\

We find that the momentum distribution of quasiparticles at the end of the pump contains valuable information about the character of the photoexcitation processes and can be used to distinguish between the four regimes discussed above. Namely, the local nature of Landau-Zener tunneling gives rise to a rather flat momentum distribution, in agreement with the prediction of \cite{oka_nonlinear_doublons}. Multiphoton excitation, on the other hand, is found to be sharply concentrated around the resonant energy contour corresponding to a specific pumping frequency, in contrast to the result of \cite{oka_nonlinear_doublons} which has carriers concentrated at the band minimum. The incoherent pathway gives a broad but nonuniform pattern, sensitive to both the energy and the slope of the quasiparticle dispersion. Therefore, probing such momentum distributions can be a valuable complementary tool for distinguishing the photoexcitation mechanisms, and mapping out the boundaries which separate them in parameter space. Under strong driving, the arrangement of carriers in the Brillouin zone becomes significantly more complex, as expected from a non-perturbative regime.\\

Finally, we employ an electron-phonon coupling of the SSH form, to analyze the generation of coherent longitudinal acoustic phonons on long timescales. This provides a microscopic connection between the behaviors of charge and lattice degrees of freedom, as a complement to the usual phenomenological analysis \cite{phonon_mechanism_review_macroscopic}. We argue that, for most regimes of experimental interest, the phonon amplitude will be closely related to the total photocarrier density. In turn, this observation provides a new tool for probing the distribution function of photoexcited quasiparticles.\\

The flexibility of our variational approach means that its generalizations would be suitable for further investigation of driven Mott insulators, with possible directions including breakdown of the insulating state over a wide range of pumping frequencies, nonequilibrium optical conductivity of photodoped Mott systems, or high harmonic generation. By employing an RPA-type analysis of spin fluctuations on top of the charge dynamics, one can study parametric driving of free and bound magnon states in the Hubbard model, spin-wave spectrum renormalization due to the presence or charge excitations, or magnetic Raman scattering in driven systems. Another promising direction is the dynamics of exciton formation in photodoped MI, as it has been argued to be closely connected with antiferomagnetism \cite{hubbard_excitons}.\\

With regard to experimental realizations, beyond the solid-state systems which have been the main focus of present calculations, we remark that ultracold atoms in optical lattices offer a promising complementary platform. Precise control over Hamiltonian engineering in these systems allows for tuning of fermion tunneling and interaction strength. Absence of disorder and phonons results in better control of decoherence for doublons and holes, while real-space measurements using quantum gas microscopy allow for their direct observation. The intrinsically longer timescales in cold atom setups enable closer investigation of real-time dynamics: for example, tracking the order parameter $S(t)$ by probing antiferromagnetic correlations would reveal its suppression under strong driving. Comparison to measurements of $n_\text{qp}(t)$ can then be used to highlight the alteration of quasiparticle production rates due to modifications of $S$. Fourier transforming the experimentally measured density-density correlations gives information about the momentum distribution of photoexcited quasiparticles, which can be compared to theoretical predictions.\\
\section{Acknowledgments}

We would like to acknowledge helpful discussions with Alex~G\'{o}mez~Salvador, Filip~Marijanovi\'{c}, Ivan~Morera~Navarro, Jonathan~B.~Curtis, Sambuddha~Chattopadhyay, Yi-Fan~Qu, Ryo~Noguchi, Gil~Refael, Denis~Gole\v{z}, Matteo~Mitrano, Elsa~Abreu, Zhiyuan~Sun, and Patrick~A.~Lee. R.A. and E.D. acknowledge funding from the Swiss National Science Foundation (project 200021 212899), ETH-C-06 21-2 Equilibrium Grant (project number 1-008831-001), and the Swiss State Secretariat for Education, Research and Innovation (contract number UeM019-1). D.H. acknowledges support from the Institute for Quantum Information and Matter (IQIM), an NSF Physics Frontiers Center (PHY-2317110), as well as the Brown Investigator Award, a program of the Brown Science Foundation.\\

\appendix

$$$$

\section*{Appendices}

In the following appendices, we present in detail the variational ansatz and its application to our system: Section \ref{sec:sm_sdw} discusses the SDW approach \cite{sdw_bag_superconductivity} to the half-filled Hubbard model, and its extension to the case of a driven system; we recover the spectrum of doublon-hole pair excitations, as well as the self-consistency condition on the gap, which will govern its renormalization under strong drive. Section \ref{sec:sm_phonons} rephrases the SSH electron-phonon coupling into the doublon-hole language, and highlights the resulting mechanisms for scattering of quasiparticles, as well as generation of coherent longitudinal acoustic phonons at late times. Having described the model, in Section \ref{sec:sm_ansatz} we formally introduce the variational ansatz, the equations of motion on the variational parameters, as well as expressions for observables of interest. Section \ref{sec:analytic_approximate_solutions} develops a perturbative treatment in the rigid-band approximation, and analytically derives the four photoexcitation regimes sketched in Figure \ref{fig:regime_sketch_2d} of the main text. Details of the numerical calculations are discussed in Section \ref{sec:sm_numerics}. Finally, the discussion of photocarrier momentum distributions from the main text is expanded in Section \ref{sec:appendix_momentum_distribution}.\\

\section{Time-dependent SDW treatment of the Hubbard model} \label{sec:sm_sdw}
\subsection{System Hamiltonian and external drive} \label{ssec:sm_intro_hamiltonian_drive}
As a prototype of strongly interacting electron systems, we consider a 2D single-band Fermi-Hubbard model at half filling, with an on-site interaction term $U$ taken to be much stronger than the nearest-neighbor hopping $\tau$. Assuming that the external driving is spatially uniform, we incorporate it in the Hamiltonian via a Peierls substitution:
\begin{widetext}
    \begin{equation} \label{eq:hubbard_hamiltonian_definition}
        H_\text{sys}(t) = - \tau \sum_{\mathbf{j}, \boldsymbol{\delta}, \spin} e^{i e \boldsymbol{\delta} \cdot \mathbf{A}(t)} c^\dag_{\mathbf{j} + \boldsymbol \delta, \spin} c_{\mathbf{j}, \spin} + U \sum_\mathbf{j} n_{\mathbf{j} \uparrow} n_{\mathbf{j} \downarrow} - \mu \sum_{\mathbf{j}, \spin} n_{\mathbf{j} \spin}.
    \end{equation}
\end{widetext}
Here, $\mathbf j$ denotes lattice sites, $\boldsymbol{\delta} \in \{\pm \hat{\mathbf x}, \pm \hat{\mathbf y}\}$ runs over nearest-neighbor directions, and $\spin \in \{\pm 1\}$ represents electronic spin. The vector potential $\mathbf{A}(t)$ will be kept general throughout this section, but should be understood to represent a typical pump-pulse in an experimental context: oscillations at an optical carrier frequency, modulated by a Gaussian envelope of width on the order of 100 fs. On the other hand, when analytically deriving photoexcitation rates in the rigid-band approximation, we will take $\mathbf{A}(t)$ to be monochromatic.\\

We can also implement the Peierls substitution as replacing $\mathbf k \to \mathbf k - e \mathbf A(t)$ in the kinetic energy:
\begin{widetext}
    \begin{eqnarray}
    H_\text{kin}(t) & = & \sum_{\mathbf k, \spin} \e_{\mathbf k}(t) \;  c^\dag_{\mathbf{k}, \spin} c_{\mathbf{k}, \spin},\\
    \e_{\mathbf k} (t) & = & -2 \tau \ob \cos \os k_x - e A_x (t) \cs + \cos \os k_y - e A_y(t) \cs \cb.
\end{eqnarray}
In this picture, the time derivative of kinetic energy is proportional to the electric field:
\begin{equation} \label{eq:kinetic_energy_time_derivative}
    \partial_t \e_{\mathbf k} = - \boldsymbol \field \cdot \nabla_{\mathbf A} \e_{\mathbf k} = 2 \tau e \ob \field_x \sin \os k_x - e A_x (t) \cs + \field_y \sin \os k_y - e A_y (t) \cs \cb.
\end{equation}
\end{widetext}
We will employ the SDW approach to the Hubbard model \footnote{At low temperatures, we expect an AFM  phase, where the Néel order assumption is justified. Above $T_N$, we retain strong short-range spin correlations even in the absence of long-range order. As the typical doublon-hole correlation length is short - on the order of a few lattice constants - they are still evolving in an effective AFM background. Therefore, present results are expected to qualitatively hold well above $T_N$.}, with ordering wavevector $\mathbf Q = (\pi, \pi)$. At all times, we have the additional property that $\e_{\mathbf k + \mathbf Q} (t) = - \e_{\mathbf k} (t)$.\\

We will also consider electron-phonon coupling, which is relevant to this work for two reasons:
\begin{itemize}
    \item Phonons provide one mechanism for scattering of electronic quasiparticles, which will influence photoexcitation rates;
    \item Microscopic consideration of electron-phonon coupling allows us to describe the displacive excitation of coherent acoustic phonons (CAP) as a result of photoexcitation. We will also relate the CAP amplitude to the density of photoexcited carriers.
\end{itemize}

The model we use for this coupling is SSH: altering the distance between neighboring nuclei will affect the hopping matrix element between those sites, as the latter depends exponentially on separation. At leading order, we obtain the coupling \footnote{During the drive, Peierls phases should also be added to the hopping terms in \eqref{eq:ssh_hamiltonian_definition}; however, we will only consider pulses of much shorter duration compared to the characteristic timescale of the low-energy acoustic phonons. This justifies ignoring the direct effect of the electric field on the SSH part of the Hamiltonian.}
\begin{widetext}
    \begin{equation} \label{eq:ssh_hamiltonian_definition}
    H_\text{SSH} = - g \sum_{\mathbf j, \spin} \os \oa c_{\mathbf j + \hat{\mathbf x}, \spin}^\dag c_{\mathbf j, \spin} + \text{h.c.} \ca \oa \mathbf x_{\mathbf j + \hat{\mathbf x}} - \mathbf x_{\mathbf j} \ca + \oa c_{\mathbf j + \hat{\mathbf y}, \spin}^\dag c_{\mathbf j, \spin} + \text{h.c.} \ca \oa \mathbf y_{\mathbf j + \hat{\mathbf y}} - \mathbf y_{\mathbf j} \ca \cs,
\end{equation}
\end{widetext}
where $g$ is a coupling constant. The full Hamiltonian of the system will then be $H \equiv H_\text{sys} + H_\text{SSH}$.

\subsection{Order parameter and factorizing the interaction term} \label{ssec:order_parameter_and_factorization}
The relevant SDW operator on the square lattice, with $\hat{\mathbf z}$ taken as the ordering axis, is \footnote{In real space, this reads $S^z_{\mathbf Q} = {1 \over N} \sum_{\mathbf j} e^{i \mathbf Q \cdot \mathbf j} \; (n_{\mathbf j \uparrow} - n_{\mathbf j \downarrow})$.}
\begin{equation}
    S^z_{\mathbf Q} \equiv {1 \over N} \sum_{{\mathbf k}, \beta, \gamma} c^\dag_{\mathbf k + \mathbf Q, \beta} \; \pauli_z^{\beta \gamma} \; c_{\mathbf k, \gamma}= {1 \over N} \sum_{\mathbf k, \spin} \spin \; c^\dag_{\mathbf k + \mathbf Q, \spin} c_{\mathbf k, \spin}.
\end{equation}
This will acquire an expectation value $S_0$ in the variational ground state $\ket{\vac}$:
\begin{equation}
    \mel{\vac}{S^z_{\mathbf Q}}{\vac} \equiv S_0.
\end{equation}
The self-consistent variational approach employed here allows the order parameter $S(t)$ to vary in time, as a consequence of the drive:
\begin{equation}
    S(t) = \trace [S^z_{\mathbf Q} \; \rho(t)],
\end{equation}
where $\rho(t)$ is the density matrix describing the system at arbitrary times.\\

In the absence of superconductivity, expectations of the form $\ex{cc}$ or $\ex{c^\dag c^\dag}$ will be taken to vanish. The SDW order spontaneously breaks the translation symmetry; assuming no other SSB in the variational ground state, the only nonzero two-point correlators will be at momentum difference $\mathbf q \in \{ 0, \mathbf Q \}$ and identical spins, i.e. of the form $\ex{c^\dag_{\mathbf k, \spin} c_{\mathbf k, \spin}}$ or $\ex{c^\dag_{\mathbf k + \mathbf Q, \spin} c_{\mathbf k, \spin}}$. Under spatially uniform driving, these are also the only nonzero correlators at later times. In consequence, within the Hubbard interaction term
\begin{equation}
    H_\text{int} = {U \over N} \sum_{\mathbf k_1, \mathbf k_2, \mathbf q} c_{\mathbf{k}_1 + \mathbf{q} \uparrow}^\dag c_{\mathbf{k}_1 \uparrow} \; c_{\mathbf{k}_2 - \mathbf{q} \downarrow}^\dag c_{\mathbf{k}_2 \downarrow},
\end{equation}
we perform the decoupling
\begin{widetext}
    \begin{equation} \label{eq:hf_decoupling}
    c_{\mathbf{k}_1 + \mathbf{q} \uparrow}^\dag c_{\mathbf{k}_1 \uparrow} \; c_{\mathbf{k}_2 - \mathbf{q} \downarrow}^\dag c_{\mathbf{k}_2 \downarrow} \to \ex{c_{\mathbf{k}_1 + \mathbf{q} \uparrow}^\dag c_{\mathbf{k}_1 \uparrow}} \; c_{\mathbf{k}_2 - \mathbf{q} \downarrow}^\dag c_{\mathbf{k}_2 \downarrow} + c_{\mathbf{k}_1 + \mathbf{q} \uparrow}^\dag c_{\mathbf{k}_1 \uparrow} \; \ex{c_{\mathbf{k}_2 - \mathbf{q} \downarrow}^\dag c_{\mathbf{k}_2 \downarrow}} - \ex{c_{\mathbf{k}_1 + \mathbf{q} \uparrow}^\dag c_{\mathbf{k}_1 \uparrow}} \; \ex{c_{\mathbf{k}_2 - \mathbf{q} \downarrow}^\dag c_{\mathbf{k}_2 \downarrow}},
\end{equation}
where only terms with $\mathbf q \in \{ 0, \mathbf Q \}$ are kept. The first term of \eqref{eq:hf_decoupling} then reads
\begin{align}
{U \over N} \sum_{\mathbf k_1, \mathbf k_2, \mathbf q} \ex{c_{\mathbf{k}_1 + \mathbf{q} \uparrow}^\dag c_{\mathbf{k}_1 \uparrow}} \; c_{\mathbf{k}_2 - \mathbf{q} \downarrow}^\dag c_{\mathbf{k}_2 \downarrow} &= {U \over N} \sum_{\mathbf k_1, \mathbf k_2} \os  \ex{c_{\mathbf{k}_1 \uparrow}^\dag c_{\mathbf{k}_1 \uparrow}} \; c_{\mathbf{k}_2 \downarrow}^\dag c_{\mathbf{k}_2 \downarrow} + \ex{c_{\mathbf{k}_1 + \mathbf{Q} \uparrow}^\dag c_{\mathbf{k}_1 \uparrow}} \; c_{\mathbf{k}_2 - \mathbf{Q} \downarrow}^\dag c_{\mathbf{k}_2 \downarrow}\cs\\
&= U \; \os {\sum_{\mathbf p} \ex{c_{\mathbf p \uparrow}^\dag c_{\mathbf p \uparrow}} \over N}  \cs \sum_{\mathbf k} c_{\mathbf{k} \downarrow}^\dag c_{\mathbf{k} \downarrow} + U \; \os {\sum_{\mathbf p} \ex{c_{\mathbf{p} + \mathbf{Q} \uparrow}^\dag c_{\mathbf{p} \uparrow}} \over N} \cs \sum_{\mathbf k} c_{\mathbf{k} + \mathbf{Q} \downarrow}^\dag c_{\mathbf{k} \downarrow}. \nonumber
\end{align}
By spin symmetry, the sum $\sum_{\mathbf p} \ex{c_{\mathbf p \uparrow}^\dag c_{\mathbf p \uparrow}}$, which counts total up-spin particles, is $N / 2$. The analogous sum in the second term, $\sum_{\mathbf p} \ex{c_{\mathbf{p} + \mathbf{Q} \uparrow}^\dag c_{\mathbf{p} \uparrow}}$, is the up-spin contribution to the Neel order parameter; also from symmetry arguments, we take it to be $NS/2$. From this we obtain
\begin{equation}
    {U \over N} \sum_{\mathbf k_1, \mathbf k_2, \mathbf q} \ex{c_{\mathbf{k}_1 + \mathbf{q} \uparrow}^\dag c_{\mathbf{k}_1 \uparrow}} \; c_{\mathbf{k}_2 - \mathbf{q} \downarrow}^\dag c_{\mathbf{k}_2 \downarrow} = {U \over 2} \sum_{\mathbf k} c_{\mathbf{k} \downarrow}^\dag c_{\mathbf{k} \downarrow} + {U S \over 2} \sum_{\mathbf k} c_{\mathbf{k} + \mathbf{Q} \downarrow}^\dag c_{\mathbf{k} \downarrow}.
\end{equation}
Analogously we find
\begin{align}
{U \over N} \sum_{\mathbf k_1, \mathbf k_2, \mathbf q} c_{\mathbf{k}_1 + \mathbf{q} \uparrow}^\dag c_{\mathbf{k}_1 \uparrow} \; \ex{c_{\mathbf{k}_2 - \mathbf{q} \downarrow}^\dag c_{\mathbf{k}_2 \downarrow}} &= {U \over 2} \sum_{\mathbf k} c_{\mathbf{k} \uparrow}^\dag c_{\mathbf{k} \uparrow} - {U S \over 2} \sum_{\mathbf k} c_{\mathbf{k} + \mathbf{Q} \uparrow}^\dag c_{\mathbf{k} \uparrow}\\
{U \over N} \sum_{\mathbf k_1, \mathbf k_2, \mathbf q} \ex{c_{\mathbf{k}_1 + \mathbf{q} \uparrow}^\dag c_{\mathbf{k}_1 \uparrow}} \; \ex{c_{\mathbf{k}_2 - \mathbf{q} \downarrow}^\dag c_{\mathbf{k}_2 \downarrow}} &= {U N \over 4} - {U N S^2 \over 4}.
\end{align}
Then, the Hubbard Hamiltonian reduces to
\begin{equation} \label{eq:decoupled_interaction_hamiltonian}
    H_\text{sys}^\text{HF} = H_\text{kin} + \op {U \over 2} - \mu \cp \sum_{\mathbf k, \spin} c_{\mathbf{k} \spin}^\dag c_{\mathbf{k} \spin} - {U S \over 2} \sum_{\mathbf k} \spin \; c_{\mathbf{k} + \mathbf{Q} \spin}^\dag c_{\mathbf{k} \spin} - {U N \over 4} + {U N S^2 \over 4}.
\end{equation}
At half filling, the chemical potential is $\mu = U / 2$, which removes the second term. We ignore the constant contribution to the energy of $-UN/4$. Furthermore, the $UNS^2/4$ term also represents a shift in the ground state energy, which can be disregarded when investigating the photoexcitation process \footnote{However, since $S(t)$ is time-dependent, this term should be considered towards conservation of energy.}.

\subsection{Solution of the decoupled Hamiltonian, and self-consistency condition} \label{ssec:static_solution_variational}
Ignoring the energy-shift terms in \eqref{eq:decoupled_interaction_hamiltonian}, we are left with
    \begin{align}
	H^\text{HF}(t) &= \sum_{\mathbf k, \spin} \e_{\mathbf k} (t) \;  c^\dag_{\mathbf{k}, \spin} c_{\mathbf{k}, \spin} - {U S(t) \over 2} \sum_{\mathbf k, \spin} \spin \; c_{\mathbf{k} + \mathbf{Q}, \spin}^\dag c_{\mathbf{k}, \spin} \nonumber \\
	&= \sum_{\substack{\mathbf k \in BZ' \\ \spin}} \op \e_{\mathbf k} (t) \;  c^\dag_{\mathbf{k}, \spin} c_{\mathbf{k}, \spin} + \e_{\mathbf k + \mathbf Q}(t) \;  c^\dag_{\mathbf{k} + \mathbf Q, \spin} c_{\mathbf{k} + \mathbf Q, \spin} - {U S(t) \over 2} \spin \; c_{\mathbf{k} + \mathbf{Q}, \spin}^\dag c_{\mathbf{k}, \spin} - {U S(t) \over 2} \spin \; c_{\mathbf{k}, \spin}^\dag c_{\mathbf{k}  + \mathbf{Q}, \spin} \cp \nonumber\\
	&= \sum_{\substack{\mathbf k \in BZ' \\ \spin}} \begin{pmatrix}
		c^\dag_{\mathbf{k}, \spin} & c^\dag_{\mathbf{k} + \mathbf Q, \spin}
	\end{pmatrix} \begin{pmatrix}
		\e_{\mathbf k}(t) & - \spin U S(t) / 2 \\ 
		- \spin U S(t) / 2 & - \e_{\mathbf k} (t)
	\end{pmatrix} \begin{pmatrix}
		c_{\mathbf{k}, \spin} \\ c_{\mathbf{k} + \mathbf Q, \spin}
	\end{pmatrix},
\end{align}
\end{widetext}
where $BZ'$ is half the original Brillouin zone of the square lattice. Specifically, we choose the diamond-shaped region defined by $|k_x| + |k_y| \le \pi / a$. Each $2 \times 2$ momentum sector in this reduced zone is straightforward to diagonalize with a Bogoliubov transformation. Make the notation $C_{\mathbf{k}, \spin} =\begin{pmatrix}
	c_{\mathbf{k}, \spin} \\ c_{\mathbf{k} + \mathbf Q, \spin}
\end{pmatrix}$, and let $\bogoliubon_{\mathbf{k}, \spin}(t) = \begin{pmatrix}
	d_{\mathbf{k}, \spin}(t) \\ h^{\dag}_{-\mathbf{k}, \spin}(t)
\end{pmatrix}$ be defined by
\begin{equation} \label{eq:bogoliubov_operators_definition}
    C_{\mathbf{k}, \spin} = B_{\mathbf{k}, \spin}(t) \; \bogoliubon_{\mathbf{k}, \spin}(t).
\end{equation}
Here, the $d^\dag_{\mathbf{k}, \spin}$ ($h^\dag_{\mathbf{k}, \spin}$) operators can be interpreted as creating a doublon (hole), with momentum $\mathbf k$ and spin $\spin$, in the upper (lower) Hubbard band. Note that, since momenta are shifted by coupling to the drive, the instantaneous definitions of these quasiparticle operators will change with time. The matrix $B_{\mathbf{k}, \spin}(t)$ is taken of the form
\begin{align} 
    B_{\mathbf{k}, \spin}(t) &= \begin{pmatrix}
	u_{\mathbf k}(t) & \spin v_{\mathbf k}(t)\\ 
	-\spin v_{\mathbf k}(t) & u_{\mathbf k}(t)
\end{pmatrix} \nonumber\\
&= u_{\mathbf k}(t) \pauli_0 + v_{\mathbf k}(t) (i \spin \pauli_y), \label{eq:bogoliubov_definition}
\end{align}
with the following coefficients:
\begin{subequations}\label{eq:bogoliubov_coefficients}
\begin{align}
u_\mathbf{k}(t) &= \os {1\over 2} \op 1 + {\e_{\mathbf k}(t) \over \sqrt{\e_{\mathbf k}^2(t) + (U S(t) / 2)^2}} \cp \cs^{1/2}, \\
v_\mathbf{k}(t) &= \os {1\over 2} \op 1 - {\e_{\mathbf k}(t) \over \sqrt{\e_{\mathbf k}^2(t) + (U S(t) / 2)^2}} \cp \cs^{1/2}.
\end{align}
\end{subequations}
Since the coefficients are real, $B_{\mathbf{k}, \spin}(t)$ is an orthogonal matrix. The diagonal form of the Hamiltonian is
\begin{widetext}
\begin{align} 
	H^\text{HF}(t) &= \sum_{\substack{\mathbf k \in BZ' \\ \spin}} E_{\mathbf k}(t) \; \bogoliubon_{\mathbf{k}, \spin}^\dag(t) \; \pauli_z \; \bogoliubon_{\mathbf{k}, \spin}(t) = \sum_{\substack{\mathbf k \in BZ' \\ \spin}} E_{\mathbf k}(t) \; \os d^{\dag}_{\mathbf{k}, \spin}(t) \; d_{\mathbf{k}, \spin}(t) + h^{\dag}_{\mathbf{k}, \spin}(t) \; h_{\mathbf{k}, \spin}(t) - 1 \cs,
\end{align} 
\end{widetext}
where the -1 inside the brackets also contributes just an energy shift. The quasiparticle dispersion is
\begin{equation} \label{eq:quasiparticle_dispersion}
    E_{\mathbf k}(t) = \sqrt{\e_{\mathbf k}^2(t) + U^2 S^2(t) / 4}.
\end{equation}
The resulting density of states is singular at the edge of the reduced BZ, as shown in Figure \ref{fig:density_of_states} below. Note that the gap is $\Delta(t) = U S(t)$, and its time dependence will be relevant if the order parameter $S$ is strongly affected by the drive. As mentioned previously, $S(t)$ must be determined self-consistently throughout the evolution. In terms of the folded BZ, the operator reads
\begin{align}
        S^z_{\mathbf Q} &= {1 \over N} \sum_{\mathbf k, \spin} \spin \; c^\dag_{\mathbf k + \mathbf Q, \spin} c_{\mathbf k, \spin} \nonumber \\
        &= {1 \over N} \sum_{\substack{\mathbf k \in BZ' \\ \spin}} \spin \; \ob c^\dag_{\mathbf{k} + \mathbf{Q}, \spin} c_{\mathbf{k}, \spin} + c^\dag_{\mathbf{k}, \spin} c_{\mathbf{k} + \mathbf{Q}, \spin} \cb \nonumber \\
        &= {1 \over N} \sum_{\substack{\mathbf k \in BZ' \\ \spin}} \spin \; C^\dag_{\mathbf{k}, \spin} \; \pauli_x \; C_{\mathbf{k}, \spin}.
\end{align}
Switching to the Bogoliubov basis via $C_{\mathbf{k}, \spin} = B_{\mathbf{k}, \spin} (t) \bogoliubon_{\mathbf{k}, \spin} (t)$, this reads
\begin{widetext}
    \begin{align}
	S^z_{\mathbf Q} &= {1 \over N} \sum_{\substack{\mathbf k \in BZ' \\ \spin}} \spin \; \bogoliubon_{\mathbf{k}, \spin}^{\dag} (t) \; B^{\dag}_{\mathbf{k}, \spin} (t) \; \pauli_x \; B_{\mathbf{k}, \spin} (t) \; \bogoliubon_{\mathbf{k}, \spin} (t) \nonumber \\
	&= {1 \over N} \sum_{\substack{\mathbf k \in BZ' \\ \spin}} {1 \over E_{\mathbf k} (t)} \; \bogoliubon_{\mathbf{k}, \spin}^{\dag} (t) \; \op \spin \e_{\mathbf k} (t) \; \pauli_x - {U S(t) \over 2} \; \pauli_z \cp \; \bogoliubon_{\mathbf{k}, \spin} (t) \nonumber\\
	&= {1 \over N} \sum_{\substack{\mathbf k \in BZ' \\ \spin}} {1 \over \sqrt{\e^2_{\mathbf k} + U^2 S^2 / 4}} \; \begin{pmatrix}
		d^{\dag}_{\mathbf{k}, \spin} & h_{-\mathbf{k}, \spin}
	\end{pmatrix}  \begin{pmatrix}
		- U S / 2 & \spin \e_{\mathbf k}\\ 
		\spin \e_{\mathbf k} & U S / 2
	\end{pmatrix} \begin{pmatrix}
		d_{\mathbf{k}, \spin} \\ h^\dag_{-\mathbf{k}, \spin}
	\end{pmatrix}. \label{eq:order_parameter_fourier_transformed}
\end{align}
In the last line, explicit time dependence was suppressed for simplicity. Expanding and normal-ordering the above, one obtains
\begin{equation} \label{eq:order_parameter_bog_basis}
    S^z_{\mathbf Q} = {1 \over N} \sum_{\substack{\mathbf k \in BZ' \\ \spin}} {1 \over \sqrt{\e^2_{\mathbf k} + U^2 S^2 / 4}} \;  \os {U S \over 2} - {U S \over 2} \op d^{\dag}_{\mathbf{k}, \spin} d_{\mathbf{k}, \spin} + h^{\dag}_{-\mathbf{k}, \spin} h_{-\mathbf{k}, \spin} \cp + \spin \e_{\mathbf k} \op d^{\dag}_{\mathbf{k}, \spin} h^{\dag}_{-\mathbf{k}, \spin} +  h_{-\mathbf{k}, \spin} d_{\mathbf{k}, \spin} \cp \cs.
\end{equation}
\end{widetext}
Before pumping, the system is in the variational ground state $\ket{\vac}$ defined by $d_{\mathbf{k}, \spin} \ket{\vac} = h_{\mathbf{k}, \spin} \ket{\vac} = 0$. Only the first term from the brackets above survives when taking the expectation $\mel{\vac}{S^z_{\mathbf Q}}{\vac}$, and imposing that it is equal to $S_0$ we find
\begin{equation}
    S_0 = \mel{\vac}{S^z_{\mathbf Q}}{\vac} = {1 \over N} \sum_{\substack{\mathbf k \in BZ' \\ \spin}} {U S_0 / 2 \over \sqrt{\e^2_{\mathbf k} + U^2 S_0^2 / 4}}.
\end{equation}
After dividing by $S_0$, performing the sum over spins to get a factor of 2, and turning the momentum sum into an integral, we arrive at the equilibrium self-consistency condition:
\begin{equation}
    \int_{BZ'}{d^2 \mathbf k \over 4 \pi^2} \; {1 \over \sqrt{(\e_{\mathbf k} / U)^2 + S_0^2 / 4}} = 1.
\end{equation}
For strong interactions $U / \tau \to \infty$, the order parameter saturates: $S_0 \to 1$. In general, a series expansion in $\tau /U$ yields
\begin{equation}
    S_0 = 1 - 8 \op{\tau \over U}\cp^2 + \OO\os \op{\tau \over U}\cp^4\cs.
\end{equation}
Throughout the time evolution, $S(t)$ is determined self-consistently from the expectation value of the full expression \eqref{eq:order_parameter_bog_basis}.

\begin{figure}
	\centering	
    \scalebox{\fullfigsize}{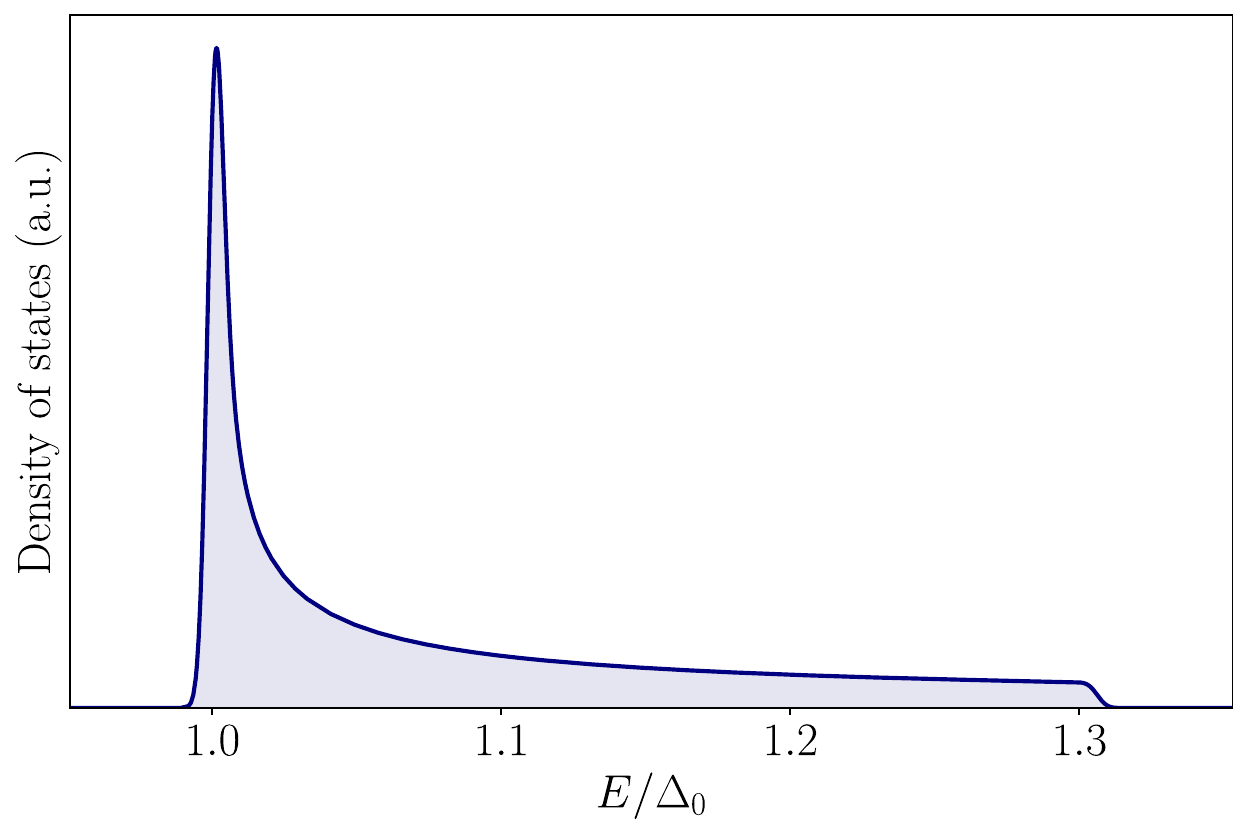}
	\caption{Typical shape for the density of states, for doublon-hole pair excitations, at zero total momentum; calculated for $U / \tau \approx 10$. The sharp peak at the bottom of the band is a consequence of assuming hopping to be nearest-neighbor only. }
	\label{fig:density_of_states}
\end{figure}
\section{Coupling to phonons} \label{sec:sm_phonons}
We aim to express the SSH coupling \eqref{eq:ssh_hamiltonian_definition} in reciprocal space. As this part of the Hamiltonian separates into $\hat{\mathbf x}$ and $\hat{\mathbf y}$ components, it suffices to consider one of them, and the other will follow by analogy. Starting with the expression
\begin{equation} 
    H_{\text{SSH}, \; \hat{\mathbf x}} = - g \sum_{\mathbf j, \spin} \oa c_{\mathbf j + \hat{\mathbf x}, \spin}^\dag c_{\mathbf j, \spin} + c_{\mathbf j, \spin}^\dag c_{\mathbf j + \hat{\mathbf x}, \spin} \ca \oa \mathbf x_{\mathbf j + \hat{\mathbf x}} - \mathbf x_{\mathbf j} \ca,
\end{equation}
the phonons can be quantized by replacing $\mathbf x_{\mathbf j} \to a_0 \oa b_{\hat{\mathbf x}, \; \mathbf j} + b_{\hat{\mathbf x}, \; \mathbf j}^\dag \ca$, where the first index of a phonon operator denotes its polarization. This yields
\begin{widetext}
    \begin{eqnarray}
    H_{\text{SSH}, \; \hat{\mathbf x}} & = & - a_0 g \sum_{\mathbf j, \spin} \oa c_{\mathbf j + \hat{\mathbf x}, \spin}^\dag c_{\mathbf j, \spin} + c_{\mathbf j, \spin}^\dag c_{\mathbf j + \hat{\mathbf x}, \spin} \ca \oa b_{\hat{\mathbf x}, \; \mathbf j + \hat{\mathbf x}} - b_{\hat{\mathbf x}, \; \mathbf j} \ca + \text{h.c.} \nonumber \\
    & = & -{2 i a_0 g \over \sqrt N} \sum_{\mathbf k_1, \mathbf k_2, \spin} c_{\mathbf k_1, \spin}^\dag c_{\mathbf k_2, \spin} \; b_{\hat{\mathbf x}, \; \mathbf k_1 - \mathbf k_2} \oa \sin k_{1x} - \sin k_{2x} \ca + \text{h.c.} .\label{eq:ssh_real_space_quantized} 
\end{eqnarray}
After explicitly including the Hermitian conjugate in the above, restoring the $\mathbf{\hat{y}}$ part of the Hamiltonian, and making the notation $g_1 \equiv -{2 i a_0 g / \sqrt N}$, we find
\begin{equation} \label{eq:ssh_momentum_definition}
    H_{\text{SSH}} = g_1 \sum_{\mathbf k_1, \mathbf k_2, \spin} c_{\mathbf k_1, \spin}^\dag c_{\mathbf k_2, \spin} \os \oa b_{\hat{\mathbf x}, \; \mathbf k_1 - \mathbf k_2} + b_{\hat{\mathbf x}, \; \mathbf k_2 - \mathbf k_1}^\dag \ca \oa \sin k_{1x} - \sin k_{2x} \ca + \oa b_{\hat{\mathbf y}, \; \mathbf k_1 - \mathbf k_2} + b_{\hat{\mathbf y}, \; \mathbf k_2 - \mathbf k_1}^\dag \ca \oa \sin k_{1y} - \sin k_{2y} \ca \cs.
\end{equation}

\subsection{Effect of phonons on short timescales} \label{ssec:phonon_scattering}
For the duration of the pump, we are interested in the phonons as a source of decoherence and relaxation of quasiparticles, so we can integrate them out to obtain jump operators. However, performing this procedure is not straightfoward in the original electron basis, due to the strong Hubbard interaction. We should therefore express \eqref{eq:ssh_momentum_definition} in terms of the doublon and hole operators. Again focusing only on the $\hat{\mathbf x}$ part for simplicity, first restrict the summation to the reduced Brillouin zone BZ':
    \begin{align}
	H_{\text{SSH}, \; \hat{\mathbf x}} = g_1 \sum_{\substack{\mathbf k_1, \mathbf k_2 \in BZ' \\ \spin}} & \bigg[ c_{\mathbf k_1, \spin}^\dag c_{\mathbf k_2, \spin} \oa b_{\hat{\mathbf x}, \; \mathbf k_1 - \mathbf k_2} + b_{\hat{\mathbf x}, \; \mathbf k_2 - \mathbf k_1}^\dag \ca \oa \sin k_{1x} - \sin k_{2x} \ca \nonumber \\
	& + c_{\mathbf k_1 + \mathbf Q, \spin}^\dag c_{\mathbf k_2, \spin} \oa b_{\hat{\mathbf x}, \; \mathbf k_1 - \mathbf k_2 + \mathbf Q} + b_{\hat{\mathbf x}, \; \mathbf k_2 - \mathbf k_1 + \mathbf Q}^\dag \ca \oa \sin (k_{1x} + \pi) - \sin k_{2x} \ca\nonumber \\
	& + c_{\mathbf k_1, \spin}^\dag c_{\mathbf k_2 + \mathbf Q, \spin} \oa b_{\hat{\mathbf x}, \; \mathbf k_1 - \mathbf k_2 + \mathbf Q} + b_{\hat{\mathbf x}, \; \mathbf k_2 - \mathbf k_1 + \mathbf Q}^\dag \ca \oa \sin k_{1x} - \sin (k_{2x} + \pi) \ca \nonumber \\
	& + c_{\mathbf k_1 + \mathbf Q, \spin}^\dag c_{\mathbf k_2 + \mathbf Q, \spin} \oa b_{\hat{\mathbf x}, \; \mathbf k_1 - \mathbf k_2} + b_{\hat{\mathbf x}, \; \mathbf k_2 - \mathbf k_1}^\dag \ca \oa \sin (k_{1x} + \pi) - \sin (k_{2x} + \pi) \ca \bigg].
\end{align}
The resulting terms can be further grouped together as follows:
\begin{align}
	H_{\text{SSH}, \; \hat{\mathbf x}} &= g_1 \sum_{\substack{\mathbf k_1, \mathbf k_2 \in BZ' \\ \spin}} \bigg[ \oa c_{\mathbf k_1, \spin}^\dag c_{\mathbf k_2, \spin} - c_{\mathbf k_1 + \mathbf Q, \spin}^\dag c_{\mathbf k_2 + \mathbf Q, \spin} \ca \oa b_{\hat{\mathbf x}, \; \mathbf k_1 - \mathbf k_2} + b_{\hat{\mathbf x}, \; \mathbf k_2 - \mathbf k_1}^\dag \ca \oa \sin k_{1x} - \sin k_{2x} \ca \nonumber\\
	& \qquad \qquad \qquad \quad + \oa c_{\mathbf k_1, \spin}^\dag c_{\mathbf k_2 + \mathbf Q, \spin} - c_{\mathbf k_1 + \mathbf Q, \spin}^\dag c_{\mathbf k_2, \spin} \ca \oa b_{\hat{\mathbf x}, \; \mathbf k_1 - \mathbf k_2 + \mathbf Q} + b_{\hat{\mathbf x}, \; \mathbf k_2 - \mathbf k_1 + \mathbf Q}^\dag \ca \oa \sin k_{1x} + \sin k_{2x} \ca \bigg] \nonumber \\
	&= g_1 \sum_{\substack{\mathbf k_1, \mathbf k_2 \in BZ' \\ \spin}} C_{\mathbf{k}_1, \spin}^\dag \bigg[ \pauli_z \oa b_{\hat{\mathbf x}, \; \mathbf k_1 - \mathbf k_2} + b_{\hat{\mathbf x}, \; \mathbf k_2 - \mathbf k_1}^\dag \ca \oa \sin k_{1x} - \sin k_{2x} \ca \nonumber \\
	& \qquad \qquad \qquad \qquad \quad + \oa i \pauli_y \ca \oa b_{\hat{\mathbf x}, \; \mathbf k_1 - \mathbf k_2 + \mathbf Q} + b_{\hat{\mathbf x}, \; \mathbf k_2 - \mathbf k_1 + \mathbf Q}^\dag \ca \oa \sin k_{1x} + \sin k_{2x} \ca \bigg] C_{\mathbf{k}_2, \spin} \nonumber \\
	&= g_1 \sum_{\substack{\mathbf k_1, \mathbf k_2 \in BZ' \\ \spin}} \bogoliubon_{\mathbf{k}_1, \spin}^\dag (t) \;  B_{\mathbf{k}_1, \spin}^\dag (t) \bigg[ \pauli_z \oa b_{\hat{\mathbf x}, \; \mathbf k_1 - \mathbf k_2} + b_{\hat{\mathbf x}, \; \mathbf k_2 - \mathbf k_1}^\dag \ca \oa \sin k_{1x} - \sin k_{2x} \ca \nonumber \\
	& \qquad \qquad \qquad \qquad + \oa i \pauli_y \ca \oa b_{\hat{\mathbf x}, \; \mathbf k_1 - \mathbf k_2 + \mathbf Q} + b_{\hat{\mathbf x}, \; \mathbf k_2 - \mathbf k_1 + \mathbf Q}^\dag \ca \oa \sin k_{1x} + \sin k_{2x} \ca \bigg] B_{\mathbf{k}_2, \spin} (t) \; \bogoliubon_{\mathbf{k}_2, \spin} (t) \nonumber \\
	&= g_1 \sum_{\substack{\mathbf k_1, \mathbf k_2 \in BZ' \\ \spin}} \bogoliubon_{\mathbf{k}_1, \spin}^\dag (t) \; P_{\mathbf k_1, \mathbf k_2, \spin} (t)  \; \bogoliubon_{\mathbf{k}_2, \spin} (t), 
\end{align}
where $P_{\mathbf k_1, \mathbf k_2, \spin} (t)$ is a $2 \times 2$ matrix. With the notations $u_j \equiv u_{\mathbf{k}_j} (t)$ and $v_j \equiv v_{\mathbf{k}_j} (t)$ for $j \in \{1,2\}$, it reads
\begin{align} \label{eq:scattering_expression_full}
	P_{\mathbf k_1, \mathbf k_2, \spin} &= (u_1 u_2 - v_1 v_2) \oa b_{\hat{\mathbf x}, \; \mathbf k_1 - \mathbf k_2} + b_{\hat{\mathbf x}, \; \mathbf k_2 - \mathbf k_1}^\dag \ca \oa \sin k_{1x} - \sin k_{2x} \ca \cdot \pauli_z \nonumber \\
	&\quad + \spin (u_1 v_2 + v_1 u_2) \oa b_{\hat{\mathbf x}, \; \mathbf k_1 - \mathbf k_2} + b_{\hat{\mathbf x}, \; \mathbf k_2 - \mathbf k_1}^\dag \ca \oa \sin k_{1x} - \sin k_{2x} \ca \cdot \pauli_x \nonumber \\
	&\quad + (u_1 u_2 + v_1 v_2) \oa b_{\hat{\mathbf x}, \; \mathbf k_1 - \mathbf k_2 + \mathbf Q} + b_{\hat{\mathbf x}, \; \mathbf k_2 - \mathbf k_1 + \mathbf Q}^\dag \ca \oa \sin k_{1x} + \sin k_{2x} \ca  \cdot \oa i \pauli_y \ca \nonumber \\
	&\quad + \spin (u_2 v_1 - u_1 v_2) \oa b_{\hat{\mathbf x}, \; \mathbf k_1 - \mathbf k_2 + \mathbf Q} + b_{\hat{\mathbf x}, \; \mathbf k_2 - \mathbf k_1 + \mathbf Q}^\dag \ca \oa \sin k_{1x} + \sin k_{2x} \ca \cdot \pauli_0 .
\end{align}
\end{widetext}
In this basis, the off-diagonal terms have physical significance of doublon-holon pair excitation or recombination; such processes involve an energy change of order $U$, and are assumed to be far off-resonance with the available phonon modes. We ignore them, and focus on terms proportional to $\pauli_z$ or $\pauli_0$, which scatter the quasiparticles within their effective bands.\\

Note that, at strong interactions $\tau / U \ll 1$, both the $u$ and $v$ coefficients of \eqref{eq:bogoliubov_coefficients} are equal to $1/\sqrt 2$ at zeroth order, with corrections appearing at higher powers of $\tau / U$. In consequence, prefactors of the form $uu - vv$ or $u v - v u$ have a vanishing zeroth-order contribution, and will be small. Specifically,
\begin{subequations}
\begin{align}
	u_1 u_2 - v_1 v_2 &= {\e_{\mathbf k_2} + \e_{\mathbf k_1} \over U S} \op 1 + \OO \os \op {\tau \over U}\cp ^2 \cs \cp, \\
	u_2 v_1 - u_1 v_2 &= {\e_{\mathbf k_2} - \e_{\mathbf k_1} \over U S} \op 1 + \OO \os \op {\tau \over U}\cp ^2 \cs \cp.
\end{align}
\end{subequations}
so that the effective coupling between quasipaticles and phonons is weaker than the original $g_1$ by a factor of $\tau / U$. The scattering of quasiparticles on phonons is then described by jump operators acting on doublons and holes, of the form
\begin{subequations}
\begin{align} \label{eq:jump_operators_form}
    L_{\mathbf k_1, \mathbf k_2, T}^d &= \sqrt{\Gamma^d_{\mathbf k_1, \mathbf k_2, T}} \; d^{\dag}_{\mathbf k_1} d_{\mathbf k_2}, \\
    L_{\mathbf k_1, \mathbf k_2, T}^h &= \sqrt{\Gamma^h_{\mathbf k_1, \mathbf k_2, T}} \; h^{\dag}_{\mathbf k_1} h_{\mathbf k_2}.
\end{align}
\end{subequations}
where the relaxation rates $\Gamma_{\mathbf k_1, \mathbf k_2, T}^{d/h}$ can be calculated from \eqref{eq:scattering_expression_full} using Fermi's golden rule, and will depend on both momenta $\mathbf k_1, \mathbf k_2$, as well as the temperature $T$ of the phonon bath.

\subsection{Coherent phonons at intermediate times} \label{ssec:coherent_phonons_analytic_introduction}
Following photoexcitation, quasiparticles can relax within their respective Hubbard bands, by transferring energy to other degrees of freedom (e.g. magnons, optical phonons, etc). After this has occurred, but on timescales still shorter than those relevant for recombination, we can ignore electronic dynamics, and replace the relevant operators by their expectation values. This will yield an effective displacement for the acoustic phonons.\\

For this part, nonuniformities in the spatial profile of the driving field are essential, as the SSH coupling between electrons and phonons vanishes at zero momentum. In an experimental pump-probe context, driving acts mainly at the edge of the system, with the electric field amplitude decreasing exponentially with depth inside the sample. Assuming the pump penetration depth to be large compared to the doublon-hole coherence length, we will work in a local-density approximation. The effect of the pumping is first calculated, assuming its strength to be uniform over a small region; then, the new electronic state is used to find an effective phonon Hamiltonian.\\

Specifically, we start from the SSH coupling term in real space \eqref{eq:ssh_real_space_quantized}, taking the expectation of electronic terms and Fourier transforming the phonon only:
\begin{widetext}
    \begin{align} 
    H_{\text{SSH}, \; \hat{\mathbf x}}^\text{eff} &= -a_0 g \sum_{\mathbf j, \spin} \ex{ c_{\mathbf j + \hat{\mathbf x}, \spin}^\dag c_{\mathbf j, \spin} + c_{\mathbf j, \spin}^\dag c_{\mathbf j + \hat{\mathbf x}, \spin}} \oa b_{\hat{\mathbf x}, \; \mathbf j + \hat{\mathbf x}}^\dag - b_{\hat{\mathbf x}, \; \mathbf j}^\dag \ca + \text{h.c.} \nonumber\\
	&= -g_1 \sum_{\mathbf q} b_{\hat{\mathbf x}, \; \mathbf q}^\dag \; \sin \op {q_x \over 2} \cp \; \os \sum_{\mathbf j,  \spin}  \ex{ c_{\mathbf j + \hat{\mathbf x}, \spin}^\dag c_{\mathbf j, \spin} + c_{\mathbf j, \spin}^\dag c_{\mathbf j + \hat{\mathbf x}, \spin}} e^{-i \mathbf q (\mathbf j + \hat{\mathbf x} / 2)} \cs + \text{h.c.}. \label{eq:late_kinetic_energy_coupling_to_phonons}
\end{align}
\end{widetext}
and a similar expression holds for the $\hat{\mathbf y}-$polarized component. The term inside the square brackets is the Fourier component, at momentum $\mathbf q$, of $\ex{ c_{\mathbf j + \hat{\boldsymbol \delta}}^\dag c_{\mathbf j} + c_{\mathbf j}^\dag c_{\mathbf j + \hat{\boldsymbol \delta}}}$, which describes the local kinetic energy density. We make the following observations:
\begin{itemize}
	\item Momenta significantly larger than $1/\lambda$ correspond to length scales that are small compared to the pump penetration depth, over which the drive strength does not vary appreciably; the Fourier component of the kinetic energy will be negligible at those $\mathbf q$.
	\item On the other hand, contributions at small wavevectors are suppressed by the SSH coupling itself, as evidenced by the factors of $\sin \op {q_{x/y} / 2} \cp$.
\end{itemize} 
In consequence, we expect the dominant contribution to come from $q \sim \lambda^{-1}$.\\

Within a local-density approximation, terms of the form $\ex{ c_{\mathbf j + \hat{\boldsymbol \delta}}^\dag c_{\mathbf j} + c_{\mathbf j}^\dag c_{\mathbf j + \hat{\boldsymbol \delta}}}$ can be related to the kinetic energy:
\begin{equation}
    \sum_{\mathbf j} \op c_{\mathbf j + \hat{\boldsymbol \delta}}^\dag c_{\mathbf j} + c_{\mathbf j}^\dag c_{\mathbf j + \hat{\boldsymbol \delta}} \cp = -{1 \over 2 \tau} \op -\tau \sum_{\mathbf j, \boldsymbol \delta} c_{\mathbf j + \hat{\boldsymbol \delta}}^\dag c_{\mathbf j}  \cp.
\end{equation}
where, in the sums, the site index $\mathbf j$ runs over a small region appropriate for LDA. The term inside the parenthesis of the RHS is the kinetic energy, whose expectation at late times in our variational state is given in eq. \eqref{eq:kinetic_energy_late_time} below. Plugging it back into \eqref{eq:late_kinetic_energy_coupling_to_phonons}, we find
\begin{widetext}
    \begin{align}
    H_{\text{SSH}, \; \hat{\mathbf x}}^\text{eff} &= - {g_1 \over 2 \tau} \sum_{\mathbf q} b_{\hat{\mathbf x}, \; \mathbf q}^\dag \; \sin \op {q_x \over 2} \cp \; \int d \depth \op {1 \over N_{BZ'}} \sum_{\mathbf k \in BZ'} {\e_{\mathbf k}^2 \over E_{\mathbf k}} \; \rho_{\mathbf k, z} \cp_{\depth} e^{-i q \depth} + \text{h.c.} \nonumber \\
    &= - {g_1 \over 4 \tau} \sum_{\mathbf q} b_{\hat{\mathbf x}, \; \mathbf q}^\dag \; \sin \op {q_x \over 2} \cp \; \int d \depth \op {1 \over N_{BZ'}} \sum_{\mathbf k \in BZ'} {\e_{\mathbf k}^2 \over E_{\mathbf k}} \; n_{\text{qp}, \mathbf k} \cp_{\depth} e^{-i q \depth} + \text{h.c.}.
\end{align}
\end{widetext}
Here, $\depth$ represents the depth inside the sample, measured from the pumped surface. After intraband relaxation, if the density of photoexcited carriers is low, the doublons (holes) will all reside at the bottom (top) of the upper (lower) Hubbard band. In this approximation, one can ignore details of quasiparticle dispersion, and the local kinetic energy per carrier will be proportional to $n_\text{qp}$. In turn, the resulting displacement of the phonon mode will be proportional to $n_\text{qp}$ as well, and so the CLAP amplitude can be used to infer the photoexcited carrier density. On the other hand, if the excitation density is larger, factors such as the shape of Hubbard bands will become relevant, complicating the expression for the effective phonon displacement. A numerical approach will be necessary in this situation, whose results are discussed in Section \ref{sec:acoustic_phonons} of the main text.\\

\section{Variational ansatz and evolution equations} \label{sec:sm_ansatz}
\subsection{Density-matrix ansatz}
As discussed in section \ref{ssec:order_parameter_and_factorization}, we employ a Gaussian ansatz which captures correlations between operators with momenta $\mathbf k$ and $\mathbf k + \mathbf Q$ and identical spins, while taking other expectations of the form $\ex{c^\dag_{\mathbf k} c_{\mathbf k + \mathbf q}}$ to vanish whenever $\mathbf q \notin \{0, \mathbf Q\}$. To include effects of quasiparticle scattering / dephasing, we formulate the problem as an open system. Therefore, we take an ansatz of the following form for the density matrix:
\begin{equation} \label{eq:density_matrix_ansatz}
    \rho_\text{var}(t) = \bigotimes_{\substack{\mathbf k \in BZ' \\ \spin}} \rho_{\mathbf k, \spin}(t)
\end{equation}
Each $\rho_{\mathbf k, \spin}$ matrix, labeled by spin $\spin$ and momentum $\mathbf k$ in the reduced Brillouin zone, describes the subspace spanned by two momentum states, and therefore has size $4 \times 4$. For simplicity, the spin label $\spin$ will be omitted moving forwards. We make the following conventions for the ordering of basis states in each such sector:
\begin{itemize}
    \item Working with the original electronic states, pick the order as $\{\ket{\mathbf k}, \ket{\mathbf k + \mathbf Q}, \ket{0}, \ket{\mathbf k + \mathbf Q, \mathbf k}\}$, where $\ket{0}$ denotes electronic vacuum (both momentum states unoccupied), and the other three states are defined via $\ket{\mathbf q} = c_{\mathbf q}^\dag \ket {0}$ and $\ket{\mathbf q_1, \mathbf q_2} = c_{\mathbf q_2}^\dag c_{\mathbf q_1}^\dag \ket {0}$ respectively.
    \item In the doublon / hole basis, $\ket{0}$ will represent a state with one hole and no doublon. In consequence, the vacuum is taken to be $\ket{\vac_{\mathbf k}} \equiv h_{-\mathbf k} \ket{0}$, and we define
    \begin{align}
        \ket{h,d} &\equiv d^{\dag}_{\mathbf k} h^{\dag}_{-\mathbf k} \ket{\vac_{\mathbf k}} = d^{\dag}_{\mathbf k} \ket{0}\\
        \ket{h} &\equiv h^{\dag}_{-\mathbf k} \ket{\vac_{\mathbf k}} = \ket{0} \nonumber \\
        \ket{d} &\equiv d^{\dag}_{\mathbf k} \ket{\vac_{\mathbf k}} = d^{\dag}_{\mathbf k} h_{-\mathbf k} \ket{0} \nonumber 
    \end{align}
    Note that the $d_{\mathbf k}$ and $h_{-\mathbf k}$ operators, defined in \eqref{eq:bogoliubov_operators_definition}, become time-dependent once the drive is applied. We choose the ordering $\{\ket{h,d}, \ket{\vac_{\mathbf k}}, \ket{h}, \ket{d}\}$. With this convention, the last two states are identical in both bases: $\ket{h} = \ket{0}$ represents no electronic occupation, and $\ket{d} = \ket{\mathbf k + \mathbf Q, \mathbf k}$ (which results from the identity $d^{\dag}_{\mathbf k} h_{-\mathbf k} = c_{\mathbf k}^\dag c_{\mathbf k + \mathbf Q}^\dag$) describes double occupation. Moreover, the first two states $\ket{h,d}$, $\ket{\vac_{\mathbf k}}$ in the doublon/hole basis are related to the original electronic $\ket{\mathbf k}$ and $\ket{\mathbf k + \mathbf Q}$ by the Bogoliubov transformation \eqref{eq:bogoliubov_definition}.
\end{itemize}

\subsection{Time evolution of density matrix elements}
Since we restrict our attention to spatially uniform driving of the system, only doublon-hole pairs at zero total momentum will be created, and the electronic occupations of each momentum sector defined above will be constant. Combined with the assumption that all sectors start in their respective $\ket{\vac_{\mathbf k}}$, we see that the drive will only couple the first two states in each basis. Scattering of doublons and holes, however, can link different momentum sectors and therefore modify their occupations; this will be implemented by terms giving matrix elements between the $\{\ket{h,d}, \ket{\vac_{\mathbf k}}\}$ and $\{\ket{h}, \ket{d}\}$ subspaces.\\

We start by focusing on the effect of the drive. After performing the decoupling discussed in Section \ref{ssec:order_parameter_and_factorization}, the system's coherent evolution, expressed in the electronic basis, will be described by
\begin{equation} \label{eq:electron_hamiltonian_evolution}
    \partial_t \rho^\text{el}_{\mathbf k, \spin} = - i \; [\mathbb{H}^\text{HF}_{\mathbf k, \spin}(t), \rho^\text{el}_{\mathbf k, \spin}]
\end{equation}
Since the sector density matrix $\rho^\text{el}_{\mathbf k, \spin}$ has size $4 \times 4$, we have to correspondingly extend the sector Hamiltonian from the $2 \times 2$ matrix $H^\text{HF}_{\mathbf k, \spin}(t)$ of Section \ref{ssec:order_parameter_and_factorization}, to the following:
\begin{align}
    \mathbb{H}^\text{HF}_{\mathbf k, \spin}(t) &= \begin{pmatrix}
	\e_{\mathbf k}(t) & - \spin U S(t) / 2 & 0 & 0\\ 
	- \spin U S(t) / 2 & - \e_{\mathbf k} (t) & 0 & 0\\
	0 & 0 & 0 & 0\\
	0 & 0 & 0 & 0
\end{pmatrix} \nonumber \\
&\equiv \begin{pmatrix}
	H^\text{HF}_{\mathbf k, \spin}(t) & 0\\ 
	0 & 0
\end{pmatrix}
\end{align}
In turn, the transformation which diagonalizes it will be
\begin{align} 
    \mathbb{B}_{\mathbf k, \spin}(t) &= \begin{pmatrix}
	u_{\mathbf k}(t) & \spin v_{\mathbf k}(t) & 0 & 0\\ 
	- \spin v_{\mathbf k}(t) & u_{\mathbf k}(t) & 0 & 0\\
	0 & 0 & 1 & 0\\
	0 & 0 & 0 & 1
\end{pmatrix} \nonumber \\ 
&\equiv \begin{pmatrix}
	B_{\mathbf k, \spin}(t) & 0\\ 
	0 & \pauli_0
\end{pmatrix} \label{eq:bogoliubov_matrix_4d_definition}
\end{align}
In the following, we work within a single $(\mathbf k, \spin)$ sector, and therefore omit both of these indices within most equations; explicit time dependence will also be suppressed for simplicity. When writing Pauli matrices in the context of a 4-dimensional subspace, they will be understood to act on the first two dimensions only, unless specified otherwise.\\

The Hamiltonian is diagonalized as $\mathbb{H}^\text{HF} = \wsenergy \cdot \mathbb{B} \pauli_z \mathbb{B}^T$, with $\wsenergy$ given by \eqref{eq:quasiparticle_dispersion}. We correspondingly transform the density matrix to the rotating basis, via $\rho^\text{el} = \mathbb{B} \rho \mathbb{B}^T$. Using \eqref{eq:electron_hamiltonian_evolution}, we write $\partial_t \rho_\text{el}$ in two ways:
\begin{equation}
    - i \wsenergy \; \mathbb{B} [\pauli_z, \rho]  \mathbb{B}^T = [\partial_t \mathbb{B}] \rho \mathbb{B}^T + \mathbb{B} [\partial_t \rho] \mathbb{B}^T + \mathbb{B} \rho [\partial_t \mathbb{B}^T]
\end{equation}
or equivalently
\begin{equation} \label{eq:hamiltonian_dm_contribution}
    \partial_t \rho = - i \wsenergy \; [\pauli_z, \rho] - \mathbb{B}^T [\partial_t \mathbb{B}] \; \rho - \rho \; [\partial_t \mathbb{B}^T] \mathbb{B}
\end{equation}
From eqs. \eqref{eq:bogoliubov_coefficients} and \eqref{eq:bogoliubov_matrix_4d_definition}, together with the definition
\begin{equation} \label{eq:bogoliubov_angle_definition}
    \bogangle_{\mathbf k}(t) \equiv \arctan \op {2 \e_{\mathbf k}(t) \over U S(t)} \cp
\end{equation}
we obtain the terms describing photoexcitation:
\begin{align}
    \mathbb{B}^T [\partial_t \mathbb{B}] &= - (\partial_t \bogangle) \; {i \spin \over 2} \; \pauli_y \nonumber \\
    [\partial_t \mathbb{B}^T] \mathbb{B} &= + (\partial_t \bogangle) \; {i \spin \over 2} \; \pauli_y
\end{align}
Note that the time dependence of $\bogangle$ can arise both from the direct effect of the external drive (modulating kinetic energy via $\e_{\mathbf k}(t) = \e_{\mathbf k - e \mathbf A(t)}$), as well as from dynamics of the order parameter $S$. Under weak subgap driving, photoexcitation is inefficient, and $S$ will not suffer significant changes; in this regime, we can use the rigid-band approximation $S(t) = S_0$ and the direct contribution to the time dependence of $\bogangle$ will dominate.\\

We turn to describing scattering processes. The generic form of jump operators, which describe the intraband relaxation of carriers, has been discussed in Section \ref{ssec:phonon_scattering}. Since in the ansatz \eqref{eq:density_matrix_ansatz} we are ignoring entanglement between different $(\mathbf k, \spin)$ sectors, when focusing on one such subspace we may trace out all the others. This results in the following jump operators acting on a single sector:
\begin{widetext}
\begin{align} \nonumber
    L_{\mathbf k, \spin}^{d, \text{in}} \equiv \sqrt{\Gamma^{d, \text{in}}_{\mathbf k, T}} \; d^\dag_{\mathbf k, \spin} &\text{ with rate } \Gamma^{d, \text{in}}_{\mathbf k, T} = \sum_{\mathbf q} \Gamma^d_{\mathbf k, \mathbf q, T} \ex{d^\dag_{\mathbf q, \spin} d_{\mathbf q, \spin}}\\
    L_{\mathbf k, \spin}^{d, \text{out}} \equiv \sqrt{\Gamma^{d, \text{out}}_{\mathbf k, T}} \; d_{\mathbf k, \spin} &\text{ with rate } \Gamma^{d, \text{out}}_{\mathbf k, T} = \sum_{\mathbf q} \Gamma^d_{\mathbf q, \mathbf k, T} \os 1 - \ex{d^\dag_{\mathbf q, \spin} d_{\mathbf q, \spin}} \cs \nonumber \\
    L_{\mathbf k, \spin}^{h, \text{in}} \equiv \sqrt{\Gamma^{h, \text{in}}_{\mathbf k, T}} \; h^\dag_{\mathbf k, \spin} &\text{ with rate } \Gamma^{h, \text{in}}_{\mathbf k, T} = \sum_{\mathbf q} \Gamma^h_{\mathbf k, \mathbf q, T} \ex{h^\dag_{\mathbf q, \spin} h_{\mathbf q, \spin}} \nonumber \\
    L_{\mathbf k, \spin}^{h, \text{out}} \equiv \sqrt{\Gamma^{h, \text{out}}_{\mathbf k, T}} \; h_{\mathbf k, \spin} &\text{ with rate } \Gamma^{h, \text{out}}_{\mathbf k, T} = \sum_{\mathbf q} \Gamma^h_{\mathbf q, \mathbf k, T} \os 1 - \ex{h^\dag_{\mathbf q, \spin} h_{\mathbf q, \spin}} \cs \label{eq:sector_jump_operators} 
\end{align}
One can further consider dephasing of doublon-hole pairs due to other processes, which will generally have the form
\begin{equation}
    L_{\mathbf k, \spin}^\text{deph} = \sqrt{\Gamma^\text{deph}_{\mathbf k, T}} \; \bogoliubon_{\mathbf k, \spin}^\dag \pauli_z \bogoliubon_{\mathbf k, \spin}
\end{equation}
Combining the action of the jump operators with eq. \eqref{eq:hamiltonian_dm_contribution}, the evolution of the sector density matrix is described by
    \begin{equation} \label{eq:density_matrix_evolution_equation}
    \partial_t \rho = - i \wsenergy \; [\pauli_z, \rho] - \mathbb{B}^T [\partial_t \mathbb{B}] \; \rho - \rho \; [\partial_t \mathbb{B}^T] \mathbb{B} + \Gamma^\text{deph} \op \pauli_z \; \rho \; \pauli_z - \rho \cp + \sum_{L} \op L \rho L^\dag - {1 \over 2} \{L^\dag L, \rho\}\cp
\end{equation}
\end{widetext}
where the sum over $L$ in the last term runs through the operators in \eqref{eq:sector_jump_operators}, and indices such as $\mathbf k, \spin, T$ as well as explicit time dependences have been omitted. We make the assumption that the relaxation rates are symmetric between doublons and holes: $\Gamma^{d, \text{in}}_{\mathbf k, T} = \Gamma^{h, \text{in}}_{\mathbf k, T} \equiv \Gamma^\text{in}_{\mathbf k, T}$, and similarly for the outgoing contributions.\\

Under these assumptions, when starting from the variational ground state $\ket{\vac}$ described in Section \ref{ssec:static_solution_variational}, the time-dependent sector density matrix only needs 4 independent entries for the system of evolution equations to close. We take 
\begin{equation} \label{eq:density_matrix_parametrization}
    \rho = \op {1 \over 2} - \rho_{t} \cp \pauli_0 + \rho_x \; \spin \pauli_x + \rho_y \; \spin \pauli_y - {\rho_z \over 2} \; \pauli_z + \rho_t \; \pauli_t
\end{equation}
where $\{\pauli_0, \pauli_x, \pauli_y, \pauli_z\}$ denote the usual Pauli matrices acting on the first 2 dimensions only, while $\pauli_t$ is defined as the identity on the last two dimensions:
\begin{equation}
    \pauli_t \equiv \begin{pmatrix}
	0 & 0\\ 
	0 & \pauli_0
\end{pmatrix}
\end{equation}
Note that the variational ground state $\ket{\vac}$ corresponds to $\rho_z = 1$, and $\rho_{x} = \rho_{y} = \rho_{t} = 0$, so these values are used as initial conditions for the time evolution.\\

Plugging the parametrization \eqref{eq:density_matrix_parametrization} into the evolution \eqref{eq:density_matrix_evolution_equation}, the resulting set of equations is
\begin{align}
    \partial_t \rho_x &= {\partial_t \bogangle \over 2} \; \rho_z - 2 \wsenergy \; \rho_y - (\Gamma^\text{in} + \Gamma^\text{out} + 2 \Gamma^\text{deph}) \; \rho_x \nonumber \\ \nonumber
	\partial_t \rho_y &= 2 \wsenergy \; \rho_x - (\Gamma^\text{in} + \Gamma^\text{out} + 2 \Gamma^\text{deph}) \; \rho_y\\ \nonumber
	\partial_t \rho_z &= - 2 (\partial_t \bogangle) \; \rho_x - \op \Gamma^\text{in} + \Gamma^\text{out} \cp \; \op \rho_z - {\Gamma^\text{out} - \Gamma^\text{in} \over \Gamma^\text{out} + \Gamma^\text{in}} \cp \\ 
	\partial_t \rho_t &= - 2 \Gamma \op \rho_t - {1 - \rho_z \; {\Gamma^\text{out} - \Gamma^\text{in} \over \Gamma^\text{out} + \Gamma^\text{in}} \over 4}\cp
\end{align}
At this point, we adopt a relaxation-time approximation, in which the momentum dependences of $\Gamma^\text{relax}_{\mathbf k, T} \equiv \Gamma^\text{in}_{\mathbf k, T} + \Gamma^\text{out}_{\mathbf k, T}$ as well as of $\Gamma^\text{deph}_{\mathbf k, T}$ are ignored, and the ratio $(\Gamma^\text{out} - \Gamma^\text{in}) / (\Gamma^\text{out} + \Gamma^\text{in})$ in the third equation is replaced by an instantaneous configuration $\rho_z^\insteq$ that intraband relaxation tends to move the system towards. Such a quasi-equilibrium, metastable configuration can be estimated at any point in time by first finding the total number of excitations present in the system 
\begin{equation} \label{eq:total_quasiparticle_number}
    N_\text{qp}(t) = \sum_{\substack{\mathbf k \in BZ' \\ \spin}} {1 - \rho_{\mathbf k, z}(t) \over 2}
\end{equation}
and then taking $\rho_z^\insteq$ to describe a thermal distribution of the $N_\text{qp}(t)$ doublons (holes) within the upper (lower) Hubbard band. If intraband relaxation is dominated by phonons, then the corresponding temperature should be that of the lattice. Note that doublon-hole recombination is ignored here, since it is expected to take place on timescales much longer than the pump duration. We arrive at the system
\begin{align} \nonumber
    \partial_t \rho_x &= {\partial_t \bogangle \over 2} \; \rho_z - 2 \wsenergy \; \rho_y - (\Gamma^\text{relax} + 2 \Gamma^\text{deph}) \; \rho_x\\ \nonumber
	\partial_t \rho_y &= 2 \wsenergy \; \rho_x - (\Gamma^\text{relax} + 2 \Gamma^\text{deph}) \; \rho_y\\ \nonumber
	\partial_t \rho_z &= - 2 (\partial_t \bogangle) \; \rho_x - \Gamma^\text{relax} \; \op \rho_z - \rho_z^\insteq \cp \\ 
	\partial_t \rho_t &= - 2 \Gamma \op \rho_t - {1 - \rho_z \; \rho_z^\insteq \over 4}\cp \label{eq:square_evolution_equations_rta}
\end{align}
After the driving has stopped, time dependence of $\bogangle$ can only arise from ringing of the gap $\Delta(t) = U S(t)$. If that effect is small, the terms involving $\partial_t \bogangle$ can be ignored in \eqref{eq:square_evolution_equations_rta}, and the conclusion will be that off-diagonal density matrix elements decay at rate $\Gamma^\text{relax} + 2 \Gamma^\text{deph}$, while the photoexcited carrier population relaxes intraband with rate $\Gamma^\text{relax}$ without recombining. If, on the other hand, the gap keeps ringing, total photocarrier density can still vary until a metastable state is reached, on a timescale set by $(\Gamma^\text{relax})^{-1}$.\\

Further simplification is possible if we specialize to pump durations much shorter than the typical intraband relaxation time: $T_\text{pump} \ll (\Gamma^\text{relax})^{-1}$. Then, transfer of photoexcited carriers between different momenta can be ignored while the pump is active. Dropping the second term in the equation for $\rho_z$, and defining\footnote{We are ignoring $\Gamma^\text{relax}$ in the equation for $\rho_z$, but keeping it in those for $\rho_{x/y}$. The reason is that, in the former case, it only leads to a (slow) redistribution of the photoexcited carriers between momentum states; in the latter, it strongly impacts excitation rates under weak drive. See Section \ref{sec:analytic_approximate_solutions} for further discussion.} a total dephasing rate $\Gamma \equiv \Gamma^\text{relax} + 2 \Gamma^\text{deph}$, we have on short timescales
\begin{align} \nonumber
    \partial_t \rho_x &= {\partial_t \bogangle \over 2} \; \rho_z - 2 E_\mathbf{k} \; \rho_y - \Gamma \rho_x \\ \nonumber
	\partial_t \rho_y &= 2 E_\mathbf{k} \; \rho_x - \Gamma \rho_y \\ 
	\partial_t \rho_z &= - 2 (\partial_t \bogangle) \; \rho_x \label{eq:square_evolution_equations_dme}
\end{align}
where the equation for $\rho_t$ has been dropped, since the three sets of matrix elements in \eqref{eq:square_evolution_equations_dme}
suffice for calculating observables of interest.

\subsection{Time evolution of order parameter}
We now turn to the evolution of $S$, which is coupled to the system \eqref{eq:square_evolution_equations_dme} via the definition \eqref{eq:bogoliubov_angle_definition} of $\bogangle$, as well as its presence in the quasiparticle dispersion \eqref{eq:quasiparticle_dispersion}. Recall the expression \eqref{eq:order_parameter_bog_basis} for $S^z_{\mathbf Q}$ in the doublon/hole basis:
\begin{equation}
    S^z_{\mathbf Q} = {1 \over N} \sum_{\substack{\mathbf k \in BZ' \\ \spin}} \bogoliubon_{\mathbf{k}, \spin}^\dag \; {\e_{\mathbf k} \; (\spin \pauli_x) - (U S /2) \; \pauli_z \over E_{\mathbf k}} \; \bogoliubon_{\mathbf{k}, \spin}
\end{equation}
We calculate the time derivative of the order parameter $S(t) = \trace [S^z_{\mathbf Q} \; \rho(t)]$:
\begin{align}
    \partial_t S &= \partial_t \trace [S^z_{\mathbf Q} \; \rho] \nonumber \\
    &= {1 \over N} \sum_{\substack{\mathbf k \in BZ' \\ \spin}} \spin \; \partial_t \trace [ \mathbb{B}^T \pauli_x \mathbb{B} \rho]
\end{align}
which, upon expansion of the time derivative, yields
\begin{widetext}
\begin{align} 
	 \partial_t S &= {1 \over N} \sum_{\substack{\mathbf k \in BZ' \\ \spin}} \spin \ob \trace [ (\partial_t \mathbb{B}^T) \pauli_x \mathbb{B} \rho] + \trace [ \mathbb{B}^T \pauli_x (\partial_t  \mathbb{B}) \rho] + \trace [ \mathbb{B}^T \pauli_x \mathbb{B} (\partial_t \rho)] \cb \label{eq:expanding_derivative_of_op}
\end{align}
The first two terms can be combined using the properties of $\mathbb{B}$ from before:
\begin{alignat}{3}
    (\partial_t \mathbb{B}^T) \pauli_x \mathbb{B} \rho &= [(\partial_t \mathbb{B}^T) \mathbb{B}] \; \mathbb{B}^T \pauli_x \mathbb{B} \rho &&= +(\partial_t \bogangle) \; {i \spin \over 2} \; \pauli_y \mathbb{B}^T \pauli_x \mathbb{B} \rho \nonumber\\
	 \mathbb{B}^T \pauli_x (\partial_t \mathbb{B}) \rho &= \mathbb{B}^T \pauli_x \mathbb{B} \; [\mathbb{B}^T (\partial_t \mathbb{B})] \; \rho &&= -(\partial_t \bogangle) \; {i \spin \over 2} \;  \mathbb{B}^T \pauli_x \mathbb{B} \pauli_y \rho 
\end{alignat}
which combine to
\begin{equation} \label{eq:combined_transition_effect_on_op}
    \trace [ (\partial_t \mathbb{B}^T) \pauli_x \mathbb{B} \rho + \mathbb{B}^T \pauli_x (\partial_t \mathbb{B}) \rho] = (\partial_t \bogangle) \; {i \spin \over 2} \; \trace \oc [\pauli_y, \mathbb{B}^T \pauli_x \mathbb{B}] \; \rho \cc
\end{equation}
Plugging the results of \eqref{eq:density_matrix_evolution_equation} and \eqref{eq:combined_transition_effect_on_op} into \eqref{eq:expanding_derivative_of_op}, we arrive at an expression which has both coherent and relaxation contributions:
\begin{equation} \label{eq:full_s_evolution_equation}
    \partial_t S = -{1 \over N_{BZ'}} \sum_{\mathbf k \in BZ'} \os 4 \e_{\mathbf k} \; \rho_{\mathbf k, y} + 2 \Gamma \; { \e_{\mathbf k} \over E_{\mathbf k}} \;  \rho_{\mathbf k, x} + \Gamma^\text{relax} \; {U S \over 2 E_{\mathbf k}} \op \rho_{\mathbf k, z} - \rho_{\mathbf k, z}^\insteq \cp \cs
\end{equation}
As before, we argue that the last term can be neglected. Observe that the first contribution in the square brackets is of order $\tau$ due to the presence of $\e_{\mathbf k}$, while the second term goes as $\Gamma \; \tau / U$. Since $E_{\mathbf k}$ is of order $U$, it may seem that the third term is of order $\Gamma$, which would make it dominate over the second in the $\tau \ll U$ limit. However, from eq. \eqref{eq:total_quasiparticle_number} and the definition of $\rho_{\mathbf k, z}^\insteq$ which follows it, one finds
\begin{equation}
    \sum_{\mathbf k \in BZ'} \rho_{\mathbf k, z} = N_{BZ'} - N_\text{qp}(t) = \sum_{\mathbf k \in BZ'} \rho_{\mathbf k, z}^\insteq
\end{equation}
leading to the identity
\begin{equation}
    \sum_{\mathbf k \in BZ'} \op \rho_{\mathbf k, z} - \rho_{\mathbf k, z}^\insteq \cp = 0
\end{equation}
so the third term, summed over all momenta, cancels to leading order. A more careful treatment, based on a series expansion of \eqref{eq:quasiparticle_dispersion} in $\tau / U$, yields
    \begin{align}
    \Gamma^\text{relax} \; \sum_{\mathbf k \in BZ'} {U S \over 2 E_{\mathbf k}} \op \rho_{\mathbf k, z} - \rho_{\mathbf k, z}^\insteq \cp &= \Gamma^\text{relax} \; \sum_{\mathbf k \in BZ'} \op 1 - 2 \op {\e_{\mathbf k} \over U S} \cp^2 + \OO \os \op {\tau \over U} \cp^4 \cs \cp \op \rho_{\mathbf k, z} - \rho_{\mathbf k, z}^\insteq \cp \nonumber \\
    &= -2 \Gamma^\text{relax} \; \sum_{\mathbf k \in BZ'} \op {\e_{\mathbf k} \over U S} \cp^2  \op \rho_{\mathbf k, z} - \rho_{\mathbf k, z}^\insteq \cp + \OO \os \Gamma^\text{relax} \op {\tau \over U} \cp^4 \cs 
\end{align}
showing that the third term of \eqref{eq:full_s_evolution_equation} is in fact of order $\Gamma \; (\tau / U)^2$, and therefore smaller than both of the other contributions. To capture the system's short-time behavior, then, we restrict to the following simplified form of \eqref{eq:full_s_evolution_equation}:
\begin{equation} \label{eq:simplified_s_evolution_equation}
    \partial_t S = -{1 \over N_{BZ'}} \sum_{\mathbf k \in BZ'} \os 4 \e_{\mathbf k} \; \rho_{\mathbf k, y} + 2 \Gamma \; { \e_{\mathbf k} \over E_{\mathbf k}} \;  \rho_{\mathbf k, x} \cs
\end{equation}
Together, the system of equations formed by \eqref{eq:square_evolution_equations_dme} and \eqref{eq:simplified_s_evolution_equation} can be used to capture the photoexcitation process. Before discussing solutions, we show how to calculate a few relevant observables within our variational ansatz.

\subsection{Computing observables} 
An observable defined on a single $(\mathbf k, \spin)$ sector can be expanded as
    \begin{equation}
    O = O_0 \; \pauli_0 + O_x \; \spin \pauli_x + O_y \; \spin \pauli_y + O_z \; \pauli_z + O_\text{double} P_\text{double} + O_\text{empty} P_\text{empty}
\end{equation}
where $P_\text{double}$ projects onto the doubly-occupied state in the sector, and $P_\text{empty}$ onto the electronic vacuum. The observable's expectation becomes
\begin{equation} \label{eq:general_observable_expectation}
    \ex{O} = \text{trace} (O \rho) = O_0 + 2 O_x \rho_x + 2 O_y \rho_y - O_z \rho_z + (O_\text{double} + O_\text{empty} - 2 O_0) \; \rho_t
\end{equation}
\end{widetext}

\subsubsection{Order parameter}
Starting from \eqref{eq:order_parameter_fourier_transformed} and using \eqref{eq:general_observable_expectation}, we find:
\begin{equation}
    S(t) = {1 \over N_{BZ'}} \sum_{\mathbf k \in BZ'} {2 \e_{\mathbf k} \; \rho_{\mathbf k, x} + (U S / 2) \; \rho_{\mathbf k, z} \over E_{\mathbf k}}
\end{equation}

\subsubsection{Total photocarrier density}
Similarly to eq. \eqref{eq:total_quasiparticle_number}, we get the density of doublons and holes as
\begin{equation} \label{eq:total_quasiparticle_density}
    n_\text{qp}(t) = {1 \over N_{BZ'}} \sum_{\mathbf k \in BZ'} {1 - \rho_{\mathbf k, z}(t) \over 2}
 \end{equation}

\subsubsection{Total energy}
The time-dependent part of the Hamiltonian is 
\begin{equation}
    H = \sum_{\substack{\mathbf k \in BZ' \\ \spin}} E_{\mathbf k} \cdot \bogoliubon_{\mathbf{k}, \spin}^\dag \; \pauli_z \; \bogoliubon_{\mathbf{k}, \spin} + {U N S^2 \over 4}
\end{equation}
yielding in turn the total energy per site
\begin{align}
    {E_\text{tot} \over N} &= {U S^2 \over 4} - {1 \over N} \sum_{\substack{\mathbf k \in BZ' \\ \spin}} E_{\mathbf k} \; \rho_{\mathbf k, z} \nonumber \\
    &= {U S^2 \over 4} - {1 \over N_{BZ'}} \sum_{\mathbf k \in BZ'} E_{\mathbf k} \; \rho_{\mathbf k, z}
\end{align}

\subsubsection{Kinetic energy}
\noindent The kinetic energy has a simple form in the electron basis; we then transform to the doublon/hole one:
\begin{align}
    H_\text{kin} &= \sum_{\substack{\mathbf k \in BZ' \\ \spin}} \e_{\mathbf k} C^\dag_{\mathbf{k}, \spin} \; \pauli_z \; C_{\mathbf{k}, \spin} \nonumber \\
    &= \sum_{\substack{\mathbf k \in BZ' \\ \spin}} {\e_{\mathbf k} \over E_{\mathbf k}} \; \bogoliubon_{\mathbf{k}, \spin}^\dag \op \e_{\mathbf k} \pauli_z + {US \over 2} \spin \pauli_x \cp \bogoliubon_{\mathbf{k}, \spin}
\end{align}
so the expectation per lattice site is
\begin{align}
    {E_\text{kin} \over N} &= {1 \over N} \sum_{\substack{\mathbf k \in BZ' \\ \spin}} {\e_{\mathbf k} \over E_{\mathbf k}} \; \ob US  \; \rho_{\mathbf k, x} - \e_{\mathbf k} \; \rho_{\mathbf k, z} \cb \nonumber \\
    &= {1 \over N_{BZ'}} \sum_{\mathbf k \in BZ'} {\e_{\mathbf k} \over E_{\mathbf k}} \; \ob US  \; \rho_{\mathbf k, x} - \e_{\mathbf k} \; \rho_{\mathbf k, z} \cb
\end{align}
In particular, long after the pulse has ended, the off-diagonal coherences $\rho_{\mathbf k, x}$ and $\rho_{\mathbf k, y}$ will decay to zero. We are left with
\begin{equation} \label{eq:kinetic_energy_late_time}
    {E_\text{kin}^\text{(late)} \over N} = - {1 \over N_{BZ'}} \sum_{\mathbf k \in BZ'} {\e_{\mathbf k}^2 \over E_{\mathbf k}} \; \rho_{\mathbf k, z}
\end{equation}

\section{Approximate analytic treatment} \label{sec:analytic_approximate_solutions}

In this section, we focus on the case of weak driving, which we understand as the situation when the order parameter $S$ is negligibly suppressed: 
\begin{equation} \label{eq:rigid_band_condition}
    S(t) \approx S_0
\end{equation}
The aim is to recover the usual multiphoton and dc tunneling excitation regimes of Ref. \cite{oka_nonlinear_doublons} in our 2D setting, but also to highlight their complex interplay beyond a simple Keldysh crossover. We additionally discuss the incoherent mechanism which dominates the low-field region of parameter space.

\subsection{Perturbative approach} \label{ssec:perturbative_approach}
As the tunneling expression \eqref{eq:maintext_tunneling_production_rate} contains an exponential dependence on $\field^{-1}$, it will not be recovered by standard perturbation theory in the electric field $\field$. Rather, the system of equations \eqref{eq:square_evolution_equations_dme} contains another natural expansion parameter in the form of $(\partial_t \bogangle) / U$. Indeed, from the definition \eqref{eq:bogoliubov_angle_definition}, the angle $\bogangle$ itself is seen to be at most of order $\tau / U \ll 1$. Its time derivative additionally brings out a factor proportional to the driving field, namely $e \field a$. By working in the regime $e \field a \leq U$, we therefore ensure that the condition $(\partial_t \bogangle) / U \ll 1$ holds throughout the evolution. Moreover, the rigid-band assumption \eqref{eq:rigid_band_condition} will additionally require $e \field a \ll U$ throughout a wide range of driving frequencies $\omega_d$; this will further ensure that $(\partial_t \bogangle) / U$ is indeed a small parameter.\\

The evolution equations as written in \eqref{eq:square_evolution_equations_dme} are particularly useful for numerics, since all the variables are real. However, for the analytic treatment it is more convenient to work with $\rho_+ \equiv \rho_x + i \rho_y$ and $\rho_- \equiv \rho_x - i \rho_y$, since this eliminates the coupling between $x$ and $y$, at the scale of $E_\mathbf{k}$. Omitting the equation for $\rho_t$ as well as the population transfer between different momentum sectors, we find
\begin{align} 
	\partial_t \rho_+ &= {\partial_t \bogangle \over 2} \; \rho_z + 2 (i E_\mathbf{k} - \relaxrate) \; \rho_+ \nonumber\\ \nonumber
	\partial_t \rho_- &= {\partial_t \bogangle \over 2} \; \rho_z + 2 (-i E_\mathbf{k} - \relaxrate) \; \rho_-\\ 
	\partial_t \rho_z &= - (\partial_t \bogangle) \; (\rho_+ + \rho_-) \label{eq:square_evolution_equations_complex}
\end{align}
It suffices to consider only the evolution of $\rho_z$ and $\rho_+$, since $\rho_-$ is the complex conjugate of the latter. At zeroth order, we have the initial conditions $\rho_z^{(0)} = 1$ and $\rho_\pm^{(0)} = 0$. It follows from the structure of the equations \eqref{eq:square_evolution_equations_complex} that $\rho_z$ will contain contributions at even order in $(\partial_t \bogangle) / U$, while $\rho_+$ will only have odd ones. Then, at first order, use $\rho_z^{(0)} = 1$ to find
\begin{equation}
    \partial_t \rho_{+}^{(1)} = {\partial_t \bogangle \over 2} + 2 i \; (E_\mathbf{k} + i \relaxrate) \; \rho_{+}^{(1)}
\end{equation}
The complication is that $E_\mathbf{k}$ is also time-dependent, since it is modulated by the vector potential. The relevant propagator necessarily depends on two times,
\begin{equation}
    G(t_1, t_2) = \exp \op 2 i \int_{t_1}^{t_2} du \; \oa E_\mathbf{k}(u) + i \relaxrate \ca \cp
\end{equation}
and yields the formal solution
\begin{equation}
    \rho_{+}^{(1)} (t) = {1 \over 2} \int_{-\infty}^t ds \; (\partial \bogangle)_s \; G(s,t) 
\end{equation}
where $(\partial \bogangle)_s$ denotes the time derivative of $\bogangle$, evaluated at time $s$. The leading nontrivial contribution to $\rho_z$ is then at second order, giving the equation
\begin{equation} \label{eq:rho_z_2nd_order}
\partial_t \rho_z^{(2)} = - (\partial \bogangle)_t \os \int_{-\infty}^t ds \; (\partial \bogangle)_s \; {G(s,t) + G^*(s,t) \over 2} \cs
\end{equation}
which in turn has the formal solution
\begin{widetext}
\begin{equation} \label{eq:formal_rhoz2_solution}
    \rho_z^{(2)} (t) = - \iint_{-\infty}^t ds \; du \; \Theta(u - s) \; e^{-2 \relaxrate (u-s)} \; (\partial \bogangle)_s \; \cos \op 2 \int_{s}^{u} dv \; E_\mathbf{k}(v) \cp  \; (\partial \bogangle)_u
\end{equation}
Here, the cosine comes from taking the real part of the oscillatory component in $G(t_1, t_2)$. We can shift the two time integration variables to temporal midpoint $x$ and time difference $\delta$, namely let $u = x + \delta/2$ and $s = x - \delta/2$. Furthermore make the notation
\begin{equation} \label{eq:definition_h_propagator_time}
    H(x, \delta) \equiv \cos \op 2 \int_{x - \delta/2}^{x + \delta/2} dv \; E_\mathbf{k}(v) \cp  = {1 \over 2 } \os \exp \op 2 i \int_{x - \delta/2}^{x + \delta/2} dv \; E_\mathbf{k}(v) \cp + \exp \op - 2 i \int_{x - \delta/2}^{x + \delta/2} dv \; E_\mathbf{k}(v) \cp \cs
\end{equation}
which has the Fourier transform
\begin{equation} \label{eq:definition_h_propagator_freq}
    H(\Omega, \omega) \equiv \int dx \; d\delta \; H(x, \delta) \; e^{-i x \Omega} \; e^{-i \delta \omega}
\end{equation}
The determinant of the Jacobian for the variable change in \eqref{eq:formal_rhoz2_solution} is equal to 1, so the new integration measure is $dx \; d\delta$. The constraint $\Theta(u - s)$ becomes $\Theta(\delta)$, so we set the lower integration bound on $\delta$ to zero. This gives
\begin{equation}
    \rho_z^{(2)} (t) = - \int_{-\infty}^t dx \int_{0}^{2(t-x)} d\delta \; e^{-2 \relaxrate \delta} \; (\partial \bogangle)_{x - \delta/2} \; H (x, \delta) \; (\partial \bogangle)_{x + \delta/2}
\end{equation}
To calculate the total number of photocarriers produced by the pump, we restrict our attention to the final state of the system. Sending $t\to \infty$ and Fourier transforming with respect to both integration times,
\begin{align}
	\rho_z^{(2)} (\infty) &= - \int_{-\infty}^\infty dx \int_{0}^{\infty} d\delta \; e^{-2 \relaxrate \delta} \int \dbar \omega_1 \; \dbar \Omega \; \dbar \omega \; \dbar \omega_2 \; (\partial \bogangle)_{\omega_1} \; H(\Omega, \omega) \; (\partial \bogangle)_{\omega_2} \; e^{i x (\omega_1 + \Omega + \omega_2)} \; e^{i  \delta (\omega + (\omega_2 - \omega_1) / 2 )}  \nonumber  \\
	&= - \int_{0}^{\infty} d\delta \; e^{-2 \relaxrate \delta} \int \dbar \omega_1 \; \dbar \Omega \; \dbar \omega \; \dbar \omega_2 \; (\partial \bogangle)_{\omega_1} \; H(\Omega, \omega) \; (\partial \bogangle)_{\omega_2} \; 2 \pi \; \delta(\omega_1 + \omega_2 + \Omega) \; e^{i  \delta (\omega + (\omega_2 - \omega_1) / 2 )} \nonumber \\
	&= - \int_{0}^{\infty} d\delta \; e^{-2 \relaxrate \delta} \int \dbar \omega_1 \; \dbar \Omega \; \dbar \omega \; (\partial \bogangle)_{\omega_1} \; H(\Omega, \omega) \; (\partial \bogangle)_{- \omega_1 - \Omega} \; e^{i \delta (\omega - \omega_1 - \Omega/2)} \nonumber \\
	&= - \int_{0}^{\infty} d\delta \; e^{-2 \relaxrate \delta} \int \dbar \omega_1 \; \dbar \Omega \; \dbar \omega \; (\partial \bogangle)_{\omega_1 - \Omega / 2} \; H(\Omega, \omega) \; (\partial \bogangle)_{- \omega_1 - \Omega / 2} \; e^{i \delta (\omega - \omega_1)}
\end{align}
where, going to the last line, $\omega_1$ was shifted down by $\Omega / 2$. We can now perform the intergral over $\delta$:
\begin{equation}
    \rho_z^{(2)} (\infty) = - \int \dbar \omega_1 \; \dbar \Omega \; \dbar \omega \; {(\partial \bogangle)_{\omega_1 - \Omega / 2} \; H(\Omega, \omega) \; (\partial \bogangle)_{- \omega_1 - \Omega / 2} \over 2 \relaxrate - i (\omega - \omega_1)}
\end{equation}
The quantity $\rho_z^{(2)} (\infty)$ is clearly real, so we can safely keep the real part of the integrand. This gives
\begin{equation} \label{eq:qpno_before_gamma_discussion}
	\rho_z^{(2)} (\infty) = - \int \dbar \omega_1 \; \dbar \Omega \; \dbar \omega \; {2 \relaxrate \over (2 \relaxrate)^2 + (\omega - \omega_1)^2} \; \os (\partial \bogangle)_{\omega_1 - \Omega / 2} \; H(\Omega, \omega) \; (\partial \bogangle)_{- \omega_1 - \Omega / 2} \cs
\end{equation}
\end{widetext}
At this point, one may consider taking $\Gamma \to 0^+$ to recover the fully coherent limit for evolution. In this case, the Lorentzian above is peaked at $\omega = \omega_1$, giving
\begin{equation}
    \lim_{\relaxrate \to 0^+} \; {2 \relaxrate \over (2 \relaxrate)^2 + (\omega - \omega_1)^2} = \pi \; \delta(\omega - \omega_1)
\end{equation}
which is just a generalized resonance condition on the drive term $\partial \bogangle$, since it can be used to eliminate $\omega_1$ from \eqref{eq:qpno_before_gamma_discussion} and yield
\begin{equation} \label{eq:no_decoherence_resonance}
\rho_z^{(2)} (\infty) = - {1 \over 2}\int \dbar \Omega \; \dbar \omega \; (\partial \bogangle)_{\omega - \Omega / 2} \; H(\Omega, \omega) \; (\partial \bogangle)_{- \omega - \Omega / 2}
\end{equation}
One can recover both the multiphoton and tunneling photoexcitation mechanisms from \eqref{eq:no_decoherence_resonance}, by investigating the regimes $e \field a \ll \hbar \omega_d$, or respectively $e \field a \gg \hbar \omega_d$, as will be discussed in Sections \ref{ssec:multiphoton_analytic} and \ref{ssec:low_frequency_analytic}. On the other hand, the presence of finite $\relaxrate$ in a realistic solid-state system will give rise to an incoherent, off-resonant pathway to be discussed in Section \ref{ssec:incoherent_excitation_mechanism}. This will dominate if both multiphoton and tunneling processes are strongly suppressed, i.e. when the frequency and amplitude of the drive are simultaneously small.\\

Before analyzing all these regimes in more detail, we simplify the driving term $\partial_t \bogangle$ within the rigid-band approximation \eqref{eq:rigid_band_condition}. From the definition \eqref{eq:bogoliubov_angle_definition} of the Bogoliubov angle, we can explicitly compute $\partial_t \bogangle$. Using \eqref{eq:kinetic_energy_time_derivative} for the time derivative of the kinetic energy, obtain the full expression
\begin{equation} \label{eq:driving_parameter_full_def}
    \partial_t \bogangle = - {US / 2 \over (U S / 2)^2 + \e^2} \op \boldsymbol{\field} \cdot \nabla_{\mathbf A} \e + \e \; {\partial_t S \over S} \cp
\end{equation}
By ignoring the $\partial_t S$ contribution, we arrive at:
\begin{equation} \label{eq:approximate_driving_term}
	\partial_t \bogangle_{\mathbf k} \approx - {US / 2 \over (U S / 2)^2 + \e_{\mathbf k}^2} \; \boldsymbol{\field} \cdot \nabla_{\mathbf A} \e_{\mathbf k}
\end{equation}
It is worth observing that the external drive appears twice in \eqref{eq:approximate_driving_term}: explicitly, through the factor of electric field $\boldsymbol{\field}$, but also implicitly via $\e_{\mathbf k}(t) = \e_{\mathbf k - \mathbf A(t)}$. The time dependence of the former term will follow the driving frequency $\omega_d$; on the other hand, the latter contribution is sensitive to both $\omega_d$ and $\field$, as we will see below. This interplay will give rise to the Keldysh crossover, as well as a cooperative regime between multiphoton and ac tunneling processes.

\subsection{High-frequency regime} \label{ssec:multiphoton_analytic}
We start by recovering the conventional multiphoton and dc tunneling regimes, in the absence of decoherence terms; let $\relaxrate \to 0^+$, and work with the expression \eqref{eq:no_decoherence_resonance}. Consider for simplicity a monochromatic drive $\boldsymbol{\field} = \boldsymbol{\field_0} \cos (\omega_d t)$, such that the vector potential is given by $\mathbf A(t) = - (\boldsymbol{\field_0} / \omega_d) \sin (\omega_d t)$. Furthermore, let the field be linearly polarized along the one of the crystal axes, $\beta$.\\

If the driving field is small compared to the frequency, $\field_0 e a \ll \hbar \omega_d$, then $\mathbf A$ never gets large in magnitude, and a series expansion in $\field_0 / \omega_d$ is natural. For every momentum sector, the shifts in $\mathbf k$ only explore a small region of the band, around the starting point; the momentum-dependent part of \eqref{eq:approximate_driving_term} can be expanded as
\begin{widetext}
    \begin{equation}
    {\partial_{A_\beta} \e_{\mathbf k}(t) \over E_{\mathbf k}^2 (t)} = - e {[\partial_{k_\beta} \e]_{\mathbf k - e \mathbf A (t)} \over E_{\mathbf k - e \mathbf A (t)}^2} = - e \sum_{n \ge 0} {1 \over n!} \; \op {e \field_{0, \beta} \over \omega_d} \sin(\omega_d t) \cp^n \partial_{k_\beta}^n \op {\partial_{k_\beta} \e_{\mathbf k} \over E_{\mathbf k, 0}^2} \cp
\end{equation}
which gives the following expression for the driving term \eqref{eq:approximate_driving_term}:
\begin{equation} \label{eq:driving_term_high_freq_rt_expansion}
    (\partial \bogangle)_t \approx {US \over 2} \; \omega_d  \op {e \field_{0, \beta} \over \omega_d} \; \cos(\omega_d t) \cp  \sum_{n \ge 0} {1 \over n!} \; \op {e \field_{0, \beta} \over \omega_d} \sin(\omega_d t) \cp^n \partial_{k_\beta}^n \op {\partial_{k_\beta} \e_{\mathbf k} \over E_{\mathbf k, 0}^2} \cp
\end{equation}
\end{widetext}

Expanding $\sin^n (\omega_d t)$ will yield contributions at multiple frequencies between $-n \omega_d$ and $n \omega_d$, and they will all carry prefactors of $(e \field_{0, \beta} / 2 \omega_d)^n$. The most relevant components are those at $\pm n \omega_d$ itself, since they cannot be obtained from lower-order terms of \eqref{eq:driving_term_high_freq_rt_expansion}. In contrast, terms oscillating at $m \omega_d$, where $|m| < n$, will also arise from expanding $\sin^{|m|} (\omega_d t)$, and carry a corresponding prefactor of $(e \field_{0, \beta} / 2 \omega_d)^{|m|}$. For $\field_{0} \ll \omega_d$, we find that the contribution that the $|m|-$th order term makes to $m \omega_d$ will dominate over any contributions from $n > |m|$. Therefore we keep only the $\pm n \omega_d$ components in the expansion of each $\sin^n (\omega_d t)$. Omitting $\omega = 0$, this yields
\begin{widetext}
\begin{equation} \label{eq:frequency_domain_drive_hf_1}
    (\partial \bogangle)_\omega \approx {US \over 2} \; \omega_d \sum_{n \ge 0} {1 \over n!} \os \partial_{k_\beta}^n \op {\partial_{k_\beta} \e_{\mathbf k} \over E_{\mathbf k, 0}^2} \cp \cs \op {e \field_{0, \beta} \over 2 \omega_d} \cp^{n+1} \ob i^{-n} \; \delta[\omega - (n+1)\omega_d] + i^n \; \delta[\omega + (n+1)\omega_d] \cb
\end{equation}
The term inside the square brackets, $\partial_{k_\beta}^n \op {\partial_{k_\beta} \e_{\mathbf k} / E_{\mathbf k, 0}^2} \cp$, is also rather complicated, due to $n$ consecutive derivatives acting on a ratio of momentum-dependent expressions. The important observation is such terms are always on the order of $\tau / U^2$, regardless of $n$. To see this, we write
    \begin{equation}
    \partial_{k_\beta} \op {\partial_{k_\beta}^m \e_{\mathbf k} \over E_{\mathbf k, 0}^2} \cp = {1 \over E_{\mathbf k, 0}^2} \op \partial_{k_\beta}^{m+1} \e_{\mathbf k} - 2 (\partial_{k_\beta}^m \e_{\mathbf k}) {\partial_{k_\beta} E_{\mathbf k, 0} \over E_{\mathbf k, 0}}\cp = {\partial_{k_\beta}^{m+1} \e_{\mathbf k} \over E_{\mathbf k, 0}^2} - 2 \; {\e_{\mathbf k} (\partial_{k_\beta} \e_{\mathbf k}) (\partial_{k_\beta}^{m} \e_{\mathbf k}) \over E_{\mathbf k, 0}^4}
\end{equation}
\end{widetext}
In the rightmost expression, the first term is of order $\tau / U^2$, while the second is $\tau^3 / U^4$. For $\tau \ll U$ the first one dominates, i.e. it is more advantageous for the momentum derivative to always act on the numerator. By induction, we find
\begin{equation}
    \partial_{k_\beta}^n \op {\partial_{k_\beta} \e_{\mathbf k} \over E_{\mathbf k, 0}^2} \cp = {\partial_{k_\beta}^{n+1} \e_{\mathbf k} \over E_{\mathbf k, 0}^2} + \mathcal{O} \op \tau^3 \over U^4 \cp
\end{equation}
and all such terms indeed go like $\tau / U^2$ to leading order. We can already see the multiphoton field-dependence expression appearing: it has been previously hinted in Section \ref{ssec:incoherent_excitation_mechanism} that $H(\Omega, \omega)$ concentrates most of its weight at $\Omega = 0$ and $\omega = 2 E_{\mathbf k, 0}$. The resonance condition \eqref{eq:no_decoherence_resonance} then requires us to calculate $(\partial \bogangle)_{2 E_{\mathbf k, 0}}$, which in \eqref{eq:frequency_domain_drive_hf_1} imposes 
\begin{equation} \label{eq:mph_resonant_intermediary_condition}
    n+1 = 2 E_{\mathbf k, 0} / \omega_d
\end{equation}
This, in turn, will bring a prefactor of $(e \field_{0, \beta} / 2 \omega_d)^{2 E_{\mathbf k, 0} / \omega_d}$, which we will need to square, as \eqref{eq:no_decoherence_resonance} contains two copies of $(\partial \bogangle)_\omega$. We find an overall field-dependence of $(e \field_{0, \beta} / 2 \omega_d)^{4 E_{\mathbf k, 0} / \omega_d}$, which resembles the usual multiphoton expression if we recall that $2 E_{\mathbf k, 0} \sim U \sim \Delta$ at leading order.\\

We remark that the implicit assumption in the condition \eqref{eq:mph_resonant_intermediary_condition} on $n$ is that $2 E_{\mathbf k, 0} / \omega_d$ is an integer at all; otherwise, there is little overlap between the factors of the integrand in \eqref{eq:no_decoherence_resonance}, which will yield negligible photoexcitation. It follows that the multiphoton mechanism is highly selective towards momenta close to the energy contour $2 E_{\mathbf k, 0} = n \omega_d$, as illustrated in Figure \ref{fig:main_text_momentum_distributions}b, and further discussed in Section \ref{sec:appendix_momentum_distribution}.\\

For completeness, however, we should perform the same expansion of $H(\Omega, \omega)$ as we did with $(\partial \bogangle)_\omega$, and ensure that any other contributions to \eqref{eq:no_decoherence_resonance} scale similarly with the applied field $\field_0$. We start by expanding the time-dependent $E_{\mathbf k}$ appearing in the definition \eqref{eq:definition_h_propagator_time} in powers of the electric field:
\begin{widetext}
    \begin{align}
    E_{\mathbf k}(v) &= E_{\mathbf k, 0} + \sum_{n > 0} {1 \over n!} \op {e \field_{0, \beta} \over \omega_d} \sin(\omega_d v) \cp^n \partial_{k_\beta}^n E_{\mathbf k, 0} \nonumber \\
    &= E_{\mathbf k, 0} + \sum_{n > 0} {1 \over n!} \op \partial_{k_\beta}^n E_{\mathbf k, 0} \cp \op {e \field_{0, \beta} \over 2 \omega_d} \cp^n \op i^{-n} e^{i n \omega_d v} + i^{n} e^{-i n \omega_d v}\cp + \text{subleading} 
\end{align}
where again subleading contributions in $\op {e \field_{0, \beta} / 2 \omega_d} \cp^n$ have been omitted. This needs to be integrated as
\begin{align} 
    2 \int_{x - \delta /2}^{x + \delta /2} dv \; E_{\mathbf k}(v) &= 2 E_{\mathbf k, 0} \; \delta + 2 \sum_{n > 0} {1 \over n!} \op \partial_{k_\beta}^n E_{\mathbf k, 0} \cp \op {e \field_{0, \beta} \over 2 \omega_d} \cp^n \int_{x - \delta /2}^{x + \delta /2} dv \op i^{-n} e^{i n \omega_d v} + i^{n} e^{-i n \omega_d v}\cp + \dots \nonumber\\
    &= 2 E_{\mathbf k, 0} \; \delta + 2 \sum_{n > 0} {1 \over n!} {\partial_{k_\beta}^n E_{\mathbf k, 0} \over n \omega_d } \op {e \field_{0, \beta} \over 2 \omega_d} \cp^n \os i^{-(n+1)} \op e^{i n \omega_d (x+\delta/2)}-e^{i n \omega_d (x-\delta/2)}\cp + \text{c.c.} \cs + \dots \label{eq:time_domain_propagator_hf_1}
\end{align}
\end{widetext}
Taking the cosine of the above will give $H(x, \delta)$; using the Jacobi-Anger expansion, it can be shown that in the frequency domain, $H(\Omega, \omega)$ will have a strong contribution at $(\Omega = 0, \omega = \pm 2 E_{\mathbf k, 0})$, which we have already discussed, as well as weaker ones at $(\Omega = \pm n_1 \omega_d, \; \omega = \pm 2 E_{\mathbf k, 0} \pm n_2 \omega_d / 2)$. These latter ones will be suppressed by a prefactor of $\op {e \field_{0, \beta} / 2 \omega_d} \cp^{n_1 + n_2}$.\\

Going back to \eqref{eq:no_decoherence_resonance}, we can analyze the various $(\Omega, \omega)$ pairs which pick up contributions both from $H$ and $\partial \bogangle$. The case with $\Omega = 0$ and $\omega = \pm 2 E_{\mathbf k, 0}$ has already been discussed above. Other possibilities are:
\begin{itemize}
    \item Choosing $\Omega = 0$ and $\omega = \pm \omega_d$ will give a strong contribution from $(\partial \bogangle)_\omega$, with the $n=0$ term in \eqref{eq:frequency_domain_drive_hf_1} bringing only one prefactor of $\op {e \field_{0, \beta} / 2 \omega_d} \cp$ for each copy of $(\partial \bogangle)_\omega$ within the resonance condition \eqref{eq:no_decoherence_resonance}, i.e. two in total. However, at the same time we require a higher-order contribution from $H(\Omega, \omega)$. Since each prefactor of $\op {e \field_{0, \beta} / 2 \omega_d} \cp$ in \eqref{eq:time_domain_propagator_hf_1} comes attached to factors of $e^{\pm i \omega_d \delta/2}$, it contributes to shifting the frequency $\omega$ away from $2 E_{\mathbf k, 0}$ by $\omega_d/2$. We require a minimal shift of $2 E_{\mathbf k, 0} - \omega_d$, which enters at order
    \begin{equation}
        n_2 = {2 E_{\mathbf k, 0} - \omega_d \over \omega_d/2} = {4 E_{\mathbf k, 0}\over \omega_d} - 2
    \end{equation}
    and therefore naturally contributes $\op {e \field_{0, \beta} / 2 \omega_d} \cp^{n_2}$ to the field dependence. Together with the two factors from the $(\partial \bogangle)_\omega$, we find a contribution at order $n_2 + 2$, that again scales like $\op {e \field_{0, \beta} / 2 \omega_d} \cp^{4 E_{\mathbf k, 0} / \omega_d}$.
    \item Taking $\Omega = 0$ and $|\omega|$ between $\omega_d$ and $2 E_{\mathbf k, 0}$ will collect nontrivial contributions from both $(\partial \bogangle)_\omega$ and $H(\Omega, \omega)$, whose orders in $\field$ can be seen, by analogy to the previous discussions, to add up to $4 E_{\mathbf k, 0} / \omega_d$. Therefore such contributions will scale identically.
    \item With $\Omega = 0$ and $|\omega| > 2 E_{\mathbf k, 0}$ we select unnecessarily high orders in $(\partial \bogangle)_{\omega}$ while also getting extra suppression from $H(\Omega, \omega)$, so these contributions will be subleading.
    \item Choosing nonzero $\Omega$ will have the effect of picking the driving contributions in \eqref{eq:no_decoherence_resonance} no longer as $(\partial \bogangle)_{\pm \omega}$, but rather as $(\partial \bogangle)_{\omega - \Omega / 2}$ and $(\partial \bogangle)_{-(\omega + \Omega / 2)}$ respectively. This raises the order in $\field$ of one term while lowering the other by the same amount, therefore requiring the same total contribution to the field dependence from the driving $(\partial \bogangle)$. However, it also brings suppression from expanding $H(\Omega, \omega)$, so this type of contribution is again subleading.  
\end{itemize}
Combining all the terms, the result is that the final density of quasiparticles obeys the field dependence
\begin{equation} \label{eq:multiphoton_result}
    n_\text{qp} \propto \int_\text{BZ'} {d^2 \mathbf k \over 2 \pi^2} \op {e \field_{0} \over 2 \omega_d} \cp^{4 E_{\mathbf k, 0} / \omega_d} 
\end{equation}
The density of states for doublon-hole pairs is highest right above the gap, $2 E_{\mathbf k, 0} \approx \Delta$, as seen in Figure \ref{fig:density_of_states}. Driving on a multiphoton resonance with this peak (i.e. choosing $\omega_d \approx \Delta / n$ with integer $n$) will produce a significant density of photocarriers, which scales as
\begin{equation}
    n_\text{qp} \propto \op {e \field_{0} \over 2 \omega_d} \cp^{2 \Delta / \omega_d} 
\end{equation}
which is the standard multiphoton expression.

\subsection{Low-frequency regime} \label{ssec:low_frequency_analytic}

Having recovered the usual multiphoton dependence from the high-frequency limit $\hbar \omega_d \gg \field_0 e a$, we turn to the opposite regime $\hbar \omega_d \ll \field_0 e a$, to investigate tunneling.

\subsubsection{DC limit}
Let us first consider $\omega_d \to 0$, which allows for the vector potential to be taken of the form $\mathbf A = - \boldsymbol{\field}_0 t$. Shifts in momentum will be linear in time, and so $\e_{\mathbf k}$ as well as $\nabla_{\mathbf A} \e_{\mathbf k}$ will have frequency components only at $\pm \field_0$. The driving term can be expanded as
\begin{align}
	(\partial \bogangle)_t &\approx - {2 \over U S} \os 1 + \op {2 \e_{\mathbf k} \over U S} \cp^2 \cs^{-1} \; \op\boldsymbol{\field}_0 \cdot \nabla_{\mathbf A} \e_{\mathbf k} \cp \nonumber\\
	&\approx - {2 \over U S} \; \op \boldsymbol{\field}_0 \cdot \nabla_{\mathbf A} \e_{\mathbf k} \cp \; \sum_{n = 0}^{\infty} \op {2 \e_{\mathbf k} \over U S} \cp^{2 n}  \label{eq:time_domain_drive_lf_1}
\end{align}
The approach of Section \ref{ssec:multiphoton_analytic} follows through analogously, only with $\field_0$ rather than $\omega_d$ dictating the time dependence of $\e_{\mathbf k}$, and therefore setting the frequency scale of the driving term $(\partial \bogangle)$. We focus on matching the higher-order terms in the drive with the main peak of $H(\Omega, \omega)$, as other relevant contributions will again have the same field scaling.\\

In contrast to the high-frequency regime, the vector potential now gets arbitrarily large, and any starting momentum state will explore an entire linecut of the Brillouin zone. Therefore, the leading contribution to $H$ will come not from the undriven quasiparticle energy $E_{\mathbf k, 0}$, but rather a time-averaged $\tilde{E}_{\mathbf k}$ along the trajectory that the state takes in momentum space. This will be on the order of the gap $\Delta$, and will have weak momentum dependence as a consequence of averaging over the entire Brillouin zone width. The resonance condition becomes $(2n+1)\field_0 = 2 \tilde{E}_{\mathbf k}$, so within the expansion \eqref{eq:time_domain_drive_lf_1} we are interested in the order $n = (\tilde{E}_{\mathbf k} / \field_0) - 1/2$. The prefactor associated with that, including the $2\tau$ from the hopping and the $2^{-2n}$ overlap between $(\sin x)^{2n}$ and $\sin(2nx)$, is
\begin{align} 
    (\partial \bogangle)_{\pm \Delta} &\sim - {2\tau \over US} \; \field_0 \cdot \op {2\tau \over US} \cp^{2 \tilde{E}_{\mathbf k} / \field_0 - 1} \nonumber \\
    &= \field_0 \; \exp \os - \ln \op {US \over 2 \tau} \cp  {2 \tilde{E}_{\mathbf k} \over \field_0} \cs \label{eq:tunneling_pre_result}
\end{align}
so, after integrating over all momenta, the density of photoexcited carriers should be proportional to
\begin{align} 
    n_\text{qp} &\propto \int_{BZ'} {d^2 \mathbf k \over 2 \pi^2} \; |(\partial \bogangle)_{\Delta}|^2 \nonumber \\ 
    &\sim \field_0^2 \; \exp \os - 2 \ln \op {US \over 2 \tau} \cp  {\Delta \over \field_0} \cs \label{eq:tunneling_result}
\end{align}
which is the usual tunneling result. In terms of the generalized tunneling expression \eqref{eq:maintext_tunneling_production_rate} of the main text, this yields an estimate for the dc tunneling strength:
\begin{equation}
    \gamma(0) \approx 2 \ln \op {US \over 2 \tau} \cp
\end{equation}
We remark that the behavior $\gamma \propto \ln (U / \tau)$ is consistent with the strongly interacting limit of the results in \cite{oka_nonlinear_doublons}.\\

Note that $\mathbf k$ enters \eqref{eq:tunneling_pre_result} only via $\tilde{E}_{\mathbf k}$, which was argued to have weak momentum dependence. For low driving fields, when the total carrier density is already strongly suppressed, the resulting momentum distribution will be rather uniform. Under stronger pumping, however, this conclusion will change; multiphoton-tunneling cooperation (to be discussed in the next subsection) will also contribute towards preferential production of carriers at particular momenta.\\ 

By analogy with the multiphoton resonances discussed in Section \ref{ssec:multiphoton_analytic}, we may a priori expect an enhancement of the tunneling rate when the order $n = (\tilde{E}_{\mathbf k} / \field_0) - 1/2$ is an integer; this would roughly correspond to the odd harmonics of the Bloch oscillation frequency matching the gap. Previous DMFT work on steady states of CW-driven Mott insulators \cite{driven_steady_state, dmft_doublon_production_heating} does indeed find enhanced doublon occupancy when $\field = U / n$, and interprets it as arising from resonant tunneling over a distance of $n$ sites. However, in our case, two factors will hinder the observation of such features:
\begin{itemize}
    \item For low fields $\field$, the necessary order $n$ is large, and so the integer requirement on $n$ is not stringent enough. The hypothetical peaks in carrier production versus $\field$ will be washed out by broadening due to scattering, details of the band structure, and the range of instantaneous field strengths present in the case of pulsed driving (i.e. the slow modulation $\field_0(t)$ that the pulse envelope produces).
    \item At strong driving, on the other hand, dynamical renormalization of the gap will become an important effect, as discussed in Sections \ref{ssec:results_1_low_frequency} and \ref{ssec:results_gap_suppresison}. The rigid-band assumptions of this section will no longer hold, and the corresponding system evolution becomes significantly more complex. In particular, the dimensionless field strength $e \field_0 a / \Delta$ changes with time not only due to the envelope effect $\field_0(t)$, but also from gap renormalization $\Delta(t)$.
\end{itemize}

\subsubsection{Finite frequency} \label{ssec:perturbative_approach_finite_frequency}
\begin{figure*}
	\centering	
    \scalebox{\fullfigsize}{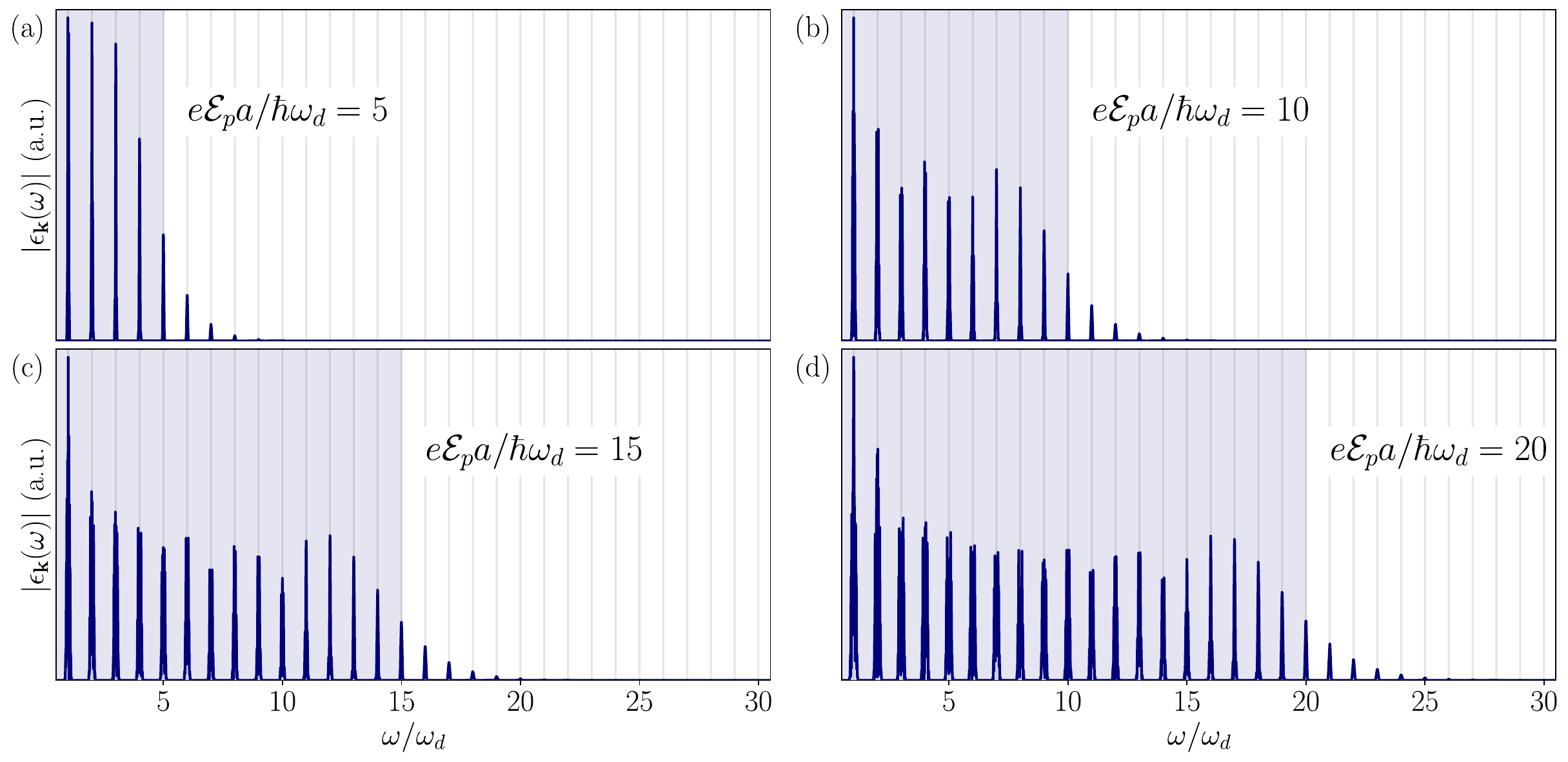}
	\caption{Typical Fourier transforms of $\e_{\mathbf k}(t)$, when pumping with a pulse in the high-field ac regime $e \field_p a \gg \hbar \omega_d > 0$. Integer multiples of the driving frequency $\omega_d$ are marked by grey vertical lines, and the frequency range below $e \field_p a / \hbar$ is shaded in blue. The spectra consist of peaks at $n \omega_d$, with $n \in \mathbb{Z}$, which get cut off around $e \field_p a / \hbar$.}
	\label{fig:ac_tunnel_spectra}
\end{figure*}

When $0 < \hbar \omega_d \ll \field e a$, the form of the driving term $\partial \bogangle$ in the frequency domain will become more complicated. Consider again a monochromatic drive $\boldsymbol{\field} = \boldsymbol{\field_0} \cos (\omega_d t)$ and $\mathbf A(t) = - (\boldsymbol{\field_0} / \omega_d) \sin (\omega_d t)$. Now the ratio $\field_0 / \omega_d$ is a large parameter, so it is not directly suitable for producing a series expansion like that of Section \ref{ssec:multiphoton_analytic}. Instead, consider the $\mathbf k-$dependent trigonometric functions appearing in the kinetic energy $\e_{\mathbf k}$, as well as in related quantities. Such functions are constructed out of $e^{\pm i A(t)}$, which in our case is $e^{\pm (i \field_0 / \omega_d) \sin (\omega_d t)}$. These can be expanded via the Jacobi-Anger identity:
\begin{equation} \label{eq:jacobi_anger_low_freq}
    e^{(i \field_0 / \omega_d) \sin (\omega_d t)} = \sum_{n = -\infty}^{\infty} J_n \op {\field_0 \over \omega_d} \cp e^{i n \omega_d t}
\end{equation}
where $J_n$ denote Bessel functions of the first kind. At small enough values of $n$, which fulfill the condition $n \ll \field_0 / \omega_d$, the following approximation holds:
\begin{equation}
    J_n \op {\field_0 \over \omega_d} \cp \approx \sqrt{2 \omega_d \over \pi \field_0} \; \cos \op {\field_0 \over \omega_d} - {n \pi \over 2} - {\pi \over 4} \cp + \dots
\end{equation}
which may be used to conclude the respective prefactors of $e^{i n \omega_d t}$ in the expansion \eqref{eq:jacobi_anger_low_freq} are all of the same order of magnitude, namely $\sqrt{2 \omega_d / \pi E_0}$. On the other hand, when $n$ is large (as defined by the condition $n \gg \field_0 / \omega_d$), the asymptotic form of the Bessel functions reads
\begin{equation} \label{eq:bessel_function_small_argument}
J_n \op {\field_0 \over \omega_d} \cp \approx {1 \over n!} \op {\field_0 \over 2 \omega_d} \cp^n
\end{equation}
which vanishes if $n \to \infty$. In practice, the prefactors of $e^{i n \omega_d t}$ terms in \eqref{eq:jacobi_anger_low_freq} are cut off rather quickly above $n_0 = \field_0 / \omega_d$, the expansion order which corresponds to an oscillation at $e^{i \field_0 t}$. Therefore, the spectrum of the time-dependent kinetic energy $\e_{\mathbf k}(t)$ looks like a comb, with peaks of comparable strength spaced by $\omega_d$, and running roughly between $-\field_0$ and $\field_0$ (Figure \ref{fig:ac_tunnel_spectra}). In particular, the largest frequency which contributes significantly ($n_0 \omega_d$) is determined by $\field_0$, as expected from Bloch oscillations; but $\omega_d$ is also a relevant quantity, since $n_0$ needs to be an integer. We therefore expect enhancements in the production rate whenever $\Delta / \omega_d$ is also integer.\\

To compute the density of photoexcited carriers, we again impose the resonance condition between higher-order contributions in $(\partial \bogangle)_\omega$, arising from quasiparticle dispersion nonlinearities, and the pole of $H(\Omega, \omega)$ situated at the gap energy. The new feature, compared to the dc case, is that $\e_{\mathbf k}$ contains multiple frequency contributions, and we can combine different ones to reach up to the gap. The generalized resonance condition will look like
\begin{equation} \label{eq:general_resonance_condition_lf1}
    \Delta = (n_1 + n_2 + \ldots + n_m) \; \omega_d
\end{equation}
where each term $n_i \omega_d$ is picked to be on the order of $\field_0$, i.e. $n_i \sim \field_0 / \omega_d$, to ensure a sizable contribution at that frequency in the expansion of $\e_{\mathbf k}$. Each such contribution will carry a prefactor of $J_{n_i}(n_i)$, which does not have a nice approximate form but is $\OO(1)$ and does not strongly depend on $n_i$ when the latter is large. On the other hand, the number of terms in \eqref{eq:general_resonance_condition_lf1} is $m \sim \Delta / \field_0 - 1$, and a factor of $2 \tau / US$ for each will lead us back to a tunneling form of field dependence, similar to \eqref{eq:tunneling_result}. Specifically, assuming for simplicity $n_1 \sim n_2 \sim \ldots \equiv n$, we find a contribution of
\begin{align} 
    n_\text{qp} &\propto \os {2\tau \over US} \; \field_0 \cdot \op {2\tau \over US} \; J_n(n) \cp^{\Delta / \field_0 - 1} \cs^2 \nonumber \\
    &\sim \field_0^2 \; \exp \os - 2 \ln \op {US \over 2 \tau J_n(n)} \cp  {\Delta \over \field_0} \cs \label{eq:multiphoton_tunneling_cooperation}
\end{align}
so the effect of finite-frequency rather than dc driving, on the field-dependence of production rates, should be a renormalization of the prefactor $\gamma(\omega_d)$ which multiplies $-(\Delta / \field_0)$ in the exponential. Since $J_n(n) < 1$ appears in a denominator of the final expression \eqref{eq:multiphoton_tunneling_cooperation}, the renormalization is expected to be an increase. However, a more precise calculation of this enhancement needs to count all the possible combinations of frequencies $\{n_i\}$ satisfying the generalized resonance, rather than just the case $n_1 \sim n_2 \sim \ldots \equiv n$ considered above.\\

The crucial aspect is that such an enhancement takes place only when \eqref{eq:general_resonance_condition_lf1} can be satisfied, i.e. $\Delta / \omega_d$ is an integer, which is reminiscent of the multiphoton condition. This gives rise, as discussed at length in Section \ref{ssec:results_1_low_frequency}, to a regime where (renormalized) tunneling behavior is visible in the field dependence, but multiphoton-like thresholds are visible in frequency. While the conventional Keldysh crossover can be understood as arising from the competition of multiphoton and tunneling pathways, this regime can be thought of as cooperation between the two mechanisms.

\subsection{Non-monochromatic driving}
In a realistic experimental setting, and especially when strong electric fields are required, the sample will be pumped with short laser pulses, which will therefore contain a finite range of frequencies. In general, we should write for the field and vector potential
\begin{align}
    \boldsymbol{\field}(t) &= \int \dbar \omega \; \boldsymbol{\field_0}(\omega) e^{i \omega t}\\
    \mathbf A(t) &= \int \dbar \omega \; {\boldsymbol{\field_0}(\omega) \over -i \omega} e^{i \omega t}
\end{align}
However, this will no longer allow for the series expansions of the previous sections. In practice, as long as the pulse contains enough cycles in real time (and is therefore relatively narrow in the frequency domain), the qualitative conclusions previously drawn about the spectral composition of $\partial \bogangle$ and $H$ will still hold; their peaks will no longer be $\delta$ functions, but will acquire a frequency width set by that of the original pulse. If on the other hand this width is large enough such that neighboring peaks overlap, as will happen for few-cycle pulses, the photoexcitation rates may deviate from the simple forms derived above; a full numerical solution of the evolution equations will be necessary.

\subsection{Incoherent excitation mechanism} \label{ssec:incoherent_excitation_mechanism}

As discussed previously, $H(\Omega, \omega)$ will have most of its weight where $\Omega$ is low, and $\omega$ is on the order of the gap $\Delta = US$. On the other hand, $(\partial \bogangle)_{\omega}$ mostly gets contributions at frequencies set by the pump field strength $\field$ (via Bloch oscillations) or the pump frequency $\omega_d$ itself. In the regime $\hbar \omega_d, \; e \field a \ll \Delta$, these two terms will be far off-resonant in \eqref{eq:no_decoherence_resonance}, yielding negligible photoexcitation. However, as soon as a finite $\relaxrate>0$ is physically present in the system, a new contribution can be obtained by letting $\omega$ and $\omega_1$ be different in \eqref{eq:qpno_before_gamma_discussion}, and paying instead a penalty given by the tail of the Lorentzian: $2 \relaxrate / (\omega - \omega_1)^2$. With $\omega \sim U$ and $\omega_1 \ll U$, the price to be paid is a prefactor on the order of $\relaxrate / U^2$; although this is small, it does lead to better scaling with the electric field of photoexcitation rates, in comparison with the highly nonlinear multiphoton or tunneling expressions.\\

Let us calculate the leading contribution due to the effect described above. For simplicity, we will neglect any time dependence of either the order parameter or of any $\mathbf k$ dependent energies, such as $\e_{\mathbf k}$ and $E_{\mathbf k}$. The former approximation is justified if the density of excited quasiparticles is low. The latter one is especially good in the high-frequency regime $\field e a \ll \hbar \omega$, when the typical magnitude of the vector potential, $A \sim \field / \omega$, which produces the shift in $\mathbf k$, is low compared to the BZ size. With $E_{\mathbf k}(t) \approx E_{\mathbf k, 0}$, the result for $H$ is simple in both time and frequency domains: 
\begin{align}
    H(x, \delta) &\approx {1 \over 2} \ob e^{2 i \delta E_{\mathbf k, 0}} + e^{-2 i \delta E_{\mathbf k, 0}} \\
    H(\Omega, \omega) &\approx 2\pi^2 \; \delta(\Omega) \; \ob \delta(\omega - 2 E_{\mathbf k, 0}) + \delta(\omega + 2 E_{\mathbf k, 0})\cb
\end{align}

\begin{figure*}
	\centering	
    \scalebox{\largehalffigsize}{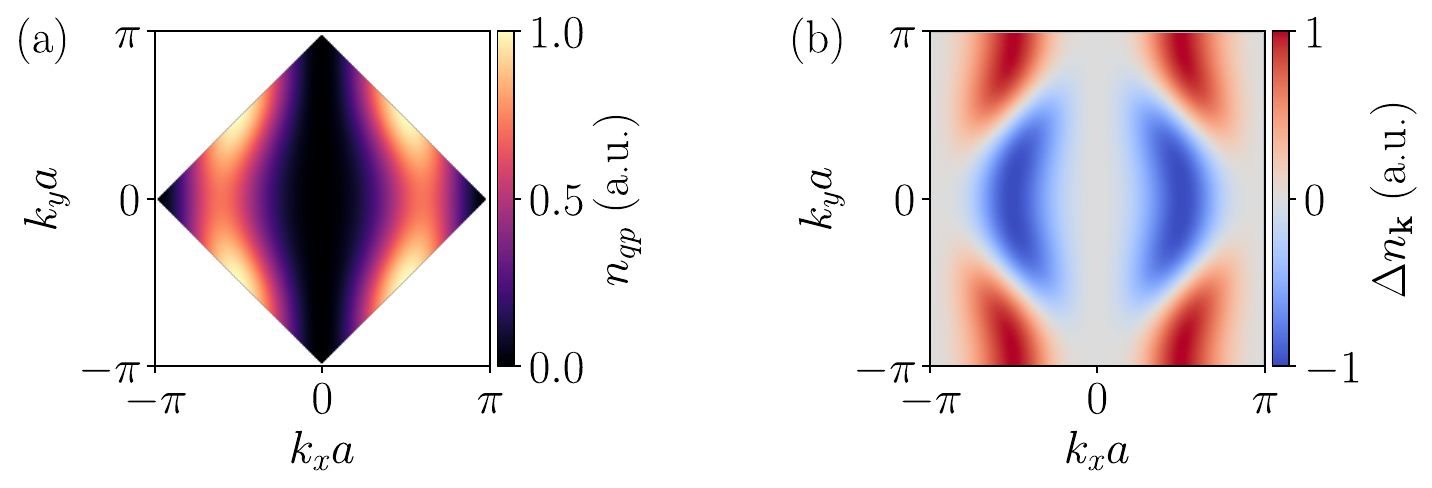}
	\caption{\textbf{(a)} Momentum distribution of photocarries in the incoherent regime, as extracted from the integrand of eq. \eqref{eq:analytic_decoherence_qp}; \textbf{(b)} The corresponding momentum-resolved change in electronic occupations. There is good agreement with the full numerical solutions depicted in panels (a) and respectively (d) of Fig. \ref{fig:main_text_momentum_distributions}.}
	\label{fig:analytic_incoherent_momentum_distribution}
\end{figure*}

This allows us to perform the integrals over both $\Omega$ and $\omega$ in \eqref{eq:qpno_before_gamma_discussion}, obtaining
\begin{widetext}
    \begin{equation} \label{eq:pump_tail_incoherent_expression}
    \rho_z^{(2)} (\infty) = - {1 \over 2} \int \dbar \omega_1 \; \os {2 \relaxrate \over (2 \relaxrate)^2 + (2 E_{\mathbf k, 0} - \omega_1)^2} + {2 \relaxrate \over (2 \relaxrate)^2 + (2 E_{\mathbf k, 0} + \omega_1)^2} \cs \; |(\partial \bogangle)_{\omega_1}|^2
\end{equation}
\end{widetext}
As mentioned previously, $(\partial \bogangle)_{\omega_1}$ has most of its weight at frequencies well below $2 E_{\mathbf k, 0}$, so we can make the approximation $\omega_1 \ll 2 E_{\mathbf k, 0}$ in the denominators. With $\relaxrate$ much smaller than both other energy scales, we arrive at the simplification
\begin{equation} \label{eq:incoherent_expand_lorentzians}
    \rho_z^{(2)} (\infty) = - {\relaxrate \over 2 E_{\mathbf k, 0}^2}\int \dbar \omega \; \op 1 + {3 \over 4} {\omega^2 \over E_{\mathbf k, 0}^2} + \dots \cp \; |(\partial \bogangle)_{\omega}|^2 
\end{equation}
where the integration variable was renamed from $\omega_1$ to $\omega$. On the other hand, recalling the time derivative of the Bogoliubov phase, as approximated in \eqref{eq:approximate_driving_term},
\begin{equation}
    (\partial \bogangle)_t \approx - {US / 2 \over E_{\mathbf k}^2} \; \boldsymbol{\field} \cdot \nabla_{\mathbf A} \e_{\mathbf k}
\end{equation}
we see that, if shifts in $\mathbf k$ are ignored at leading order, the time dependence comes just from $\boldsymbol \field$ itself. The Fourier transform is then straightforward,
\begin{align}
    (\partial \bogangle)_\omega &\approx - {US / 2 \over E_{\mathbf k, 0}^2} \; \boldsymbol{\field}(\omega) \cdot (\nabla_{\mathbf A} \e_{\mathbf k})_0 \nonumber \\
    &= - \field(\omega) \; {US / 2 \over E_{\mathbf k, 0}^2} \; \hat{\field} \cdot (\nabla_{\mathbf A} \e_{\mathbf k})_0
\end{align}
where $\boldsymbol \field$ is assumed to be linearly polarized, with amplitude $\field$ and orientation $\hat \field$. Keeping only the leading order of \eqref{eq:incoherent_expand_lorentzians}, we have for $\rho_{\mathbf k, z}^{(2)} (\infty)$ at a given momentum
\begin{widetext}
    \begin{equation}
    \rho_{\mathbf k, z}^{(2)} (\infty) = - {\relaxrate \over 2 E_{\mathbf k, 0}^2} \; \op {(US / 2)  \; [\hat{\field} \cdot (\nabla_{\mathbf A} \e_{\mathbf k})_0] \over E_{\mathbf k, 0}^2} \cp^2 \int \dbar \omega \; |\field(\omega)|^2
\end{equation}
\end{widetext}
The frequency integral just gives the pump fluence, times the free space impedance $\eta_0$: 
\begin{equation}
    \rho_z^{(2)} (\infty) = - {\relaxrate \eta_0 F \over 2} \; {(US / 2)^2  \; [\hat{\field} \cdot (\nabla_{\mathbf A} \e_{\mathbf k})_0]^2 \over E_{\mathbf k, 0}^6}
\end{equation}
After the pulse, the total density of excited quasiparticles in the band is 
\begin{equation}
    n_\text{qp} = \int_{BZ'} {d^2 \mathbf k \over 2 \pi^2} \; {1 - \rho_{\mathbf k ,z}(\infty) \over 2} \approx - {1 \over 2} \int_{BZ'} {d^2 \mathbf k \over 2 \pi^2} \; \rho_{\mathbf k, z}^{(2)} (\infty)
\end{equation}
where the constant factor of 1 in the integrand's numerator cancels with $\rho_z^{(0)} \equiv 1$. We obtain a result linear in both fluence $F$ and the rate $\relaxrate$:
\begin{equation} \label{eq:analytic_decoherence_qp}
n_\text{qp} = 4 \relaxrate F \; {\eta_0 \tau^2 \over (U S)^4} \; \int_{BZ'} {d^2 \mathbf k \over 2 \pi^2} \; {[\hat{\field} \cdot (\nabla_{\mathbf A} \e_{\mathbf k})_0 / \tau]^2 \over (2 E_{\mathbf k, 0} / U S)^6}
\end{equation}
The momentum integral will in fact be independent on the orientation of the pump pulse. To see this, take an arbitrary polarization $\hat{\field} = \cos \theta \; \hat x + \sin \theta \; \hat y$. Then, from arguments similar to those of \eqref{eq:kinetic_energy_time_derivative} we find
\begin{equation} \label{eq:analytic_decoherence_polarization_investigation}
    [\hat \field \cdot (\nabla_{\mathbf A} \e_{\mathbf k})_0 / \tau]^2 = 4 \; \oa \cos \theta \sin k_x + \sin \theta \sin k_y \ca^2
\end{equation}
The denominator of the integrand in \eqref{eq:analytic_decoherence_qp} is invariant under $k_x \leftrightarrow k_y$, and even in both variables. Therefore, the cross term proportional to $\sin k_x \sin k_y$ in the numerator \eqref{eq:analytic_decoherence_polarization_investigation} will not contribute, by virtue of being odd in both variables. Furthermore, the $\sin^2 k_x$ and $\sin^2 k_y$ contributions will be equal by symmetry; so the final result will have polarization dependence that goes like $\sin^2 \theta + \cos^2\theta \equiv 1$, i.e. will be independent of $\theta$.\\

The integral in \eqref{eq:analytic_decoherence_qp} is still tricky to evaluate due to the band dispersion details, but can be expressed as a series in $\tau/ U$. The first few terms are:
\begin{widetext}
    \begin{equation} \label{eq:incoherent_series_expansion}
    n_\text{qp} \approx 8 \relaxrate F {\eta_0 \tau^2 \over (U S)^4} \os 1 - 36 \op {\tau \over US} \cp^2 + 1920 \op {\tau \over US} \cp^4 -112000 \op {\tau \over US} \cp^6 + 6773760 \op {\tau \over US} \cp^8 + \dots \cs
\end{equation}
\end{widetext}
The series converges for $\tau / U \sim 10^{-1}$, but does so very slowly. Alternatively, one can also perform the momentum integral in \eqref{eq:analytic_decoherence_qp} numerically. Its value is, in any case, of order 1 when $\tau / U \sim 10^{-1}$. For example, the terms inside the square brackets of \eqref{eq:incoherent_series_expansion} add up to about $0.786$, while the corresponding numerical integral yields $0.7459$.\\

Momentum distribution of the photoexcited carriers is also contained within the expression \eqref{eq:analytic_decoherence_qp}, as its integrand sets relative doublon-hole occupations at different momenta. The result is plotted in Figure \ref{fig:analytic_incoherent_momentum_distribution} below, and the result is in good agreement with the full numerical solution presented in Figure \ref{fig:main_text_momentum_distributions}a,d.\\

\section{Details of numerical calculations} \label{sec:sm_numerics}

As mentioned in the main text, throughout the paper we fix the Hamiltonian parameters to be $U = 1.93$ eV and $\tau = 189$ meV, also assuming a low decoherence rate $\relaxrate = 1.64$ meV. Driving is implemented as an electric field of variable carrier frequency $\omega_d$, modulated by a Gaussian envelope of width $141$ fs; equivalently, the envelope of the intensity $\field^2$ has a width of $T_\text{pump} = 100$ fs. We calculate the evolution for a square lattice with $N \times N$ sites, where $N$ is chosen large enough that finite-size effects are negligible, and convergence of the relevant observables is achieved. For momentum-integrated observables, such as total photocarrier density, we use $N = 91$, while for momentum-resolved plots we take $N = 251$. All computations were performed on the Euler cluster of ETH Z\"urich.\\
\section{Momentum distribution of photocarriers} \label{sec:appendix_momentum_distribution}

We have argued in Section \ref{ssec:results_1_momentum_distribution}, as well as illustrated in Figures \ref{fig:main_text_momentum_distributions},\ref{fig:qp_momentum_variance} of the main text, that the momentum distribution of photoexcited carriers is an extremely valuable indicator of their generation mechanism. Indeed, while the electric field dependence of the production rate displays a smooth crossover between different regimes, the resulting momentum distributions are markedly different. In this section, we investigate the full doublon-hole momentum distribution in various regions of pump parameter space, with an emphasis on boundaries between regions with different excitation mechanisms.\\

\begin{figure}[h]
	\centering	
    \scalebox{\halffigsize}{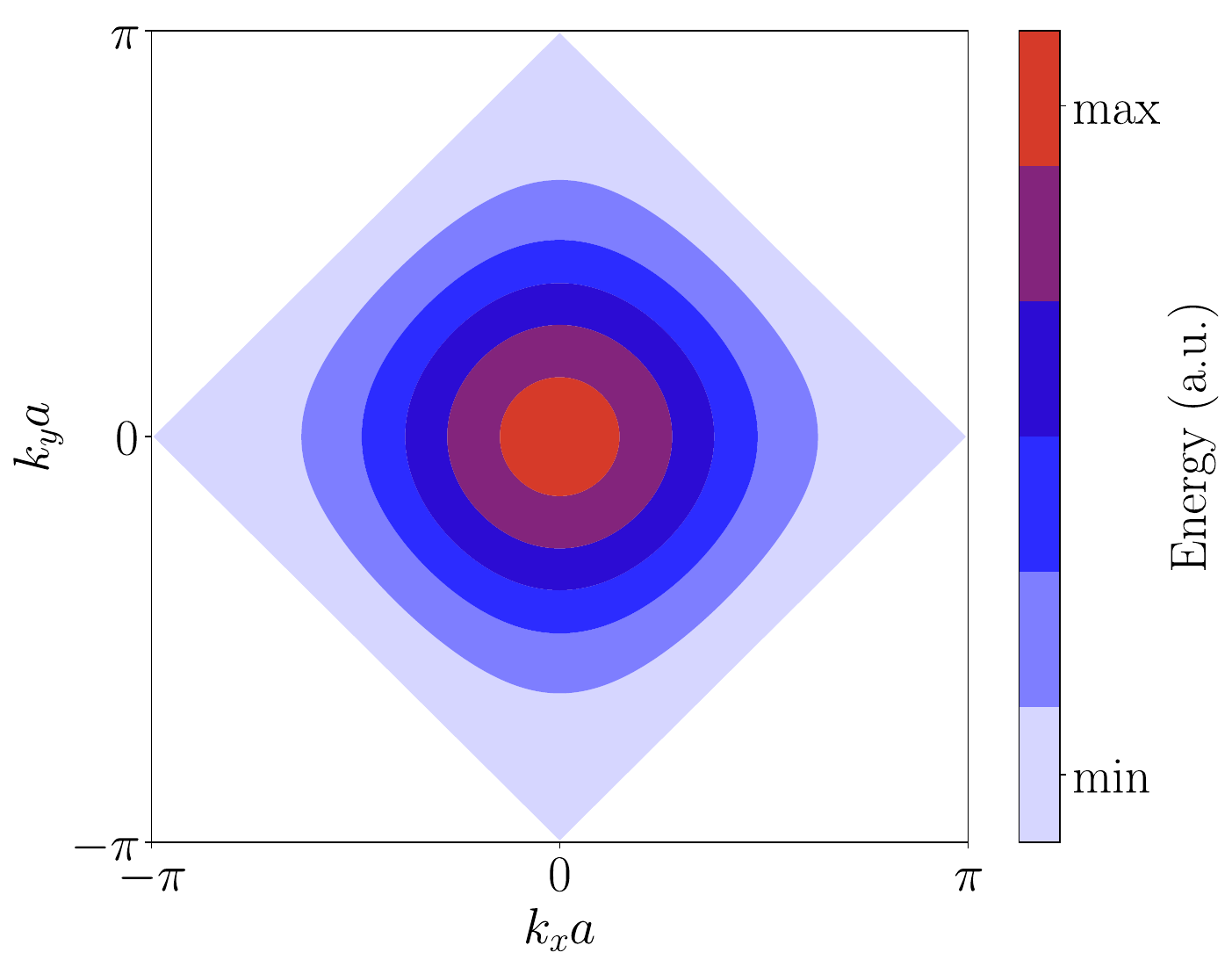}
	\caption{Constant energy contours in the magnetic BZ, as a function of relative momentum, for optically generated doublon-hole pairs (with zero total momentum).}
	\label{fig:qp_energy_contours}
\end{figure}

\begin{figure*}
	\centering	
    \scalebox{\halffigsize}{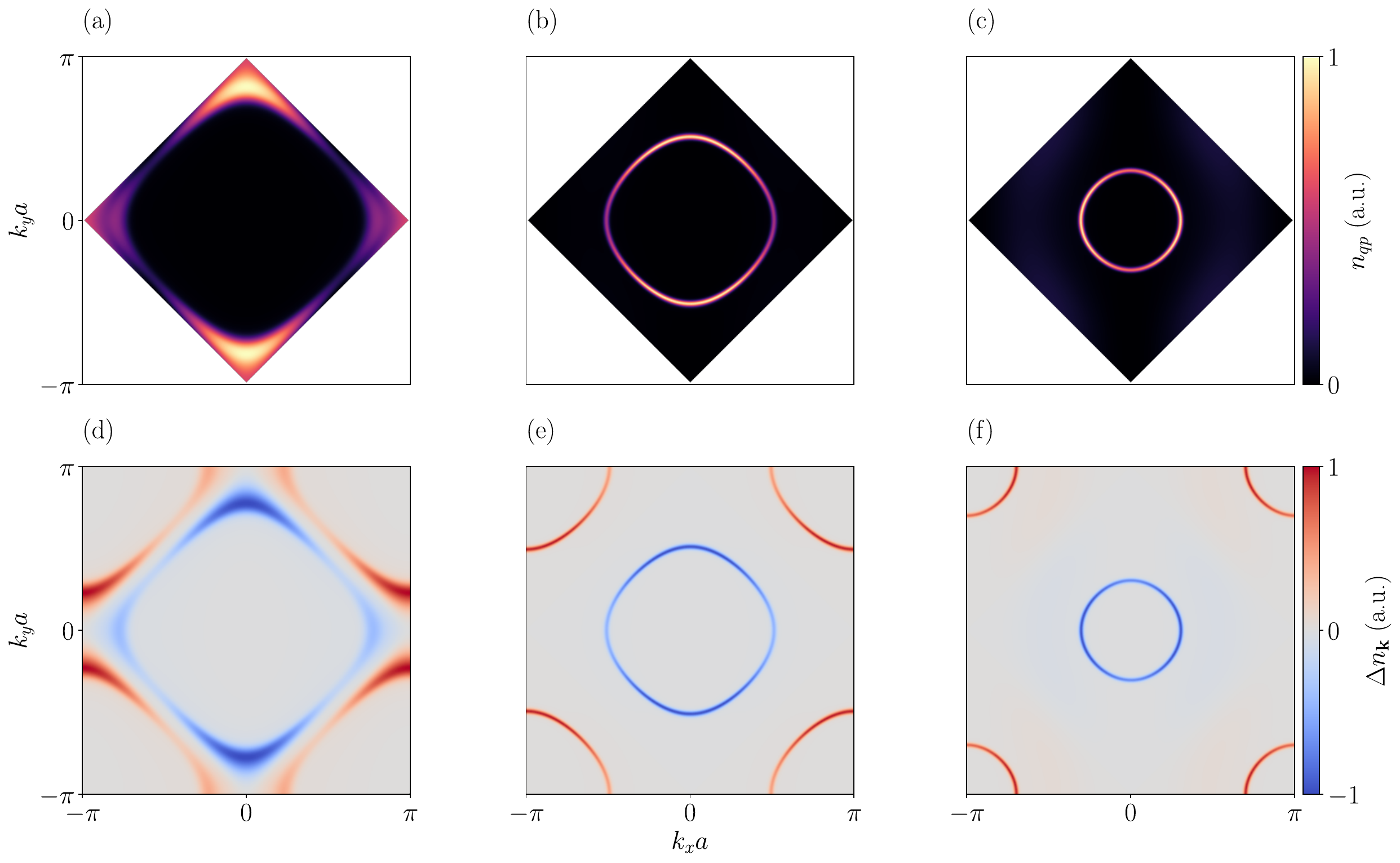}
	\caption{For a weak electric field, tuning the pump frequency moves the region in the BZ resonant with a multiphoton transition. Here, we show the case of a 2-photon process: calculated for $\field / \Delta_0 = 0.06$ and $\omega_d / \Delta_0$ of $0.5$ \textbf{(a, d)}, $0.54$ \textbf{(b, e)}, and respectively $0.6$ \textbf{(c, f)}.}
	\label{fig:low_field_frequency_tuning}
\end{figure*}

The energy \eqref{eq:maintext_quasiparticle_dispersion} of doublon-hole pairs as a function of relative momentum is shown in Fig. \ref{fig:qp_energy_contours}. The energy minimum is reached on the diamond marking the boundary of the magnetic BZ, while the top of the band lies at the $\Gamma$ point, around which the dispersion is approximately quadratic. Multiphoton processes will yield momentum distributions closely following dispersion contours, as seen from the resonance condition $2 E_\mathbf{k} / \omega_d \in \mathbb{Z}$, eq. \eqref{eq:mph_resonant_intermediary_condition}. We illustrate this in Figure \ref{fig:low_field_frequency_tuning} by fixing a relatively weak driving field $\field$, and tuning the pump frequency $\omega_d$ above the two-photon resonance; higher $\omega_d$ will produce photocarriers on contours closer to the $\Gamma$ point.\\

\begin{figure*}
	\centering	
    \scalebox{\halffigsize}{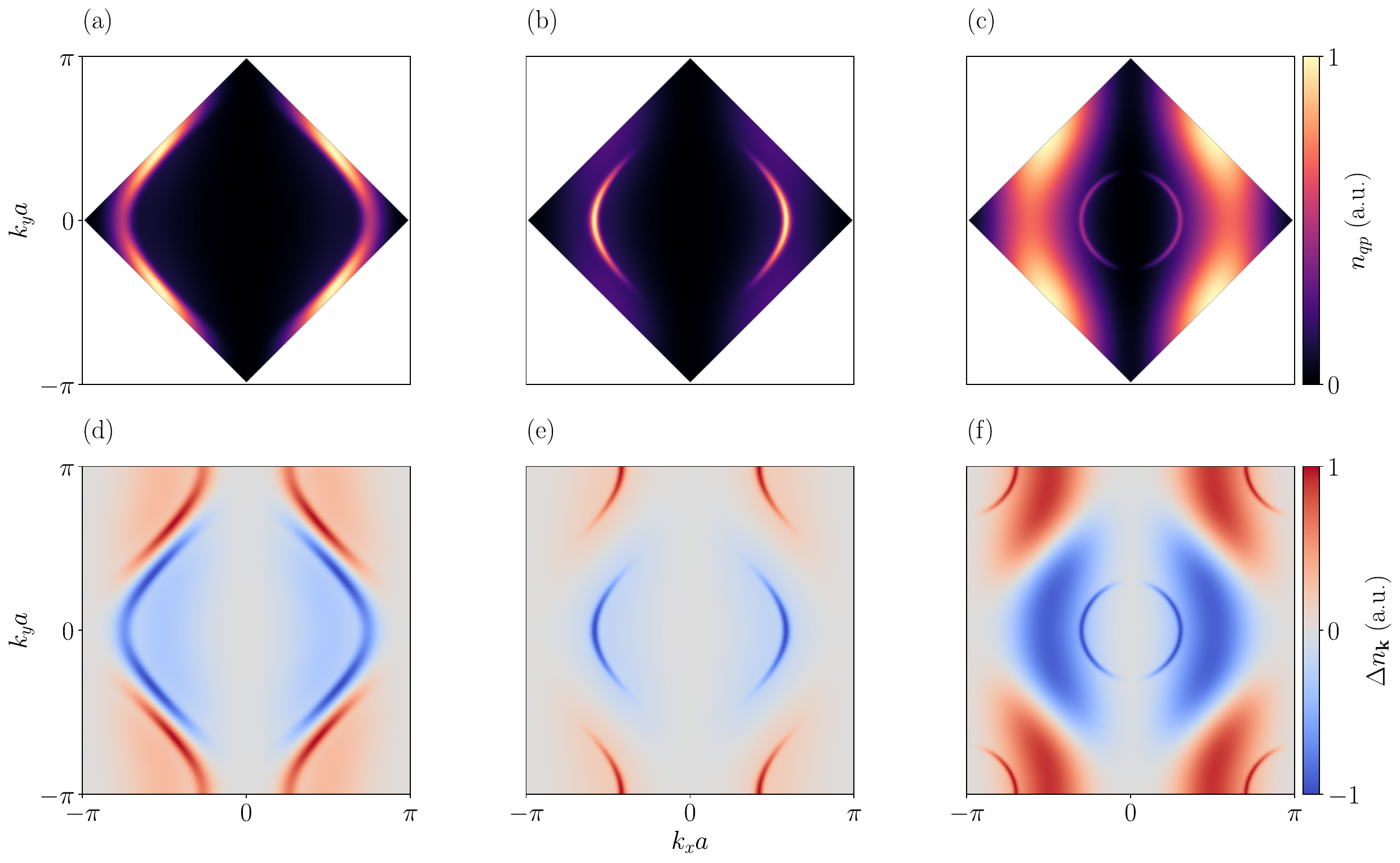}
	\caption{Crossing the boundary between photoexcitation regimes by tuning pump frequency: above the 3-photon threshold at low electric fields, the multiphoton and incoherent pathways compete. Increasing $\omega_d$ favors the latter, due to the reduced density of states for doublon-hole pairs at the top of the band. Calculated for $\field / \Delta_0 = 0.08$ and $\omega_d / \Delta_0$ of $0.335$ \textbf{(a, d)}, $0.35$ \textbf{(b, e)}, and respectively $0.4$ \textbf{(c, f)}.}
	\label{fig:incoherent_vs_multiphoton}
\end{figure*}

When the multiphoton mechanism competes with a different excitation pathway, tuning the drive frequency will affect the efficiency of the former. Indeed, the multiphoton production rate \eqref{eq:maintext_mph_freq_dep} is sensitive to the density of states shown in Fig. \ref{fig:density_of_states}, which is peaked at the bottom of the band. Therefore, increasing $\omega_d$ beyond that resonance may favor the competing excitation mechanism, which we illustrate in Fig. \ref{fig:incoherent_vs_multiphoton} by exploring a crossover between multiphoton- and incoherent-dominated regions in parameter space.\\

Regime boundaries can be traversed by tuning not only the pump frequency $\omega_d$, but also the field $\field$. Taking again the example of multiphoton versus incoherent competition, the production rate for the multiphoton mechanism scales with $\field^{2n}$, where $n$ is the number of necessary photons, while the incoherent one is proportional to $\field^2$. Stronger driving will therefore favor the former, which we show in Figure \ref{fig:incoherent_multiphoton_crossover} below.\\

\begin{figure*}
	\centering	
    \scalebox{\halffigsize}{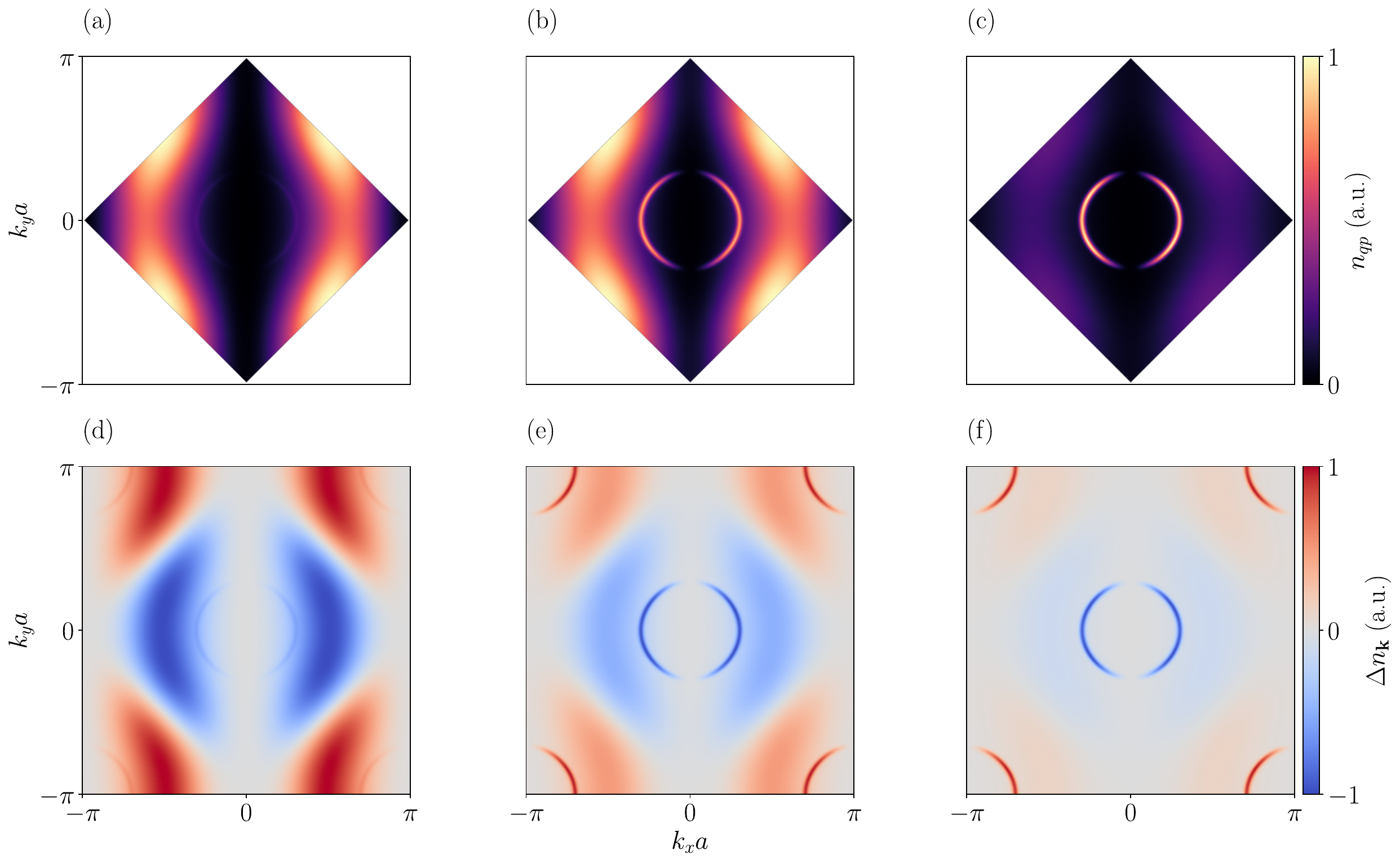}
	\caption{Crossing the incoherent / multiphoton regime boundary by increasing the driving field, for a fixed frequency above the 3-photon threshold. The higher power of the electric field in multiphoton production rates, compared to incoherent ones, favors the former at strong driving. Calculated for $\omega_d/ \Delta_0 = 0.4$ and $\field / \Delta_0$ of $0.05$ \textbf{(a, d)}, $0.1$ \textbf{(b, e)}, and respectively $0.15$ \textbf{(c, f)}.}
	\label{fig:incoherent_multiphoton_crossover}
\end{figure*}

Besides favoring excitation pathways that scale with higher powers thereof, strong fields will also renormalize the photocarrier dispersion, invalidating the rigid-band approximation. When this effect is weak, the main result will be a slight distortion of the usual energy contours; for the multiphoton pathway, an example is shown in Figure \ref{fig:multiphoton_gap_suppression} below. On the other hand, strong gap renormalization will completely shift the resonant contours throughout the evolution, giving rise to a broader momentum distribution, which tracks the entire trajectory of the contour (Fig. \ref{fig:continuous_excitation_suppression}).

\begin{figure*}
	\centering	
    \scalebox{\halffigsize}{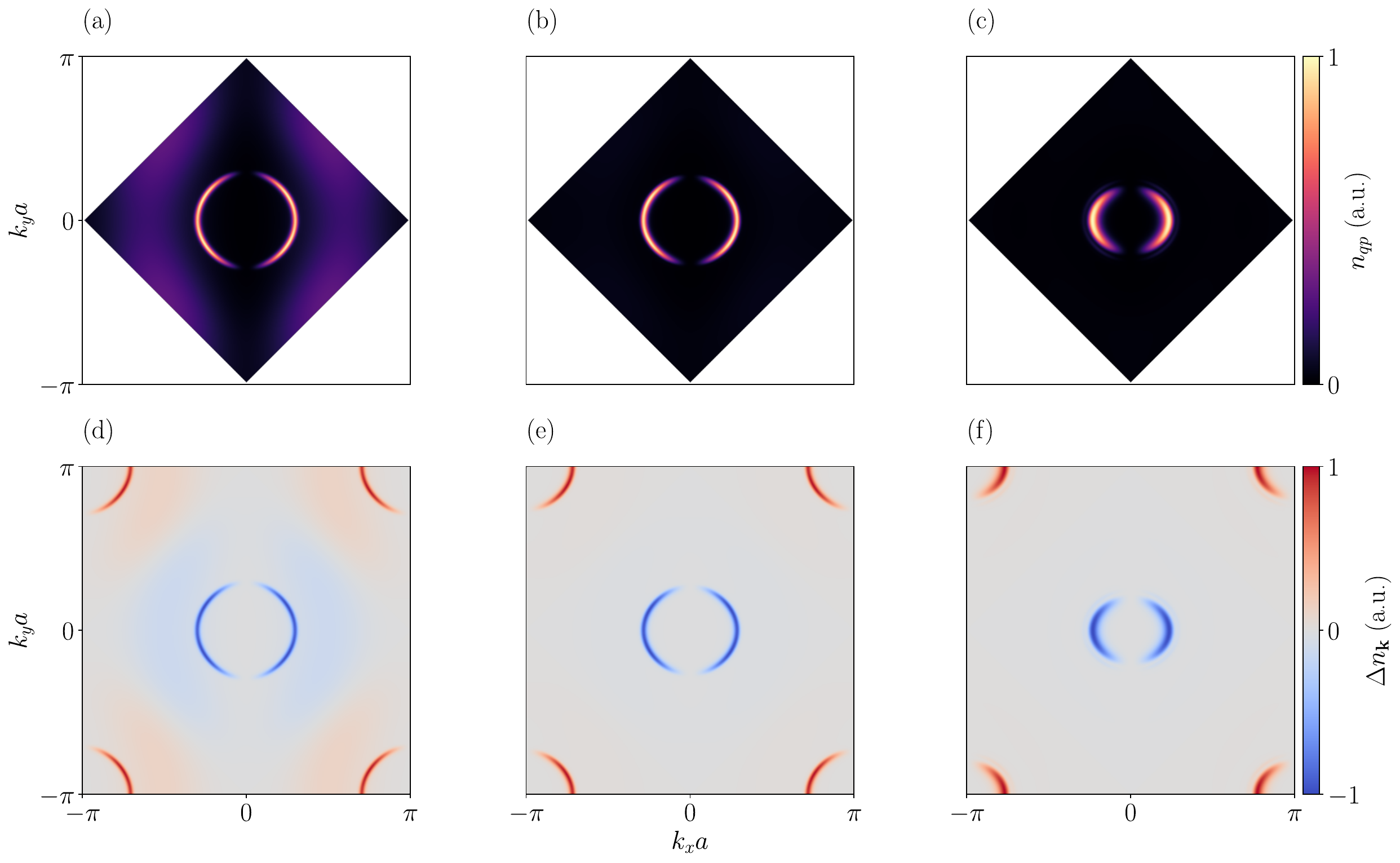}
	\caption{Momentum distributions in the 3-photon regime, with slight gap suppression. Calculated for $\omega_d/ \Delta_0 = 0.4$ and $\field / \Delta_0$ of $0.15$ \textbf{(a, d)}, $0.25$ \textbf{(b, e)}, and respectively $0.375$ \textbf{(c, f)}.}
	\label{fig:multiphoton_gap_suppression}
\end{figure*}

\begin{figure*}
	\centering	
    \scalebox{\fullfigsize}{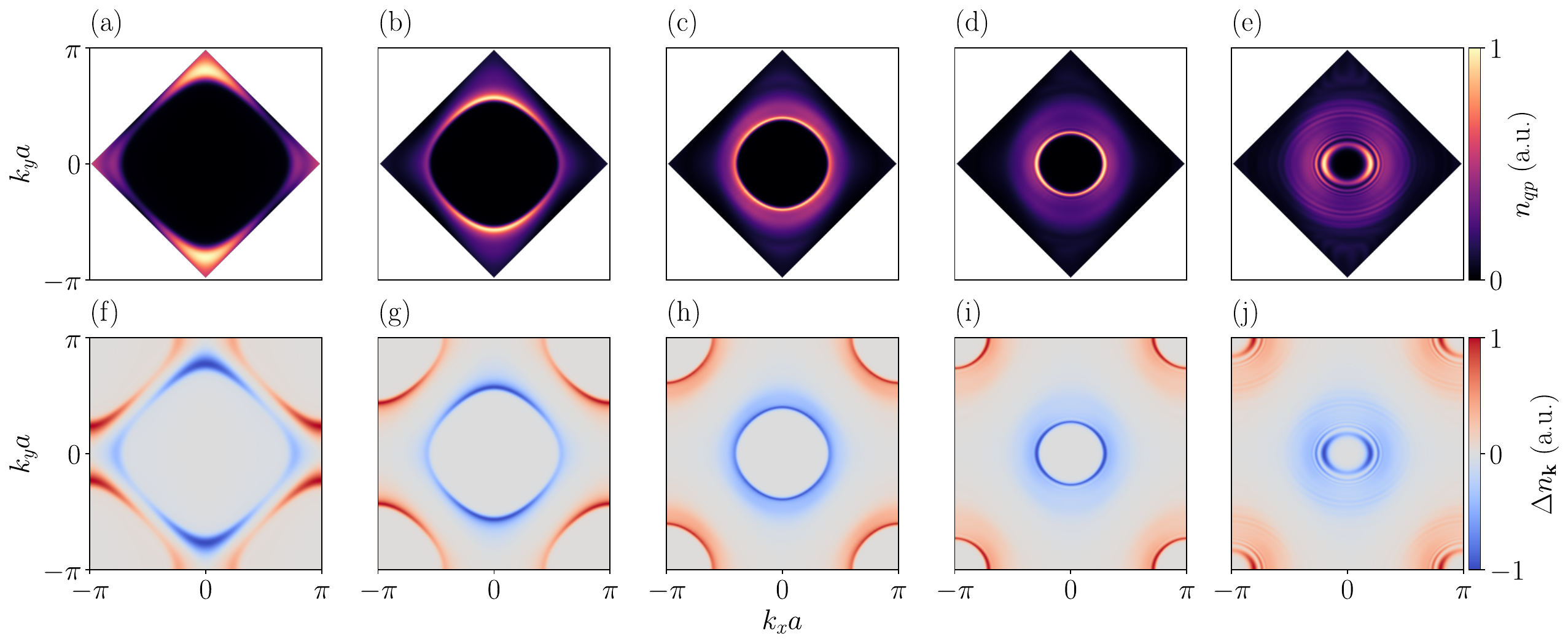}
	\caption{Momentum distributions in the 2-photon regime, with stronger gap suppression; although the driving frequency is fixed, high fields $\field$ access contours closer to the $\Gamma$ point. Continuous photoexcitation during the gap suppression gives rise to a broader distribution of carriers. Calculated for $\omega_d/ \Delta_0 = 0.501$ and $\field / \Delta_0$ of $0.05$ \textbf{(a, f)}, $0.15$ \textbf{(b, g)}, $0.25$ \textbf{(c, h)}, $0.35$ \textbf{(d, i)}, and $0.45$ \textbf{(e, j)}.}
	\label{fig:continuous_excitation_suppression}
\end{figure*}


\begin{thebibliography}{73}%
	\makeatletter
	\providecommand \@ifxundefined [1]{%
		\@ifx{#1\undefined}
	}%
	\providecommand \@ifnum [1]{%
		\ifnum #1\expandafter \@firstoftwo
		\else \expandafter \@secondoftwo
		\fi
	}%
	\providecommand \@ifx [1]{%
		\ifx #1\expandafter \@firstoftwo
		\else \expandafter \@secondoftwo
		\fi
	}%
	\providecommand \natexlab [1]{#1}%
	\providecommand \enquote  [1]{``#1''}%
	\providecommand \bibnamefont  [1]{#1}%
	\providecommand \bibfnamefont [1]{#1}%
	\providecommand \citenamefont [1]{#1}%
	\providecommand \href@noop [0]{\@secondoftwo}%
	\providecommand \href [0]{\begingroup \@sanitize@url \@href}%
	\providecommand \@href[1]{\@@startlink{#1}\@@href}%
	\providecommand \@@href[1]{\endgroup#1\@@endlink}%
	\providecommand \@sanitize@url [0]{\catcode `\\12\catcode `\$12\catcode
		`\&12\catcode `\#12\catcode `\^12\catcode `\_12\catcode `\%12\relax}%
	\providecommand \@@startlink[1]{}%
	\providecommand \@@endlink[0]{}%
	\providecommand \url  [0]{\begingroup\@sanitize@url \@url }%
	\providecommand \@url [1]{\endgroup\@href {#1}{\urlprefix }}%
	\providecommand \urlprefix  [0]{URL }%
	\providecommand \Eprint [0]{\href }%
	\providecommand \doibase [0]{https://doi.org/}%
	\providecommand \selectlanguage [0]{\@gobble}%
	\providecommand \bibinfo  [0]{\@secondoftwo}%
	\providecommand \bibfield  [0]{\@secondoftwo}%
	\providecommand \translation [1]{[#1]}%
	\providecommand \BibitemOpen [0]{}%
	\providecommand \bibitemStop [0]{}%
	\providecommand \bibitemNoStop [0]{.\EOS\space}%
	\providecommand \EOS [0]{\spacefactor3000\relax}%
	\providecommand \BibitemShut  [1]{\csname bibitem#1\endcsname}%
	\let\auto@bib@innerbib\@empty
	\bibitem [{\citenamefont {Basov}\ \emph {et~al.}(2017)\citenamefont {Basov},
		\citenamefont {Averitt},\ and\ \citenamefont {Hsieh}}]{basov_towards_2017}%
	\BibitemOpen
	\bibfield  {author} {\bibinfo {author} {\bibfnamefont {D.~N.}\ \bibnamefont
			{Basov}}, \bibinfo {author} {\bibfnamefont {R.~D.}\ \bibnamefont {Averitt}},\
		and\ \bibinfo {author} {\bibfnamefont {D.}~\bibnamefont {Hsieh}},\ }\bibfield
	{title} {\bibinfo {title} {Towards properties on demand in quantum
			materials},\ }\href {https://doi.org/10.1038/nmat5017} {\bibfield  {journal}
		{\bibinfo  {journal} {Nature Materials}\ }\textbf {\bibinfo {volume} {16}},\
		\bibinfo {pages} {1077} (\bibinfo {year} {2017})},\ \bibinfo {note}
	{publisher: Nature Publishing Group}\BibitemShut {NoStop}%
	\bibitem [{\citenamefont {Giannetti}\ \emph {et~al.}(2016)\citenamefont
		{Giannetti}, \citenamefont {~}, \citenamefont {~}, \citenamefont {~},
		\citenamefont {~}, ,\ and\ \citenamefont
		{Mihailovic}}]{giannetti_ultrafast_2016}%
	\BibitemOpen
	\bibfield  {author} {\bibinfo {author} {\bibfnamefont {C.}~\bibnamefont
			{Giannetti}}, \bibinfo {author} {\bibfnamefont {C.}~\bibnamefont {~},
			\bibfnamefont {Massimo}}, \bibinfo {author} {\bibfnamefont {F.}~\bibnamefont
			{~}, \bibfnamefont {Daniele}}, \bibinfo {author} {\bibfnamefont
			{F.}~\bibnamefont {~}, \bibfnamefont {Michele}}, \bibinfo {author}
		{\bibfnamefont {P.}~\bibnamefont {~}, \bibfnamefont {Fulvio}}, ,\ and\
		\bibinfo {author} {\bibfnamefont {D.}~\bibnamefont {Mihailovic}},\ }\bibfield
	{title} {\bibinfo {title} {Ultrafast optical spectroscopy of strongly
			correlated materials and high-temperature superconductors: a non-equilibrium
			approach},\ }\href {https://doi.org/10.1080/00018732.2016.1194044} {\bibfield
		{journal} {\bibinfo  {journal} {Advances in Physics}\ }\textbf {\bibinfo
			{volume} {65}},\ \bibinfo {pages} {58} (\bibinfo {year} {2016})},\ \bibinfo
	{note} {publisher: Taylor \& Francis \_eprint:
		https://doi.org/10.1080/00018732.2016.1194044}\BibitemShut {NoStop}%
	\bibitem [{\citenamefont {Baykusheva}\ \emph {et~al.}(2022)\citenamefont
		{Baykusheva}, \citenamefont {Jang}, \citenamefont {Husain}, \citenamefont
		{Lee}, \citenamefont {TenHuisen}, \citenamefont {Zhou}, \citenamefont {Park},
		\citenamefont {Kim}, \citenamefont {Kim}, \citenamefont {Kim}, \citenamefont
		{Kim}, \citenamefont {Park}, \citenamefont {Abbamonte}, \citenamefont {Kim},
		\citenamefont {Gu}, \citenamefont {Wang},\ and\ \citenamefont
		{Mitrano}}]{renormalizing_u_lco_ultrafast}%
	\BibitemOpen
	\bibfield  {author} {\bibinfo {author} {\bibfnamefont {D.~R.}\ \bibnamefont
			{Baykusheva}}, \bibinfo {author} {\bibfnamefont {H.}~\bibnamefont {Jang}},
		\bibinfo {author} {\bibfnamefont {A.~A.}\ \bibnamefont {Husain}}, \bibinfo
		{author} {\bibfnamefont {S.}~\bibnamefont {Lee}}, \bibinfo {author}
		{\bibfnamefont {S.~F.~R.}\ \bibnamefont {TenHuisen}}, \bibinfo {author}
		{\bibfnamefont {P.}~\bibnamefont {Zhou}}, \bibinfo {author} {\bibfnamefont
			{S.}~\bibnamefont {Park}}, \bibinfo {author} {\bibfnamefont {H.}~\bibnamefont
			{Kim}}, \bibinfo {author} {\bibfnamefont {J.-K.}\ \bibnamefont {Kim}},
		\bibinfo {author} {\bibfnamefont {H.-D.}\ \bibnamefont {Kim}}, \bibinfo
		{author} {\bibfnamefont {M.}~\bibnamefont {Kim}}, \bibinfo {author}
		{\bibfnamefont {S.-Y.}\ \bibnamefont {Park}}, \bibinfo {author}
		{\bibfnamefont {P.}~\bibnamefont {Abbamonte}}, \bibinfo {author}
		{\bibfnamefont {B.~J.}\ \bibnamefont {Kim}}, \bibinfo {author} {\bibfnamefont
			{G.~D.}\ \bibnamefont {Gu}}, \bibinfo {author} {\bibfnamefont
			{Y.}~\bibnamefont {Wang}},\ and\ \bibinfo {author} {\bibfnamefont
			{M.}~\bibnamefont {Mitrano}},\ }\bibfield  {title} {\bibinfo {title}
		{Ultrafast renormalization of the on-site coulomb repulsion in a cuprate
			superconductor},\ }\href {https://doi.org/10.1103/PhysRevX.12.011013}
	{\bibfield  {journal} {\bibinfo  {journal} {Phys. Rev. X}\ }\textbf {\bibinfo
			{volume} {12}},\ \bibinfo {pages} {011013} (\bibinfo {year}
		{2022})}\BibitemShut {NoStop}%
	\bibitem [{\citenamefont {Singla}\ \emph {et~al.}(2015)\citenamefont {Singla},
		\citenamefont {Cotugno}, \citenamefont {Kaiser}, \citenamefont {Först},
		\citenamefont {Mitrano}, \citenamefont {Liu}, \citenamefont {Cartella},
		\citenamefont {Manzoni}, \citenamefont {Okamoto}, \citenamefont {Hasegawa},
		\citenamefont {Clark}, \citenamefont {Jaksch},\ and\ \citenamefont
		{Cavalleri}}]{singla_thz_frequency_2015}%
	\BibitemOpen
	\bibfield  {author} {\bibinfo {author} {\bibfnamefont {R.}~\bibnamefont
			{Singla}}, \bibinfo {author} {\bibfnamefont {G.}~\bibnamefont {Cotugno}},
		\bibinfo {author} {\bibfnamefont {S.}~\bibnamefont {Kaiser}}, \bibinfo
		{author} {\bibfnamefont {M.}~\bibnamefont {Först}}, \bibinfo {author}
		{\bibfnamefont {M.}~\bibnamefont {Mitrano}}, \bibinfo {author} {\bibfnamefont
			{H.}~\bibnamefont {Liu}}, \bibinfo {author} {\bibfnamefont {A.}~\bibnamefont
			{Cartella}}, \bibinfo {author} {\bibfnamefont {C.}~\bibnamefont {Manzoni}},
		\bibinfo {author} {\bibfnamefont {H.}~\bibnamefont {Okamoto}}, \bibinfo
		{author} {\bibfnamefont {T.}~\bibnamefont {Hasegawa}}, \bibinfo {author}
		{\bibfnamefont {S.}~\bibnamefont {Clark}}, \bibinfo {author} {\bibfnamefont
			{D.}~\bibnamefont {Jaksch}},\ and\ \bibinfo {author} {\bibfnamefont
			{A.}~\bibnamefont {Cavalleri}},\ }\bibfield  {title} {\bibinfo {title}
		{{THz}-{Frequency} {Modulation} of the {Hubbard} \${U}\$ in an {Organic}
			{Mott} {Insulator}},\ }\href {https://doi.org/10.1103/PhysRevLett.115.187401}
	{\bibfield  {journal} {\bibinfo  {journal} {Physical Review Letters}\
		}\textbf {\bibinfo {volume} {115}},\ \bibinfo {pages} {187401} (\bibinfo
		{year} {2015})},\ \bibinfo {note} {publisher: American Physical
		Society}\BibitemShut {NoStop}%
	\bibitem [{\citenamefont {Lojewski}\ \emph {et~al.}(2024)\citenamefont
		{Lojewski}, \citenamefont {Golež}, \citenamefont {Ollefs}, \citenamefont
		{Le~Guyader}, \citenamefont {Kämmerer}, \citenamefont {Rothenbach},
		\citenamefont {Engel}, \citenamefont {Miedema}, \citenamefont {Beye},
		\citenamefont {Chiuzbăian}, \citenamefont {Carley}, \citenamefont {Gort},
		\citenamefont {Van~Kuiken}, \citenamefont {Mercurio}, \citenamefont
		{Schlappa}, \citenamefont {Yaroslavtsev}, \citenamefont {Scherz},
		\citenamefont {Döring}, \citenamefont {David}, \citenamefont {Wende},
		\citenamefont {Bovensiepen}, \citenamefont {Eckstein}, \citenamefont
		{Werner},\ and\ \citenamefont {Eschenlohr}}]{nio_renormalization}%
	\BibitemOpen
	\bibfield  {author} {\bibinfo {author} {\bibfnamefont {T.}~\bibnamefont
			{Lojewski}}, \bibinfo {author} {\bibfnamefont {D.}~\bibnamefont {Golež}},
		\bibinfo {author} {\bibfnamefont {K.}~\bibnamefont {Ollefs}}, \bibinfo
		{author} {\bibfnamefont {L.}~\bibnamefont {Le~Guyader}}, \bibinfo {author}
		{\bibfnamefont {L.}~\bibnamefont {Kämmerer}}, \bibinfo {author}
		{\bibfnamefont {N.}~\bibnamefont {Rothenbach}}, \bibinfo {author}
		{\bibfnamefont {R.~Y.}\ \bibnamefont {Engel}}, \bibinfo {author}
		{\bibfnamefont {P.~S.}\ \bibnamefont {Miedema}}, \bibinfo {author}
		{\bibfnamefont {M.}~\bibnamefont {Beye}}, \bibinfo {author} {\bibfnamefont
			{G.~S.}\ \bibnamefont {Chiuzbăian}}, \bibinfo {author} {\bibfnamefont
			{R.}~\bibnamefont {Carley}}, \bibinfo {author} {\bibfnamefont
			{R.}~\bibnamefont {Gort}}, \bibinfo {author} {\bibfnamefont {B.~E.}\
			\bibnamefont {Van~Kuiken}}, \bibinfo {author} {\bibfnamefont
			{G.}~\bibnamefont {Mercurio}}, \bibinfo {author} {\bibfnamefont
			{J.}~\bibnamefont {Schlappa}}, \bibinfo {author} {\bibfnamefont
			{A.}~\bibnamefont {Yaroslavtsev}}, \bibinfo {author} {\bibfnamefont
			{A.}~\bibnamefont {Scherz}}, \bibinfo {author} {\bibfnamefont
			{F.}~\bibnamefont {Döring}}, \bibinfo {author} {\bibfnamefont
			{C.}~\bibnamefont {David}}, \bibinfo {author} {\bibfnamefont
			{H.}~\bibnamefont {Wende}}, \bibinfo {author} {\bibfnamefont
			{U.}~\bibnamefont {Bovensiepen}}, \bibinfo {author} {\bibfnamefont
			{M.}~\bibnamefont {Eckstein}}, \bibinfo {author} {\bibfnamefont
			{P.}~\bibnamefont {Werner}},\ and\ \bibinfo {author} {\bibfnamefont
			{A.}~\bibnamefont {Eschenlohr}},\ }\bibfield  {title} {\bibinfo {title}
		{Photoinduced charge transfer renormalization in {NiO}},\ }\href
	{https://doi.org/10.1103/PhysRevB.110.245120} {\bibfield  {journal} {\bibinfo
			{journal} {Physical Review B}\ }\textbf {\bibinfo {volume} {110}},\ \bibinfo
		{pages} {245120} (\bibinfo {year} {2024})},\ \bibinfo {note} {publisher:
		American Physical Society}\BibitemShut {NoStop}%
	\bibitem [{\citenamefont {Beaulieu}\ \emph {et~al.}(2021)\citenamefont
		{Beaulieu}, \citenamefont {Dong}, \citenamefont {Tancogne-Dejean},
		\citenamefont {Dendzik}, \citenamefont {Pincelli}, \citenamefont {Maklar},
		\citenamefont {Xian}, \citenamefont {Sentef}, \citenamefont {Wolf},
		\citenamefont {Rubio}, \citenamefont {Rettig},\ and\ \citenamefont
		{Ernstorfer}}]{beaulieu_ultrafast_2021}%
	\BibitemOpen
	\bibfield  {author} {\bibinfo {author} {\bibfnamefont {S.}~\bibnamefont
			{Beaulieu}}, \bibinfo {author} {\bibfnamefont {S.}~\bibnamefont {Dong}},
		\bibinfo {author} {\bibfnamefont {N.}~\bibnamefont {Tancogne-Dejean}},
		\bibinfo {author} {\bibfnamefont {M.}~\bibnamefont {Dendzik}}, \bibinfo
		{author} {\bibfnamefont {T.}~\bibnamefont {Pincelli}}, \bibinfo {author}
		{\bibfnamefont {J.}~\bibnamefont {Maklar}}, \bibinfo {author} {\bibfnamefont
			{R.~P.}\ \bibnamefont {Xian}}, \bibinfo {author} {\bibfnamefont {M.~A.}\
			\bibnamefont {Sentef}}, \bibinfo {author} {\bibfnamefont {M.}~\bibnamefont
			{Wolf}}, \bibinfo {author} {\bibfnamefont {A.}~\bibnamefont {Rubio}},
		\bibinfo {author} {\bibfnamefont {L.}~\bibnamefont {Rettig}},\ and\ \bibinfo
		{author} {\bibfnamefont {R.}~\bibnamefont {Ernstorfer}},\ }\bibfield  {title}
	{\bibinfo {title} {Ultrafast dynamical {Lifshitz} transition},\ }\href
	{https://doi.org/10.1126/sciadv.abd9275} {\bibfield  {journal} {\bibinfo
			{journal} {Science Advances}\ }\textbf {\bibinfo {volume} {7}},\ \bibinfo
		{pages} {eabd9275} (\bibinfo {year} {2021})},\ \bibinfo {note} {publisher:
		American Association for the Advancement of Science}\BibitemShut {NoStop}%
	\bibitem [{\citenamefont {Gray}\ \emph {et~al.}(2018)\citenamefont {Gray},
		\citenamefont {Hoffmann}, \citenamefont {Jeong}, \citenamefont {Aetukuri},
		\citenamefont {Zhu}, \citenamefont {Hwang}, \citenamefont {Brandt},
		\citenamefont {Wen}, \citenamefont {Sternbach}, \citenamefont {Bonetti},
		\citenamefont {Reid}, \citenamefont {Kukreja}, \citenamefont {Graves},
		\citenamefont {Wang}, \citenamefont {Granitzka}, \citenamefont {Chen},
		\citenamefont {Higley}, \citenamefont {Chase}, \citenamefont {Jal},
		\citenamefont {Abreu}, \citenamefont {Liu}, \citenamefont {Weng},
		\citenamefont {Sokaras}, \citenamefont {Nordlund}, \citenamefont {Chollet},
		\citenamefont {Alonso-Mori}, \citenamefont {Lemke}, \citenamefont {Glownia},
		\citenamefont {Trigo}, \citenamefont {Zhu}, \citenamefont {Ohldag},
		\citenamefont {Freeland}, \citenamefont {Samant}, \citenamefont {Berakdar},
		\citenamefont {Averitt}, \citenamefont {Nelson}, \citenamefont {Parkin},\
		and\ \citenamefont {Dürr}}]{gray_ultrafast_2018}%
	\BibitemOpen
	\bibfield  {author} {\bibinfo {author} {\bibfnamefont {A.~X.}\ \bibnamefont
			{Gray}}, \bibinfo {author} {\bibfnamefont {M.~C.}\ \bibnamefont {Hoffmann}},
		\bibinfo {author} {\bibfnamefont {J.}~\bibnamefont {Jeong}}, \bibinfo
		{author} {\bibfnamefont {N.~P.}\ \bibnamefont {Aetukuri}}, \bibinfo {author}
		{\bibfnamefont {D.}~\bibnamefont {Zhu}}, \bibinfo {author} {\bibfnamefont
			{H.~Y.}\ \bibnamefont {Hwang}}, \bibinfo {author} {\bibfnamefont {N.~C.}\
			\bibnamefont {Brandt}}, \bibinfo {author} {\bibfnamefont {H.}~\bibnamefont
			{Wen}}, \bibinfo {author} {\bibfnamefont {A.~J.}\ \bibnamefont {Sternbach}},
		\bibinfo {author} {\bibfnamefont {S.}~\bibnamefont {Bonetti}}, \bibinfo
		{author} {\bibfnamefont {A.~H.}\ \bibnamefont {Reid}}, \bibinfo {author}
		{\bibfnamefont {R.}~\bibnamefont {Kukreja}}, \bibinfo {author} {\bibfnamefont
			{C.}~\bibnamefont {Graves}}, \bibinfo {author} {\bibfnamefont
			{T.}~\bibnamefont {Wang}}, \bibinfo {author} {\bibfnamefont {P.}~\bibnamefont
			{Granitzka}}, \bibinfo {author} {\bibfnamefont {Z.}~\bibnamefont {Chen}},
		\bibinfo {author} {\bibfnamefont {D.~J.}\ \bibnamefont {Higley}}, \bibinfo
		{author} {\bibfnamefont {T.}~\bibnamefont {Chase}}, \bibinfo {author}
		{\bibfnamefont {E.}~\bibnamefont {Jal}}, \bibinfo {author} {\bibfnamefont
			{E.}~\bibnamefont {Abreu}}, \bibinfo {author} {\bibfnamefont {M.~K.}\
			\bibnamefont {Liu}}, \bibinfo {author} {\bibfnamefont {T.-C.}\ \bibnamefont
			{Weng}}, \bibinfo {author} {\bibfnamefont {D.}~\bibnamefont {Sokaras}},
		\bibinfo {author} {\bibfnamefont {D.}~\bibnamefont {Nordlund}}, \bibinfo
		{author} {\bibfnamefont {M.}~\bibnamefont {Chollet}}, \bibinfo {author}
		{\bibfnamefont {R.}~\bibnamefont {Alonso-Mori}}, \bibinfo {author}
		{\bibfnamefont {H.}~\bibnamefont {Lemke}}, \bibinfo {author} {\bibfnamefont
			{J.~M.}\ \bibnamefont {Glownia}}, \bibinfo {author} {\bibfnamefont
			{M.}~\bibnamefont {Trigo}}, \bibinfo {author} {\bibfnamefont
			{Y.}~\bibnamefont {Zhu}}, \bibinfo {author} {\bibfnamefont {H.}~\bibnamefont
			{Ohldag}}, \bibinfo {author} {\bibfnamefont {J.~W.}\ \bibnamefont
			{Freeland}}, \bibinfo {author} {\bibfnamefont {M.~G.}\ \bibnamefont
			{Samant}}, \bibinfo {author} {\bibfnamefont {J.}~\bibnamefont {Berakdar}},
		\bibinfo {author} {\bibfnamefont {R.~D.}\ \bibnamefont {Averitt}}, \bibinfo
		{author} {\bibfnamefont {K.~A.}\ \bibnamefont {Nelson}}, \bibinfo {author}
		{\bibfnamefont {S.~S.~P.}\ \bibnamefont {Parkin}},\ and\ \bibinfo {author}
		{\bibfnamefont {H.~A.}\ \bibnamefont {Dürr}},\ }\bibfield  {title} {\bibinfo
		{title} {Ultrafast terahertz field control of electronic and structural
			interactions in vanadium dioxide},\ }\href
	{https://doi.org/10.1103/PhysRevB.98.045104} {\bibfield  {journal} {\bibinfo
			{journal} {Physical Review B}\ }\textbf {\bibinfo {volume} {98}},\ \bibinfo
		{pages} {045104} (\bibinfo {year} {2018})},\ \bibinfo {note} {publisher:
		American Physical Society}\BibitemShut {NoStop}%
	\bibitem [{\citenamefont {Kaiser}\ \emph {et~al.}(2014)\citenamefont {Kaiser},
		\citenamefont {Clark}, \citenamefont {Nicoletti}, \citenamefont {Cotugno},
		\citenamefont {Tobey}, \citenamefont {Dean}, \citenamefont {Lupi},
		\citenamefont {Okamoto}, \citenamefont {Hasegawa}, \citenamefont {Jaksch},\
		and\ \citenamefont {Cavalleri}}]{kaiser_optical_2014}%
	\BibitemOpen
	\bibfield  {author} {\bibinfo {author} {\bibfnamefont {S.}~\bibnamefont
			{Kaiser}}, \bibinfo {author} {\bibfnamefont {S.~R.}\ \bibnamefont {Clark}},
		\bibinfo {author} {\bibfnamefont {D.}~\bibnamefont {Nicoletti}}, \bibinfo
		{author} {\bibfnamefont {G.}~\bibnamefont {Cotugno}}, \bibinfo {author}
		{\bibfnamefont {R.~I.}\ \bibnamefont {Tobey}}, \bibinfo {author}
		{\bibfnamefont {N.}~\bibnamefont {Dean}}, \bibinfo {author} {\bibfnamefont
			{S.}~\bibnamefont {Lupi}}, \bibinfo {author} {\bibfnamefont {H.}~\bibnamefont
			{Okamoto}}, \bibinfo {author} {\bibfnamefont {T.}~\bibnamefont {Hasegawa}},
		\bibinfo {author} {\bibfnamefont {D.}~\bibnamefont {Jaksch}},\ and\ \bibinfo
		{author} {\bibfnamefont {A.}~\bibnamefont {Cavalleri}},\ }\bibfield  {title}
	{\bibinfo {title} {Optical {Properties} of a {Vibrationally} {Modulated}
			{Solid} {State} {Mott} {Insulator}},\ }\href
	{https://doi.org/10.1038/srep03823} {\bibfield  {journal} {\bibinfo
			{journal} {Scientific Reports}\ }\textbf {\bibinfo {volume} {4}},\ \bibinfo
		{pages} {3823} (\bibinfo {year} {2014})},\ \bibinfo {note} {publisher: Nature
		Publishing Group}\BibitemShut {NoStop}%
	\bibitem [{\citenamefont {Novelli}\ \emph {et~al.}(2014)\citenamefont
		{Novelli}, \citenamefont {De~Filippis}, \citenamefont {Cataudella},
		\citenamefont {Esposito}, \citenamefont {Vergara}, \citenamefont {Cilento},
		\citenamefont {Sindici}, \citenamefont {Amaricci}, \citenamefont {Giannetti},
		\citenamefont {Prabhakaran}, \citenamefont {Wall}, \citenamefont {Perucchi},
		\citenamefont {Dal~Conte}, \citenamefont {Cerullo}, \citenamefont {Capone},
		\citenamefont {Mishchenko}, \citenamefont {Gr{\"u}ninger}, \citenamefont
		{Nagaosa}, \citenamefont {Parmigiani},\ and\ \citenamefont
		{Fausti}}]{dressed_quasiparticle_formation}%
	\BibitemOpen
	\bibfield  {author} {\bibinfo {author} {\bibfnamefont {F.}~\bibnamefont
			{Novelli}}, \bibinfo {author} {\bibfnamefont {G.}~\bibnamefont
			{De~Filippis}}, \bibinfo {author} {\bibfnamefont {V.}~\bibnamefont
			{Cataudella}}, \bibinfo {author} {\bibfnamefont {M.}~\bibnamefont
			{Esposito}}, \bibinfo {author} {\bibfnamefont {I.}~\bibnamefont {Vergara}},
		\bibinfo {author} {\bibfnamefont {F.}~\bibnamefont {Cilento}}, \bibinfo
		{author} {\bibfnamefont {E.}~\bibnamefont {Sindici}}, \bibinfo {author}
		{\bibfnamefont {A.}~\bibnamefont {Amaricci}}, \bibinfo {author}
		{\bibfnamefont {C.}~\bibnamefont {Giannetti}}, \bibinfo {author}
		{\bibfnamefont {D.}~\bibnamefont {Prabhakaran}}, \bibinfo {author}
		{\bibfnamefont {S.}~\bibnamefont {Wall}}, \bibinfo {author} {\bibfnamefont
			{A.}~\bibnamefont {Perucchi}}, \bibinfo {author} {\bibfnamefont
			{S.}~\bibnamefont {Dal~Conte}}, \bibinfo {author} {\bibfnamefont
			{G.}~\bibnamefont {Cerullo}}, \bibinfo {author} {\bibfnamefont
			{M.}~\bibnamefont {Capone}}, \bibinfo {author} {\bibfnamefont
			{A.}~\bibnamefont {Mishchenko}}, \bibinfo {author} {\bibfnamefont
			{M.}~\bibnamefont {Gr{\"u}ninger}}, \bibinfo {author} {\bibfnamefont
			{N.}~\bibnamefont {Nagaosa}}, \bibinfo {author} {\bibfnamefont
			{F.}~\bibnamefont {Parmigiani}},\ and\ \bibinfo {author} {\bibfnamefont
			{D.}~\bibnamefont {Fausti}},\ }\bibfield  {title} {\bibinfo {title}
		{Witnessing the formation and relaxation of dressed quasi-particles in a
			strongly correlated electron system},\ }\href
	{https://doi.org/10.1038/ncomms6112} {\bibfield  {journal} {\bibinfo
			{journal} {Nature Communications}\ }\textbf {\bibinfo {volume} {5}},\
		\bibinfo {pages} {5112} (\bibinfo {year} {2014})}\BibitemShut {NoStop}%
	\bibitem [{\citenamefont {Wang}\ \emph {et~al.}(2022)\citenamefont {Wang},
		\citenamefont {Engel}, \citenamefont {Vaskivskyi}, \citenamefont {Turenne},
		\citenamefont {Shokeen}, \citenamefont {Yaroslavtsev}, \citenamefont
		{Grånäs}, \citenamefont {Knut}, \citenamefont {Schunck}, \citenamefont
		{Dziarzhytski}, \citenamefont {Brenner}, \citenamefont {Wang}, \citenamefont
		{Kuhlmann}, \citenamefont {Kuschewski}, \citenamefont {Bronsch},
		\citenamefont {Schüßler-Langeheine}, \citenamefont {Styervoyedov},
		\citenamefont {Parkin}, \citenamefont {Parmigiani}, \citenamefont {Eriksson},
		\citenamefont {Beye},\ and\ \citenamefont {Dürr}}]{wang_ultrafast_2022}%
	\BibitemOpen
	\bibfield  {author} {\bibinfo {author} {\bibfnamefont {X.}~\bibnamefont
			{Wang}}, \bibinfo {author} {\bibfnamefont {R.~Y.}\ \bibnamefont {Engel}},
		\bibinfo {author} {\bibfnamefont {I.}~\bibnamefont {Vaskivskyi}}, \bibinfo
		{author} {\bibfnamefont {D.}~\bibnamefont {Turenne}}, \bibinfo {author}
		{\bibfnamefont {V.}~\bibnamefont {Shokeen}}, \bibinfo {author} {\bibfnamefont
			{A.}~\bibnamefont {Yaroslavtsev}}, \bibinfo {author} {\bibfnamefont
			{O.}~\bibnamefont {Grånäs}}, \bibinfo {author} {\bibfnamefont
			{R.}~\bibnamefont {Knut}}, \bibinfo {author} {\bibfnamefont {J.~O.}\
			\bibnamefont {Schunck}}, \bibinfo {author} {\bibfnamefont {S.}~\bibnamefont
			{Dziarzhytski}}, \bibinfo {author} {\bibfnamefont {G.}~\bibnamefont
			{Brenner}}, \bibinfo {author} {\bibfnamefont {R.-P.}\ \bibnamefont {Wang}},
		\bibinfo {author} {\bibfnamefont {M.}~\bibnamefont {Kuhlmann}}, \bibinfo
		{author} {\bibfnamefont {F.}~\bibnamefont {Kuschewski}}, \bibinfo {author}
		{\bibfnamefont {W.}~\bibnamefont {Bronsch}}, \bibinfo {author} {\bibfnamefont
			{C.}~\bibnamefont {Schüßler-Langeheine}}, \bibinfo {author} {\bibfnamefont
			{A.}~\bibnamefont {Styervoyedov}}, \bibinfo {author} {\bibfnamefont
			{S.~S.~P.}\ \bibnamefont {Parkin}}, \bibinfo {author} {\bibfnamefont
			{F.}~\bibnamefont {Parmigiani}}, \bibinfo {author} {\bibfnamefont
			{O.}~\bibnamefont {Eriksson}}, \bibinfo {author} {\bibfnamefont
			{M.}~\bibnamefont {Beye}},\ and\ \bibinfo {author} {\bibfnamefont {H.~A.}\
			\bibnamefont {Dürr}},\ }\bibfield  {title} {\bibinfo {title} {Ultrafast
			manipulation of the {NiO} antiferromagnetic order via sub-gap optical
			excitation},\ }\href {https://doi.org/10.1039/D2FD00005A} {\bibfield
		{journal} {\bibinfo  {journal} {Faraday Discussions}\ }\textbf {\bibinfo
			{volume} {237}},\ \bibinfo {pages} {300} (\bibinfo {year} {2022})},\ \bibinfo
	{note} {publisher: The Royal Society of Chemistry}\BibitemShut {NoStop}%
	\bibitem [{\citenamefont {Hortensius}\ \emph {et~al.}(2021)\citenamefont
		{Hortensius}, \citenamefont {Afanasiev}, \citenamefont {Matthiesen},
		\citenamefont {Leenders}, \citenamefont {Citro}, \citenamefont {Kimel},
		\citenamefont {Mikhaylovskiy}, \citenamefont {Ivanov},\ and\ \citenamefont
		{Caviglia}}]{hortensius_coherent_2021}%
	\BibitemOpen
	\bibfield  {author} {\bibinfo {author} {\bibfnamefont {J.~R.}\ \bibnamefont
			{Hortensius}}, \bibinfo {author} {\bibfnamefont {D.}~\bibnamefont
			{Afanasiev}}, \bibinfo {author} {\bibfnamefont {M.}~\bibnamefont
			{Matthiesen}}, \bibinfo {author} {\bibfnamefont {R.}~\bibnamefont
			{Leenders}}, \bibinfo {author} {\bibfnamefont {R.}~\bibnamefont {Citro}},
		\bibinfo {author} {\bibfnamefont {A.~V.}\ \bibnamefont {Kimel}}, \bibinfo
		{author} {\bibfnamefont {R.~V.}\ \bibnamefont {Mikhaylovskiy}}, \bibinfo
		{author} {\bibfnamefont {B.~A.}\ \bibnamefont {Ivanov}},\ and\ \bibinfo
		{author} {\bibfnamefont {A.~D.}\ \bibnamefont {Caviglia}},\ }\bibfield
	{title} {\bibinfo {title} {Coherent spin-wave transport in an
			antiferromagnet},\ }\href {https://doi.org/10.1038/s41567-021-01290-4}
	{\bibfield  {journal} {\bibinfo  {journal} {Nature Physics}\ }\textbf
		{\bibinfo {volume} {17}},\ \bibinfo {pages} {1001} (\bibinfo {year}
		{2021})},\ \bibinfo {note} {publisher: Nature Publishing Group}\BibitemShut
	{NoStop}%
	\bibitem [{\citenamefont {Radovskaia}\ \emph {et~al.}(2025)\citenamefont
		{Radovskaia}, \citenamefont {Andrei}, \citenamefont {Hortensius},
		\citenamefont {Mikhaylovskiy}, \citenamefont {Citro}, \citenamefont
		{Chattopadhyay}, \citenamefont {Na}, \citenamefont {Ivanov}, \citenamefont
		{Demler}, \citenamefont {Kimel}, \citenamefont {Caviglia},\ and\
		\citenamefont {Afanasiev}}]{radovskaia_photoengineering_2025}%
	\BibitemOpen
	\bibfield  {author} {\bibinfo {author} {\bibfnamefont {V.}~\bibnamefont
			{Radovskaia}}, \bibinfo {author} {\bibfnamefont {R.}~\bibnamefont {Andrei}},
		\bibinfo {author} {\bibfnamefont {J.~R.}\ \bibnamefont {Hortensius}},
		\bibinfo {author} {\bibfnamefont {R.~V.}\ \bibnamefont {Mikhaylovskiy}},
		\bibinfo {author} {\bibfnamefont {R.}~\bibnamefont {Citro}}, \bibinfo
		{author} {\bibfnamefont {S.}~\bibnamefont {Chattopadhyay}}, \bibinfo {author}
		{\bibfnamefont {M.~X.}\ \bibnamefont {Na}}, \bibinfo {author} {\bibfnamefont
			{B.~A.}\ \bibnamefont {Ivanov}}, \bibinfo {author} {\bibfnamefont
			{E.}~\bibnamefont {Demler}}, \bibinfo {author} {\bibfnamefont {A.~V.}\
			\bibnamefont {Kimel}}, \bibinfo {author} {\bibfnamefont {A.~D.}\ \bibnamefont
			{Caviglia}},\ and\ \bibinfo {author} {\bibfnamefont {D.}~\bibnamefont
			{Afanasiev}},\ }\href {https://doi.org/10.48550/arXiv.2505.00459} {\bibinfo
		{title} {Photoengineering the {Magnon} {Spectrum} in an {Insulating}
			{Antiferromagnet}}} (\bibinfo {year} {2025}),\ \bibinfo {note}
	{arXiv:2505.00459 [cond-mat]}\BibitemShut {NoStop}%
	\bibitem [{\citenamefont {Perfetti}\ \emph {et~al.}(2006)\citenamefont
		{Perfetti}, \citenamefont {Loukakos}, \citenamefont {Lisowski}, \citenamefont
		{Bovensiepen}, \citenamefont {Berger}, \citenamefont {Biermann},
		\citenamefont {Cornaglia}, \citenamefont {Georges},\ and\ \citenamefont
		{Wolf}}]{perfetti_time_2006}%
	\BibitemOpen
	\bibfield  {author} {\bibinfo {author} {\bibfnamefont {L.}~\bibnamefont
			{Perfetti}}, \bibinfo {author} {\bibfnamefont {P.~A.}\ \bibnamefont
			{Loukakos}}, \bibinfo {author} {\bibfnamefont {M.}~\bibnamefont {Lisowski}},
		\bibinfo {author} {\bibfnamefont {U.}~\bibnamefont {Bovensiepen}}, \bibinfo
		{author} {\bibfnamefont {H.}~\bibnamefont {Berger}}, \bibinfo {author}
		{\bibfnamefont {S.}~\bibnamefont {Biermann}}, \bibinfo {author}
		{\bibfnamefont {P.~S.}\ \bibnamefont {Cornaglia}}, \bibinfo {author}
		{\bibfnamefont {A.}~\bibnamefont {Georges}},\ and\ \bibinfo {author}
		{\bibfnamefont {M.}~\bibnamefont {Wolf}},\ }\bibfield  {title} {\bibinfo
		{title} {Time {Evolution} of the {Electronic} {Structure} of
			\${1T}{\textbackslash}mathrm\{{\textbackslash}text\{{\textbackslash}ensuremath\{-\}\}\}\{{\textbackslash}mathrm\{{TaS}\}\}\_\{2\}\$
			through the {Insulator}-{Metal} {Transition}},\ }\href
	{https://doi.org/10.1103/PhysRevLett.97.067402} {\bibfield  {journal}
		{\bibinfo  {journal} {Physical Review Letters}\ }\textbf {\bibinfo {volume}
			{97}},\ \bibinfo {pages} {067402} (\bibinfo {year} {2006})},\ \bibinfo {note}
	{publisher: American Physical Society}\BibitemShut {NoStop}%
	\bibitem [{\citenamefont {Gillmeister}\ \emph {et~al.}(2020)\citenamefont
		{Gillmeister}, \citenamefont {Golež}, \citenamefont {Chiang}, \citenamefont
		{Bittner}, \citenamefont {Pavlyukh}, \citenamefont {Berakdar}, \citenamefont
		{Werner},\ and\ \citenamefont {Widdra}}]{gillmeister_ultrafast_2020}%
	\BibitemOpen
	\bibfield  {author} {\bibinfo {author} {\bibfnamefont {K.}~\bibnamefont
			{Gillmeister}}, \bibinfo {author} {\bibfnamefont {D.}~\bibnamefont {Golež}},
		\bibinfo {author} {\bibfnamefont {C.-T.}\ \bibnamefont {Chiang}}, \bibinfo
		{author} {\bibfnamefont {N.}~\bibnamefont {Bittner}}, \bibinfo {author}
		{\bibfnamefont {Y.}~\bibnamefont {Pavlyukh}}, \bibinfo {author}
		{\bibfnamefont {J.}~\bibnamefont {Berakdar}}, \bibinfo {author}
		{\bibfnamefont {P.}~\bibnamefont {Werner}},\ and\ \bibinfo {author}
		{\bibfnamefont {W.}~\bibnamefont {Widdra}},\ }\bibfield  {title} {\bibinfo
		{title} {Ultrafast coupled charge and spin dynamics in strongly correlated
			{NiO}},\ }\href {https://doi.org/10.1038/s41467-020-17925-8} {\bibfield
		{journal} {\bibinfo  {journal} {Nature Communications}\ }\textbf {\bibinfo
			{volume} {11}},\ \bibinfo {pages} {4095} (\bibinfo {year} {2020})},\ \bibinfo
	{note} {publisher: Nature Publishing Group}\BibitemShut {NoStop}%
	\bibitem [{\citenamefont {Dean}\ \emph {et~al.}(2016)\citenamefont {Dean},
		\citenamefont {Cao}, \citenamefont {Liu}, \citenamefont {Wall}, \citenamefont
		{Zhu}, \citenamefont {Mankowsky}, \citenamefont {Thampy}, \citenamefont
		{Chen}, \citenamefont {Vale}, \citenamefont {Casa}, \citenamefont {Kim},
		\citenamefont {Said}, \citenamefont {Juhas}, \citenamefont {Alonso-Mori},
		\citenamefont {Glownia}, \citenamefont {Robert}, \citenamefont {Robinson},
		\citenamefont {Sikorski}, \citenamefont {Song}, \citenamefont {Kozina},
		\citenamefont {Lemke}, \citenamefont {Patthey}, \citenamefont {Owada},
		\citenamefont {Katayama}, \citenamefont {Yabashi}, \citenamefont {Tanaka},
		\citenamefont {Togashi}, \citenamefont {Liu}, \citenamefont {Rayan~Serrao},
		\citenamefont {Kim}, \citenamefont {Huber}, \citenamefont {Chang},
		\citenamefont {McMorrow}, \citenamefont {Först},\ and\ \citenamefont
		{Hill}}]{dean_ultrafast_2016}%
	\BibitemOpen
	\bibfield  {author} {\bibinfo {author} {\bibfnamefont {M.~P.~M.}\
			\bibnamefont {Dean}}, \bibinfo {author} {\bibfnamefont {Y.}~\bibnamefont
			{Cao}}, \bibinfo {author} {\bibfnamefont {X.}~\bibnamefont {Liu}}, \bibinfo
		{author} {\bibfnamefont {S.}~\bibnamefont {Wall}}, \bibinfo {author}
		{\bibfnamefont {D.}~\bibnamefont {Zhu}}, \bibinfo {author} {\bibfnamefont
			{R.}~\bibnamefont {Mankowsky}}, \bibinfo {author} {\bibfnamefont
			{V.}~\bibnamefont {Thampy}}, \bibinfo {author} {\bibfnamefont {X.~M.}\
			\bibnamefont {Chen}}, \bibinfo {author} {\bibfnamefont {J.~G.}\ \bibnamefont
			{Vale}}, \bibinfo {author} {\bibfnamefont {D.}~\bibnamefont {Casa}}, \bibinfo
		{author} {\bibfnamefont {J.}~\bibnamefont {Kim}}, \bibinfo {author}
		{\bibfnamefont {A.~H.}\ \bibnamefont {Said}}, \bibinfo {author}
		{\bibfnamefont {P.}~\bibnamefont {Juhas}}, \bibinfo {author} {\bibfnamefont
			{R.}~\bibnamefont {Alonso-Mori}}, \bibinfo {author} {\bibfnamefont {J.~M.}\
			\bibnamefont {Glownia}}, \bibinfo {author} {\bibfnamefont {A.}~\bibnamefont
			{Robert}}, \bibinfo {author} {\bibfnamefont {J.}~\bibnamefont {Robinson}},
		\bibinfo {author} {\bibfnamefont {M.}~\bibnamefont {Sikorski}}, \bibinfo
		{author} {\bibfnamefont {S.}~\bibnamefont {Song}}, \bibinfo {author}
		{\bibfnamefont {M.}~\bibnamefont {Kozina}}, \bibinfo {author} {\bibfnamefont
			{H.}~\bibnamefont {Lemke}}, \bibinfo {author} {\bibfnamefont
			{L.}~\bibnamefont {Patthey}}, \bibinfo {author} {\bibfnamefont
			{S.}~\bibnamefont {Owada}}, \bibinfo {author} {\bibfnamefont
			{T.}~\bibnamefont {Katayama}}, \bibinfo {author} {\bibfnamefont
			{M.}~\bibnamefont {Yabashi}}, \bibinfo {author} {\bibfnamefont
			{Y.}~\bibnamefont {Tanaka}}, \bibinfo {author} {\bibfnamefont
			{T.}~\bibnamefont {Togashi}}, \bibinfo {author} {\bibfnamefont
			{J.}~\bibnamefont {Liu}}, \bibinfo {author} {\bibfnamefont {C.}~\bibnamefont
			{Rayan~Serrao}}, \bibinfo {author} {\bibfnamefont {B.~J.}\ \bibnamefont
			{Kim}}, \bibinfo {author} {\bibfnamefont {L.}~\bibnamefont {Huber}}, \bibinfo
		{author} {\bibfnamefont {C.-L.}\ \bibnamefont {Chang}}, \bibinfo {author}
		{\bibfnamefont {D.~F.}\ \bibnamefont {McMorrow}}, \bibinfo {author}
		{\bibfnamefont {M.}~\bibnamefont {Först}},\ and\ \bibinfo {author}
		{\bibfnamefont {J.~P.}\ \bibnamefont {Hill}},\ }\bibfield  {title} {\bibinfo
		{title} {Ultrafast energy- and momentum-resolved dynamics of magnetic
			correlations in the photo-doped {Mott} insulator {Sr2IrO4}},\ }\href
	{https://doi.org/10.1038/nmat4641} {\bibfield  {journal} {\bibinfo  {journal}
			{Nature Materials}\ }\textbf {\bibinfo {volume} {15}},\ \bibinfo {pages}
		{601} (\bibinfo {year} {2016})},\ \bibinfo {note} {publisher: Nature
		Publishing Group}\BibitemShut {NoStop}%
	\bibitem [{\citenamefont {Choi}\ \emph {et~al.}(2024)\citenamefont {Choi},
		\citenamefont {Yue}, \citenamefont {Azoury}, \citenamefont {Porter},
		\citenamefont {Chen}, \citenamefont {Petocchi}, \citenamefont {Baldini},
		\citenamefont {Lv}, \citenamefont {Mogi}, \citenamefont {Su}, \citenamefont
		{Wilson}, \citenamefont {Eckstein}, \citenamefont {Werner},\ and\
		\citenamefont {Gedik}}]{choi_light_induced_2024}%
	\BibitemOpen
	\bibfield  {author} {\bibinfo {author} {\bibfnamefont {D.}~\bibnamefont
			{Choi}}, \bibinfo {author} {\bibfnamefont {C.}~\bibnamefont {Yue}}, \bibinfo
		{author} {\bibfnamefont {D.}~\bibnamefont {Azoury}}, \bibinfo {author}
		{\bibfnamefont {Z.}~\bibnamefont {Porter}}, \bibinfo {author} {\bibfnamefont
			{J.}~\bibnamefont {Chen}}, \bibinfo {author} {\bibfnamefont {F.}~\bibnamefont
			{Petocchi}}, \bibinfo {author} {\bibfnamefont {E.}~\bibnamefont {Baldini}},
		\bibinfo {author} {\bibfnamefont {B.}~\bibnamefont {Lv}}, \bibinfo {author}
		{\bibfnamefont {M.}~\bibnamefont {Mogi}}, \bibinfo {author} {\bibfnamefont
			{Y.}~\bibnamefont {Su}}, \bibinfo {author} {\bibfnamefont {S.~D.}\
			\bibnamefont {Wilson}}, \bibinfo {author} {\bibfnamefont {M.}~\bibnamefont
			{Eckstein}}, \bibinfo {author} {\bibfnamefont {P.}~\bibnamefont {Werner}},\
		and\ \bibinfo {author} {\bibfnamefont {N.}~\bibnamefont {Gedik}},\ }\bibfield
	{title} {\bibinfo {title} {Light-induced insulator–metal transition in
			{Sr2IrO4} reveals the nature of the insulating ground state},\ }\href
	{https://doi.org/10.1073/pnas.2323013121} {\bibfield  {journal} {\bibinfo
			{journal} {Proceedings of the National Academy of Sciences}\ }\textbf
		{\bibinfo {volume} {121}},\ \bibinfo {pages} {e2323013121} (\bibinfo {year}
		{2024})},\ \bibinfo {note} {publisher: Proceedings of the National Academy of
		Sciences}\BibitemShut {NoStop}%
	\bibitem [{\citenamefont {Stojchevska}\ \emph {et~al.}(2014)\citenamefont
		{Stojchevska}, \citenamefont {Vaskivskyi}, \citenamefont {Mertelj},
		\citenamefont {Kusar}, \citenamefont {Svetin}, \citenamefont {Brazovskii},\
		and\ \citenamefont {Mihailovic}}]{stojchevska_ultrafast_2014}%
	\BibitemOpen
	\bibfield  {author} {\bibinfo {author} {\bibfnamefont {L.}~\bibnamefont
			{Stojchevska}}, \bibinfo {author} {\bibfnamefont {I.}~\bibnamefont
			{Vaskivskyi}}, \bibinfo {author} {\bibfnamefont {T.}~\bibnamefont {Mertelj}},
		\bibinfo {author} {\bibfnamefont {P.}~\bibnamefont {Kusar}}, \bibinfo
		{author} {\bibfnamefont {D.}~\bibnamefont {Svetin}}, \bibinfo {author}
		{\bibfnamefont {S.}~\bibnamefont {Brazovskii}},\ and\ \bibinfo {author}
		{\bibfnamefont {D.}~\bibnamefont {Mihailovic}},\ }\bibfield  {title}
	{\bibinfo {title} {Ultrafast {Switching} to a {Stable} {Hidden} {Quantum}
			{State} in an {Electronic} {Crystal}},\ }\href
	{https://doi.org/10.1126/science.1241591} {\bibfield  {journal} {\bibinfo
			{journal} {Science}\ }\textbf {\bibinfo {volume} {344}},\ \bibinfo {pages}
		{177} (\bibinfo {year} {2014})},\ \bibinfo {note} {publisher: American
		Association for the Advancement of Science}\BibitemShut {NoStop}%
	\bibitem [{\citenamefont {Janod}\ \emph {et~al.}(2015)\citenamefont {Janod},
		\citenamefont {Tranchant}, \citenamefont {Corraze}, \citenamefont {Querré},
		\citenamefont {Stoliar}, \citenamefont {Rozenberg}, \citenamefont {Cren},
		\citenamefont {Roditchev}, \citenamefont {Phuoc}, \citenamefont {Besland},\
		and\ \citenamefont {Cario}}]{janod_resistive_2015}%
	\BibitemOpen
	\bibfield  {author} {\bibinfo {author} {\bibfnamefont {E.}~\bibnamefont
			{Janod}}, \bibinfo {author} {\bibfnamefont {J.}~\bibnamefont {Tranchant}},
		\bibinfo {author} {\bibfnamefont {B.}~\bibnamefont {Corraze}}, \bibinfo
		{author} {\bibfnamefont {M.}~\bibnamefont {Querré}}, \bibinfo {author}
		{\bibfnamefont {P.}~\bibnamefont {Stoliar}}, \bibinfo {author} {\bibfnamefont
			{M.}~\bibnamefont {Rozenberg}}, \bibinfo {author} {\bibfnamefont
			{T.}~\bibnamefont {Cren}}, \bibinfo {author} {\bibfnamefont {D.}~\bibnamefont
			{Roditchev}}, \bibinfo {author} {\bibfnamefont {V.~T.}\ \bibnamefont
			{Phuoc}}, \bibinfo {author} {\bibfnamefont {M.-P.}\ \bibnamefont {Besland}},\
		and\ \bibinfo {author} {\bibfnamefont {L.}~\bibnamefont {Cario}},\ }\bibfield
	{title} {\bibinfo {title} {Resistive {Switching} in {Mott} {Insulators} and
			{Correlated} {Systems}},\ }\href {https://doi.org/10.1002/adfm.201500823}
	{\bibfield  {journal} {\bibinfo  {journal} {Advanced Functional Materials}\
		}\textbf {\bibinfo {volume} {25}},\ \bibinfo {pages} {6287} (\bibinfo {year}
		{2015})},\ \bibinfo {note} {\_eprint:
		https://onlinelibrary.wiley.com/doi/pdf/10.1002/adfm.201500823}\BibitemShut
	{NoStop}%
	\bibitem [{\citenamefont {de~la Torre}\ \emph {et~al.}(2021)\citenamefont
		{de~la Torre}, \citenamefont {Kennes}, \citenamefont {Claassen},
		\citenamefont {Gerber}, \citenamefont {McIver},\ and\ \citenamefont
		{Sentef}}]{de_la_torre_colloquium_2021}%
	\BibitemOpen
	\bibfield  {author} {\bibinfo {author} {\bibfnamefont {A.}~\bibnamefont
			{de~la Torre}}, \bibinfo {author} {\bibfnamefont {D.~M.}\ \bibnamefont
			{Kennes}}, \bibinfo {author} {\bibfnamefont {M.}~\bibnamefont {Claassen}},
		\bibinfo {author} {\bibfnamefont {S.}~\bibnamefont {Gerber}}, \bibinfo
		{author} {\bibfnamefont {J.~W.}\ \bibnamefont {McIver}},\ and\ \bibinfo
		{author} {\bibfnamefont {M.~A.}\ \bibnamefont {Sentef}},\ }\bibfield  {title}
	{\bibinfo {title} {Colloquium: {Nonthermal} pathways to ultrafast control in
			quantum materials},\ }\href {https://doi.org/10.1103/RevModPhys.93.041002}
	{\bibfield  {journal} {\bibinfo  {journal} {Reviews of Modern Physics}\
		}\textbf {\bibinfo {volume} {93}},\ \bibinfo {pages} {041002} (\bibinfo
		{year} {2021})},\ \bibinfo {note} {publisher: American Physical
		Society}\BibitemShut {NoStop}%
	\bibitem [{\citenamefont {Murakami}\ \emph {et~al.}(2023)\citenamefont
		{Murakami}, \citenamefont {Gole\v{z}}, \citenamefont {Eckstein},\ and\
		\citenamefont {Werner}}]{photoinduced_nonequilibrium_review}%
	\BibitemOpen
	\bibfield  {author} {\bibinfo {author} {\bibfnamefont {Y.}~\bibnamefont
			{Murakami}}, \bibinfo {author} {\bibfnamefont {D.}~\bibnamefont {Gole\v{z}}},
		\bibinfo {author} {\bibfnamefont {M.}~\bibnamefont {Eckstein}},\ and\
		\bibinfo {author} {\bibfnamefont {P.}~\bibnamefont {Werner}},\ }\href@noop {}
	{\bibinfo {title} {Photo-induced nonequilibrium states in mott insulators}}
	(\bibinfo {year} {2023}),\ \Eprint {https://arxiv.org/abs/2310.05201}
	{arXiv:2310.05201 [cond-mat.str-el]} \BibitemShut {NoStop}%
	\bibitem [{\citenamefont {Keldysh}(1965)}]{keldysh_original_ionization}%
	\BibitemOpen
	\bibfield  {author} {\bibinfo {author} {\bibfnamefont {L.~V.}\ \bibnamefont
			{Keldysh}},\ }\bibfield  {title} {\bibinfo {title} {{Ionization in the Field
				of a Strong Electromagnetic Wave}},\ }\href@noop {} {\bibfield  {journal}
		{\bibinfo  {journal} {J. Exp. Theor. Phys.}\ }\textbf {\bibinfo {volume}
			{20}},\ \bibinfo {pages} {1307} (\bibinfo {year} {1965})}\BibitemShut
	{NoStop}%
	\bibitem [{Note1()}]{Note1}%
	\BibitemOpen
	\bibinfo {note} {In the limit of strong interactions $U / \tau \gg 1$ (see
		Section \ref {sec:theoretical_approach} for definitions of the interaction
		$U$ and electron hopping $\tau $), the correlation length $\xi $ is on the
		order of the lattice constant $a$. For the detailed dependence of $\xi / a$
		on $U/\tau $ in the 1D Hubbard model, see also Ref. \cite
		{millis_correlation_length} and Figure 3a of Ref. \cite
		{oka_nonlinear_doublons}.}\BibitemShut {Stop}%
	\bibitem [{\citenamefont {Oka}(2012)}]{oka_nonlinear_doublons}%
	\BibitemOpen
	\bibfield  {author} {\bibinfo {author} {\bibfnamefont {T.}~\bibnamefont
			{Oka}},\ }\bibfield  {title} {\bibinfo {title} {Nonlinear doublon production
			in a mott insulator: Landau-dykhne method applied to an integrable model},\
	}\href {https://doi.org/10.1103/PhysRevB.86.075148} {\bibfield  {journal}
		{\bibinfo  {journal} {Phys. Rev. B}\ }\textbf {\bibinfo {volume} {86}},\
		\bibinfo {pages} {075148} (\bibinfo {year} {2012})}\BibitemShut {NoStop}%
	\bibitem [{\citenamefont {Eckstein}\ \emph {et~al.}(2010)\citenamefont
		{Eckstein}, \citenamefont {Oka},\ and\ \citenamefont
		{Werner}}]{dmft_dc_dielectric_breakdown}%
	\BibitemOpen
	\bibfield  {author} {\bibinfo {author} {\bibfnamefont {M.}~\bibnamefont
			{Eckstein}}, \bibinfo {author} {\bibfnamefont {T.}~\bibnamefont {Oka}},\ and\
		\bibinfo {author} {\bibfnamefont {P.}~\bibnamefont {Werner}},\ }\bibfield
	{title} {\bibinfo {title} {Dielectric breakdown of mott insulators in
			dynamical mean-field theory},\ }\href
	{https://doi.org/10.1103/PhysRevLett.105.146404} {\bibfield  {journal}
		{\bibinfo  {journal} {Phys. Rev. Lett.}\ }\textbf {\bibinfo {volume} {105}},\
		\bibinfo {pages} {146404} (\bibinfo {year} {2010})}\BibitemShut {NoStop}%
	\bibitem [{\citenamefont {Aron}(2012)}]{dmft_dc_usa_dielectric_breakdown}%
	\BibitemOpen
	\bibfield  {author} {\bibinfo {author} {\bibfnamefont {C.}~\bibnamefont
			{Aron}},\ }\bibfield  {title} {\bibinfo {title} {Dielectric breakdown of a
			mott insulator},\ }\href {https://doi.org/10.1103/PhysRevB.86.085127}
	{\bibfield  {journal} {\bibinfo  {journal} {Phys. Rev. B}\ }\textbf {\bibinfo
			{volume} {86}},\ \bibinfo {pages} {085127} (\bibinfo {year}
		{2012})}\BibitemShut {NoStop}%
	\bibitem [{\citenamefont {Eckstein}\ and\ \citenamefont
		{Werner}(2013)}]{dmft_doublon_production_heating}%
	\BibitemOpen
	\bibfield  {author} {\bibinfo {author} {\bibfnamefont {M.}~\bibnamefont
			{Eckstein}}\ and\ \bibinfo {author} {\bibfnamefont {P.}~\bibnamefont
			{Werner}},\ }\bibfield  {title} {\bibinfo {title} {Dielectric breakdown of
			mott insulators – doublon production and doublon heating},\ }\href
	{https://doi.org/10.1088/1742-6596/427/1/012005} {\bibfield  {journal}
		{\bibinfo  {journal} {Journal of Physics: Conference Series}\ }\textbf
		{\bibinfo {volume} {427}},\ \bibinfo {pages} {012005} (\bibinfo {year}
		{2013})}\BibitemShut {NoStop}%
	\bibitem [{\citenamefont {Lee}\ and\ \citenamefont
		{Park}(2014)}]{dmft_two_dielectric_breakdowns}%
	\BibitemOpen
	\bibfield  {author} {\bibinfo {author} {\bibfnamefont {W.-R.}\ \bibnamefont
			{Lee}}\ and\ \bibinfo {author} {\bibfnamefont {K.}~\bibnamefont {Park}},\
	}\bibfield  {title} {\bibinfo {title} {Dielectric breakdown via emergent
			nonequilibrium steady states of the electric-field-driven mott insulator},\
	}\href {https://doi.org/10.1103/PhysRevB.89.205126} {\bibfield  {journal}
		{\bibinfo  {journal} {Phys. Rev. B}\ }\textbf {\bibinfo {volume} {89}},\
		\bibinfo {pages} {205126} (\bibinfo {year} {2014})}\BibitemShut {NoStop}%
	\bibitem [{\citenamefont {Li}\ \emph {et~al.}(2015)\citenamefont {Li},
		\citenamefont {Aron}, \citenamefont {Kotliar},\ and\ \citenamefont
		{Han}}]{resistive_switching_hysteresis}%
	\BibitemOpen
	\bibfield  {author} {\bibinfo {author} {\bibfnamefont {J.}~\bibnamefont
			{Li}}, \bibinfo {author} {\bibfnamefont {C.}~\bibnamefont {Aron}}, \bibinfo
		{author} {\bibfnamefont {G.}~\bibnamefont {Kotliar}},\ and\ \bibinfo {author}
		{\bibfnamefont {J.~E.}\ \bibnamefont {Han}},\ }\bibfield  {title} {\bibinfo
		{title} {Electric-field-driven resistive switching in the dissipative hubbard
			model},\ }\href {https://doi.org/10.1103/PhysRevLett.114.226403} {\bibfield
		{journal} {\bibinfo  {journal} {Phys. Rev. Lett.}\ }\textbf {\bibinfo
			{volume} {114}},\ \bibinfo {pages} {226403} (\bibinfo {year}
		{2015})}\BibitemShut {NoStop}%
	\bibitem [{\citenamefont {Mazza}\ \emph {et~al.}(2016)\citenamefont {Mazza},
		\citenamefont {Amaricci}, \citenamefont {Capone},\ and\ \citenamefont
		{Fabrizio}}]{gap_collapse_and_switching}%
	\BibitemOpen
	\bibfield  {author} {\bibinfo {author} {\bibfnamefont {G.}~\bibnamefont
			{Mazza}}, \bibinfo {author} {\bibfnamefont {A.}~\bibnamefont {Amaricci}},
		\bibinfo {author} {\bibfnamefont {M.}~\bibnamefont {Capone}},\ and\ \bibinfo
		{author} {\bibfnamefont {M.}~\bibnamefont {Fabrizio}},\ }\bibfield  {title}
	{\bibinfo {title} {Field-driven mott gap collapse and resistive switch in
			correlated insulators},\ }\href
	{https://doi.org/10.1103/PhysRevLett.117.176401} {\bibfield  {journal}
		{\bibinfo  {journal} {Phys. Rev. Lett.}\ }\textbf {\bibinfo {volume} {117}},\
		\bibinfo {pages} {176401} (\bibinfo {year} {2016})}\BibitemShut {NoStop}%
	\bibitem [{\citenamefont {Udono}\ \emph {et~al.}(2023)\citenamefont {Udono},
		\citenamefont {Kaneko},\ and\ \citenamefont {Sugimoto}}]{iTEBD_study}%
	\BibitemOpen
	\bibfield  {author} {\bibinfo {author} {\bibfnamefont {M.}~\bibnamefont
			{Udono}}, \bibinfo {author} {\bibfnamefont {T.}~\bibnamefont {Kaneko}},\ and\
		\bibinfo {author} {\bibfnamefont {K.}~\bibnamefont {Sugimoto}},\ }\bibfield
	{title} {\bibinfo {title} {Wannier-stark ladders and stark shifts of excitons
			in mott insulators},\ }\href {https://doi.org/10.1103/PhysRevB.108.L081304}
	{\bibfield  {journal} {\bibinfo  {journal} {Phys. Rev. B}\ }\textbf {\bibinfo
			{volume} {108}},\ \bibinfo {pages} {L081304} (\bibinfo {year}
		{2023})}\BibitemShut {NoStop}%
	\bibitem [{\citenamefont {Tsuji}\ \emph {et~al.}(2008)\citenamefont {Tsuji},
		\citenamefont {Oka},\ and\ \citenamefont {Aoki}}]{floquet_dmft_introduction}%
	\BibitemOpen
	\bibfield  {author} {\bibinfo {author} {\bibfnamefont {N.}~\bibnamefont
			{Tsuji}}, \bibinfo {author} {\bibfnamefont {T.}~\bibnamefont {Oka}},\ and\
		\bibinfo {author} {\bibfnamefont {H.}~\bibnamefont {Aoki}},\ }\bibfield
	{title} {\bibinfo {title} {Correlated electron systems periodically driven
			out of equilibrium: Floquet + dmft formalism},\ }\href
	{https://doi.org/10.1103/PhysRevB.78.235124} {\bibfield  {journal} {\bibinfo
			{journal} {Phys. Rev. B}\ }\textbf {\bibinfo {volume} {78}},\ \bibinfo
		{pages} {235124} (\bibinfo {year} {2008})}\BibitemShut {NoStop}%
	\bibitem [{\citenamefont {Herrmann}\ \emph {et~al.}(2018)\citenamefont
		{Herrmann}, \citenamefont {Murakami}, \citenamefont {Eckstein},\ and\
		\citenamefont {Werner}}]{floquet_prethermalization_resonant_hubbard}%
	\BibitemOpen
	\bibfield  {author} {\bibinfo {author} {\bibfnamefont {A.}~\bibnamefont
			{Herrmann}}, \bibinfo {author} {\bibfnamefont {Y.}~\bibnamefont {Murakami}},
		\bibinfo {author} {\bibfnamefont {M.}~\bibnamefont {Eckstein}},\ and\
		\bibinfo {author} {\bibfnamefont {P.}~\bibnamefont {Werner}},\ }\bibfield
	{title} {\bibinfo {title} {Floquet prethermalization in the resonantly driven
			hubbard model},\ }\href {https://doi.org/10.1209/0295-5075/120/57001}
	{\bibfield  {journal} {\bibinfo  {journal} {Europhysics Letters}\ }\textbf
		{\bibinfo {volume} {120}},\ \bibinfo {pages} {57001} (\bibinfo {year}
		{2018})}\BibitemShut {NoStop}%
	\bibitem [{\citenamefont {Murakami}\ and\ \citenamefont
		{Werner}(2018)}]{driven_steady_state}%
	\BibitemOpen
	\bibfield  {author} {\bibinfo {author} {\bibfnamefont {Y.}~\bibnamefont
			{Murakami}}\ and\ \bibinfo {author} {\bibfnamefont {P.}~\bibnamefont
			{Werner}},\ }\bibfield  {title} {\bibinfo {title} {Nonequilibrium steady
			states of electric field driven mott insulators},\ }\href
	{https://doi.org/10.1103/PhysRevB.98.075102} {\bibfield  {journal} {\bibinfo
			{journal} {Phys. Rev. B}\ }\textbf {\bibinfo {volume} {98}},\ \bibinfo
		{pages} {075102} (\bibinfo {year} {2018})}\BibitemShut {NoStop}%
	\bibitem [{\citenamefont {Li}\ \emph {et~al.}(2022)\citenamefont {Li},
		\citenamefont {Ning}, \citenamefont {Mehio}, \citenamefont {Zhao},
		\citenamefont {Lee}, \citenamefont {Kim}, \citenamefont {Nakamura},
		\citenamefont {Maeno}, \citenamefont {Cao},\ and\ \citenamefont
		{Hsieh}}]{dave_experimental_crossover}%
	\BibitemOpen
	\bibfield  {author} {\bibinfo {author} {\bibfnamefont {X.}~\bibnamefont
			{Li}}, \bibinfo {author} {\bibfnamefont {H.}~\bibnamefont {Ning}}, \bibinfo
		{author} {\bibfnamefont {O.}~\bibnamefont {Mehio}}, \bibinfo {author}
		{\bibfnamefont {H.}~\bibnamefont {Zhao}}, \bibinfo {author} {\bibfnamefont
			{M.-C.}\ \bibnamefont {Lee}}, \bibinfo {author} {\bibfnamefont
			{K.}~\bibnamefont {Kim}}, \bibinfo {author} {\bibfnamefont {F.}~\bibnamefont
			{Nakamura}}, \bibinfo {author} {\bibfnamefont {Y.}~\bibnamefont {Maeno}},
		\bibinfo {author} {\bibfnamefont {G.}~\bibnamefont {Cao}},\ and\ \bibinfo
		{author} {\bibfnamefont {D.}~\bibnamefont {Hsieh}},\ }\bibfield  {title}
	{\bibinfo {title} {Keldysh space control of charge dynamics in a strongly
			driven mott insulator},\ }\href
	{https://doi.org/10.1103/PhysRevLett.128.187402} {\bibfield  {journal}
		{\bibinfo  {journal} {Phys. Rev. Lett.}\ }\textbf {\bibinfo {volume} {128}},\
		\bibinfo {pages} {187402} (\bibinfo {year} {2022})}\BibitemShut {NoStop}%
	\bibitem [{\citenamefont {Gole\v{z}}\ \emph {et~al.}(2015)\citenamefont
		{Gole\v{z}}, \citenamefont {Eckstein},\ and\ \citenamefont
		{Werner}}]{dynamics_screening_photodoped}%
	\BibitemOpen
	\bibfield  {author} {\bibinfo {author} {\bibfnamefont {D.}~\bibnamefont
			{Gole\v{z}}}, \bibinfo {author} {\bibfnamefont {M.}~\bibnamefont
			{Eckstein}},\ and\ \bibinfo {author} {\bibfnamefont {P.}~\bibnamefont
			{Werner}},\ }\bibfield  {title} {\bibinfo {title} {Dynamics of screening in
			photodoped mott insulators},\ }\href
	{https://doi.org/10.1103/PhysRevB.92.195123} {\bibfield  {journal} {\bibinfo
			{journal} {Phys. Rev. B}\ }\textbf {\bibinfo {volume} {92}},\ \bibinfo
		{pages} {195123} (\bibinfo {year} {2015})}\BibitemShut {NoStop}%
	\bibitem [{\citenamefont {Gole\v{z}}\ \emph {et~al.}(2019)\citenamefont
		{Gole\v{z}}, \citenamefont {Eckstein},\ and\ \citenamefont
		{Werner}}]{multiband_photodoped_screening}%
	\BibitemOpen
	\bibfield  {author} {\bibinfo {author} {\bibfnamefont {D.}~\bibnamefont
			{Gole\v{z}}}, \bibinfo {author} {\bibfnamefont {M.}~\bibnamefont
			{Eckstein}},\ and\ \bibinfo {author} {\bibfnamefont {P.}~\bibnamefont
			{Werner}},\ }\bibfield  {title} {\bibinfo {title} {Multiband nonequilibrium
			gw + edmft formalism for correlated insulators},\ }\href
	{https://doi.org/10.1103/PhysRevB.100.235117} {\bibfield  {journal} {\bibinfo
			{journal} {Phys. Rev. B}\ }\textbf {\bibinfo {volume} {100}},\ \bibinfo
		{pages} {235117} (\bibinfo {year} {2019})}\BibitemShut {NoStop}%
	\bibitem [{\citenamefont {Golež}\ \emph {et~al.}(2025)\citenamefont {Golež},
		\citenamefont {Paprotzki}, \citenamefont {Werner},\ and\ \citenamefont
		{Eckstein}}]{golez_theory_2025}%
	\BibitemOpen
	\bibfield  {author} {\bibinfo {author} {\bibfnamefont {D.}~\bibnamefont
			{Golež}}, \bibinfo {author} {\bibfnamefont {E.}~\bibnamefont {Paprotzki}},
		\bibinfo {author} {\bibfnamefont {P.}~\bibnamefont {Werner}},\ and\ \bibinfo
		{author} {\bibfnamefont {M.}~\bibnamefont {Eckstein}},\ }\bibfield  {title}
	{\bibinfo {title} {Theory of ultrafast screening of \${U}\$ in driven
			charge-transfer insulators: {A} time-resolved x-ray absorption study},\
	}\href {https://doi.org/10.1103/PhysRevB.111.045147} {\bibfield  {journal}
		{\bibinfo  {journal} {Physical Review B}\ }\textbf {\bibinfo {volume}
			{111}},\ \bibinfo {pages} {045147} (\bibinfo {year} {2025})},\ \bibinfo
	{note} {publisher: American Physical Society}\BibitemShut {NoStop}%
	\bibitem [{\citenamefont {Hackl}\ \emph {et~al.}(2020)\citenamefont {Hackl},
		\citenamefont {Guaita}, \citenamefont {Shi}, \citenamefont {Haegeman},
		\citenamefont {Demler},\ and\ \citenamefont
		{Cirac}}]{geometry_variational_methods}%
	\BibitemOpen
	\bibfield  {author} {\bibinfo {author} {\bibfnamefont {L.}~\bibnamefont
			{Hackl}}, \bibinfo {author} {\bibfnamefont {T.}~\bibnamefont {Guaita}},
		\bibinfo {author} {\bibfnamefont {T.}~\bibnamefont {Shi}}, \bibinfo {author}
		{\bibfnamefont {J.}~\bibnamefont {Haegeman}}, \bibinfo {author}
		{\bibfnamefont {E.}~\bibnamefont {Demler}},\ and\ \bibinfo {author}
		{\bibfnamefont {J.~I.}\ \bibnamefont {Cirac}},\ }\bibfield  {title} {\bibinfo
		{title} {{Geometry of variational methods: dynamics of closed quantum
				systems}},\ }\href {https://doi.org/10.21468/SciPostPhys.9.4.048} {\bibfield
		{journal} {\bibinfo  {journal} {SciPost Phys.}\ }\textbf {\bibinfo {volume}
			{9}},\ \bibinfo {pages} {048} (\bibinfo {year} {2020})}\BibitemShut {NoStop}%
	\bibitem [{\citenamefont {Shi}\ \emph {et~al.}(2018)\citenamefont {Shi},
		\citenamefont {Demler},\ and\ \citenamefont {{Ignacio
				Cirac}}}]{variational_non_gaussian}%
	\BibitemOpen
	\bibfield  {author} {\bibinfo {author} {\bibfnamefont {T.}~\bibnamefont
			{Shi}}, \bibinfo {author} {\bibfnamefont {E.}~\bibnamefont {Demler}},\ and\
		\bibinfo {author} {\bibfnamefont {J.}~\bibnamefont {{Ignacio Cirac}}},\
	}\bibfield  {title} {\bibinfo {title} {Variational study of fermionic and
			bosonic systems with non-gaussian states: Theory and applications},\ }\href
	{https://doi.org/https://doi.org/10.1016/j.aop.2017.11.014} {\bibfield
		{journal} {\bibinfo  {journal} {Annals of Physics}\ }\textbf {\bibinfo
			{volume} {390}},\ \bibinfo {pages} {245} (\bibinfo {year}
		{2018})}\BibitemShut {NoStop}%
	\bibitem [{\citenamefont
		{Stephen}(1965)}]{transport_equations_superconductors}%
	\BibitemOpen
	\bibfield  {author} {\bibinfo {author} {\bibfnamefont {M.~J.}\ \bibnamefont
			{Stephen}},\ }\bibfield  {title} {\bibinfo {title} {Transport equations for
			superconductors},\ }\href {https://doi.org/10.1103/PhysRev.139.A197}
	{\bibfield  {journal} {\bibinfo  {journal} {Phys. Rev.}\ }\textbf {\bibinfo
			{volume} {139}},\ \bibinfo {pages} {A197} (\bibinfo {year}
		{1965})}\BibitemShut {NoStop}%
	\bibitem [{\citenamefont {Betbeder-Matibet}\ and\ \citenamefont
		{Nozieres}(1969)}]{transport_equations_clean_superconductors}%
	\BibitemOpen
	\bibfield  {author} {\bibinfo {author} {\bibfnamefont {O.}~\bibnamefont
			{Betbeder-Matibet}}\ and\ \bibinfo {author} {\bibfnamefont {P.}~\bibnamefont
			{Nozieres}},\ }\bibfield  {title} {\bibinfo {title} {Transport equations in
			clean superconductors},\ }\href
	{https://doi.org/https://doi.org/10.1016/0003-4916(69)90136-5} {\bibfield
		{journal} {\bibinfo  {journal} {Annals of Physics}\ }\textbf {\bibinfo
			{volume} {51}},\ \bibinfo {pages} {392} (\bibinfo {year} {1969})}\BibitemShut
	{NoStop}%
	\bibitem [{\citenamefont {Barankov}\ \emph {et~al.}(2004)\citenamefont
		{Barankov}, \citenamefont {Levitov},\ and\ \citenamefont
		{Spivak}}]{bcs_formation_solitons}%
	\BibitemOpen
	\bibfield  {author} {\bibinfo {author} {\bibfnamefont {R.~A.}\ \bibnamefont
			{Barankov}}, \bibinfo {author} {\bibfnamefont {L.~S.}\ \bibnamefont
			{Levitov}},\ and\ \bibinfo {author} {\bibfnamefont {B.~Z.}\ \bibnamefont
			{Spivak}},\ }\bibfield  {title} {\bibinfo {title} {Collective rabi
			oscillations and solitons in a time-dependent bcs pairing problem},\ }\href
	{https://doi.org/10.1103/PhysRevLett.93.160401} {\bibfield  {journal}
		{\bibinfo  {journal} {Phys. Rev. Lett.}\ }\textbf {\bibinfo {volume} {93}},\
		\bibinfo {pages} {160401} (\bibinfo {year} {2004})}\BibitemShut {NoStop}%
	\bibitem [{\citenamefont {Yuzbashyan}\ \emph
		{et~al.}(2005{\natexlab{a}})\citenamefont {Yuzbashyan}, \citenamefont
		{Altshuler}, \citenamefont {Kuznetsov},\ and\ \citenamefont
		{Enolskii}}]{solution_bcs_central_spin}%
	\BibitemOpen
	\bibfield  {author} {\bibinfo {author} {\bibfnamefont {E.~A.}\ \bibnamefont
			{Yuzbashyan}}, \bibinfo {author} {\bibfnamefont {B.~L.}\ \bibnamefont
			{Altshuler}}, \bibinfo {author} {\bibfnamefont {V.~B.}\ \bibnamefont
			{Kuznetsov}},\ and\ \bibinfo {author} {\bibfnamefont {V.~Z.}\ \bibnamefont
			{Enolskii}},\ }\bibfield  {title} {\bibinfo {title} {Solution for the
			dynamics of the bcs and central spin problems},\ }\href
	{https://doi.org/10.1088/0305-4470/38/36/003} {\bibfield  {journal} {\bibinfo
			{journal} {Journal of Physics A: Mathematical and General}\ }\textbf
		{\bibinfo {volume} {38}},\ \bibinfo {pages} {7831} (\bibinfo {year}
		{2005}{\natexlab{a}})}\BibitemShut {NoStop}%
	\bibitem [{\citenamefont {Yuzbashyan}\ \emph
		{et~al.}(2005{\natexlab{b}})\citenamefont {Yuzbashyan}, \citenamefont
		{Altshuler}, \citenamefont {Kuznetsov},\ and\ \citenamefont
		{Enolskii}}]{nonequilibrium_cooper_nonadiabatic}%
	\BibitemOpen
	\bibfield  {author} {\bibinfo {author} {\bibfnamefont {E.~A.}\ \bibnamefont
			{Yuzbashyan}}, \bibinfo {author} {\bibfnamefont {B.~L.}\ \bibnamefont
			{Altshuler}}, \bibinfo {author} {\bibfnamefont {V.~B.}\ \bibnamefont
			{Kuznetsov}},\ and\ \bibinfo {author} {\bibfnamefont {V.~Z.}\ \bibnamefont
			{Enolskii}},\ }\bibfield  {title} {\bibinfo {title} {Nonequilibrium cooper
			pairing in the nonadiabatic regime},\ }\href
	{https://doi.org/10.1103/PhysRevB.72.220503} {\bibfield  {journal} {\bibinfo
			{journal} {Phys. Rev. B}\ }\textbf {\bibinfo {volume} {72}},\ \bibinfo
		{pages} {220503} (\bibinfo {year} {2005}{\natexlab{b}})}\BibitemShut
	{NoStop}%
	\bibitem [{\citenamefont {Yuzbashyan}\ \emph {et~al.}(2006)\citenamefont
		{Yuzbashyan}, \citenamefont {Tsyplyatyev},\ and\ \citenamefont
		{Altshuler}}]{quenching_bcs_no_eom}%
	\BibitemOpen
	\bibfield  {author} {\bibinfo {author} {\bibfnamefont {E.~A.}\ \bibnamefont
			{Yuzbashyan}}, \bibinfo {author} {\bibfnamefont {O.}~\bibnamefont
			{Tsyplyatyev}},\ and\ \bibinfo {author} {\bibfnamefont {B.~L.}\ \bibnamefont
			{Altshuler}},\ }\bibfield  {title} {\bibinfo {title} {Relaxation and
			persistent oscillations of the order parameter in fermionic condensates},\
	}\href {https://doi.org/10.1103/PhysRevLett.96.097005} {\bibfield  {journal}
		{\bibinfo  {journal} {Phys. Rev. Lett.}\ }\textbf {\bibinfo {volume} {96}},\
		\bibinfo {pages} {097005} (\bibinfo {year} {2006})}\BibitemShut {NoStop}%
	\bibitem [{\citenamefont {Chern}\ and\ \citenamefont
		{Barros}(2019)}]{nonequilibrium_hubbard_sc_controversial}%
	\BibitemOpen
	\bibfield  {author} {\bibinfo {author} {\bibfnamefont {G.-W.}\ \bibnamefont
			{Chern}}\ and\ \bibinfo {author} {\bibfnamefont {K.}~\bibnamefont {Barros}},\
	}\bibfield  {title} {\bibinfo {title} {Nonequilibrium dynamics of
			superconductivity in the attractive hubbard model},\ }\href
	{https://doi.org/10.1103/PhysRevB.99.035162} {\bibfield  {journal} {\bibinfo
			{journal} {Phys. Rev. B}\ }\textbf {\bibinfo {volume} {99}},\ \bibinfo
		{pages} {035162} (\bibinfo {year} {2019})}\BibitemShut {NoStop}%
	\bibitem [{\citenamefont {Schrieffer}\ \emph {et~al.}(1989)\citenamefont
		{Schrieffer}, \citenamefont {Wen},\ and\ \citenamefont
		{Zhang}}]{sdw_bag_superconductivity}%
	\BibitemOpen
	\bibfield  {author} {\bibinfo {author} {\bibfnamefont {J.~R.}\ \bibnamefont
			{Schrieffer}}, \bibinfo {author} {\bibfnamefont {X.~G.}\ \bibnamefont
			{Wen}},\ and\ \bibinfo {author} {\bibfnamefont {S.~C.}\ \bibnamefont
			{Zhang}},\ }\bibfield  {title} {\bibinfo {title} {Dynamic spin fluctuations
			and the bag mechanism of high-${T}_{c}$ superconductivity},\ }\href
	{https://doi.org/10.1103/PhysRevB.39.11663} {\bibfield  {journal} {\bibinfo
			{journal} {Phys. Rev. B}\ }\textbf {\bibinfo {volume} {39}},\ \bibinfo
		{pages} {11663} (\bibinfo {year} {1989})}\BibitemShut {NoStop}%
	\bibitem [{\citenamefont {Thomsen}\ \emph {et~al.}(1986)\citenamefont
		{Thomsen}, \citenamefont {Grahn}, \citenamefont {Maris},\ and\ \citenamefont
		{Tauc}}]{reflected_phonons_thin_film}%
	\BibitemOpen
	\bibfield  {author} {\bibinfo {author} {\bibfnamefont {C.}~\bibnamefont
			{Thomsen}}, \bibinfo {author} {\bibfnamefont {H.~T.}\ \bibnamefont {Grahn}},
		\bibinfo {author} {\bibfnamefont {H.~J.}\ \bibnamefont {Maris}},\ and\
		\bibinfo {author} {\bibfnamefont {J.}~\bibnamefont {Tauc}},\ }\bibfield
	{title} {\bibinfo {title} {Surface generation and detection of phonons by
			picosecond light pulses},\ }\href {https://doi.org/10.1103/PhysRevB.34.4129}
	{\bibfield  {journal} {\bibinfo  {journal} {Phys. Rev. B}\ }\textbf {\bibinfo
			{volume} {34}},\ \bibinfo {pages} {4129} (\bibinfo {year}
		{1986})}\BibitemShut {NoStop}%
	\bibitem [{\citenamefont {Bozovic}\ \emph {et~al.}(2004)\citenamefont
		{Bozovic}, \citenamefont {Schneider}, \citenamefont {Xu}, \citenamefont
		{Sobolewski}, \citenamefont {Ren}, \citenamefont {L\"upke}, \citenamefont
		{Demsar}, \citenamefont {Taylor},\ and\ \citenamefont
		{Onellion}}]{long_lived_oscillations_thin_film}%
	\BibitemOpen
	\bibfield  {author} {\bibinfo {author} {\bibfnamefont {I.}~\bibnamefont
			{Bozovic}}, \bibinfo {author} {\bibfnamefont {M.}~\bibnamefont {Schneider}},
		\bibinfo {author} {\bibfnamefont {Y.}~\bibnamefont {Xu}}, \bibinfo {author}
		{\bibfnamefont {R.}~\bibnamefont {Sobolewski}}, \bibinfo {author}
		{\bibfnamefont {Y.~H.}\ \bibnamefont {Ren}}, \bibinfo {author} {\bibfnamefont
			{G.}~\bibnamefont {L\"upke}}, \bibinfo {author} {\bibfnamefont
			{J.}~\bibnamefont {Demsar}}, \bibinfo {author} {\bibfnamefont {A.~J.}\
			\bibnamefont {Taylor}},\ and\ \bibinfo {author} {\bibfnamefont
			{M.}~\bibnamefont {Onellion}},\ }\bibfield  {title} {\bibinfo {title}
		{Long-lived coherent acoustic waves generated by femtosecond light pulses},\
	}\href {https://doi.org/10.1103/PhysRevB.69.132503} {\bibfield  {journal}
		{\bibinfo  {journal} {Phys. Rev. B}\ }\textbf {\bibinfo {volume} {69}},\
		\bibinfo {pages} {132503} (\bibinfo {year} {2004})}\BibitemShut {NoStop}%
	\bibitem [{\citenamefont {Ruello}\ and\ \citenamefont
		{Gusev}(2015)}]{phonon_mechanism_review_macroscopic}%
	\BibitemOpen
	\bibfield  {author} {\bibinfo {author} {\bibfnamefont {P.}~\bibnamefont
			{Ruello}}\ and\ \bibinfo {author} {\bibfnamefont {V.~E.}\ \bibnamefont
			{Gusev}},\ }\bibfield  {title} {\bibinfo {title} {Physical mechanisms of
			coherent acoustic phonons generation by ultrafast laser action},\ }\href
	{https://doi.org/https://doi.org/10.1016/j.ultras.2014.06.004} {\bibfield
		{journal} {\bibinfo  {journal} {Ultrasonics}\ }\textbf {\bibinfo {volume}
			{56}},\ \bibinfo {pages} {21} (\bibinfo {year} {2015})}\BibitemShut {NoStop}%
	\bibitem [{\citenamefont {Su}\ \emph {et~al.}(1979)\citenamefont {Su},
		\citenamefont {Schrieffer},\ and\ \citenamefont {Heeger}}]{ssh_original}%
	\BibitemOpen
	\bibfield  {author} {\bibinfo {author} {\bibfnamefont {W.~P.}\ \bibnamefont
			{Su}}, \bibinfo {author} {\bibfnamefont {J.~R.}\ \bibnamefont {Schrieffer}},\
		and\ \bibinfo {author} {\bibfnamefont {A.~J.}\ \bibnamefont {Heeger}},\
	}\bibfield  {title} {\bibinfo {title} {Solitons in polyacetylene},\ }\href
	{https://doi.org/10.1103/PhysRevLett.42.1698} {\bibfield  {journal} {\bibinfo
			{journal} {Phys. Rev. Lett.}\ }\textbf {\bibinfo {volume} {42}},\ \bibinfo
		{pages} {1698} (\bibinfo {year} {1979})}\BibitemShut {NoStop}%
	\bibitem [{\citenamefont {Johnston}\ \emph {et~al.}(2010)\citenamefont
		{Johnston}, \citenamefont {Vernay}, \citenamefont {Moritz}, \citenamefont
		{Shen}, \citenamefont {Nagaosa}, \citenamefont {Zaanen},\ and\ \citenamefont
		{Devereaux}}]{phonons_in_cuprates}%
	\BibitemOpen
	\bibfield  {author} {\bibinfo {author} {\bibfnamefont {S.}~\bibnamefont
			{Johnston}}, \bibinfo {author} {\bibfnamefont {F.}~\bibnamefont {Vernay}},
		\bibinfo {author} {\bibfnamefont {B.}~\bibnamefont {Moritz}}, \bibinfo
		{author} {\bibfnamefont {Z.-X.}\ \bibnamefont {Shen}}, \bibinfo {author}
		{\bibfnamefont {N.}~\bibnamefont {Nagaosa}}, \bibinfo {author} {\bibfnamefont
			{J.}~\bibnamefont {Zaanen}},\ and\ \bibinfo {author} {\bibfnamefont {T.~P.}\
			\bibnamefont {Devereaux}},\ }\bibfield  {title} {\bibinfo {title} {Systematic
			study of electron-phonon coupling to oxygen modes across the cuprates},\
	}\href {https://doi.org/10.1103/PhysRevB.82.064513} {\bibfield  {journal}
		{\bibinfo  {journal} {Phys. Rev. B}\ }\textbf {\bibinfo {volume} {82}},\
		\bibinfo {pages} {064513} (\bibinfo {year} {2010})}\BibitemShut {NoStop}%
	\bibitem [{\citenamefont {Sheshadri}\ \emph {et~al.}(2023)\citenamefont
		{Sheshadri}, \citenamefont {Malterre}, \citenamefont {Fujimori},\ and\
		\citenamefont {Chainani}}]{cuprate_hubbard_models_review}%
	\BibitemOpen
	\bibfield  {author} {\bibinfo {author} {\bibfnamefont {K.}~\bibnamefont
			{Sheshadri}}, \bibinfo {author} {\bibfnamefont {D.}~\bibnamefont {Malterre}},
		\bibinfo {author} {\bibfnamefont {A.}~\bibnamefont {Fujimori}},\ and\
		\bibinfo {author} {\bibfnamefont {A.}~\bibnamefont {Chainani}},\ }\bibfield
	{title} {\bibinfo {title} {Connecting the one-band and three-band hubbard
			models of cuprates via spectroscopy and scattering experiments},\ }\href
	{https://doi.org/10.1103/PhysRevB.107.085125} {\bibfield  {journal} {\bibinfo
			{journal} {Phys. Rev. B}\ }\textbf {\bibinfo {volume} {107}},\ \bibinfo
		{pages} {085125} (\bibinfo {year} {2023})}\BibitemShut {NoStop}%
	\bibitem [{Note2()}]{Note2}%
	\BibitemOpen
	\bibinfo {note} {Note that only the short-range part of AFM correlations is
		essential here, so the results will also be qualitatively valid above $T_N$;
		see Appendix \ref {ssec:sm_intro_hamiltonian_drive} for further
		discussion.}\BibitemShut {Stop}%
	\bibitem [{Note3()}]{Note3}%
	\BibitemOpen
	\bibinfo {note} {Note that Fourier transforming eq. \protect \eqref
		{eq:maintext_approximate_driving_term} directly is much more convenient than
		computing $\varphi _{\protect \mathbf k}(\omega )$ and multiplying by $i
		\omega $; hence the unusual notation. The resonance condition \protect \eqref
		{eq:maintext_no_decoherence_resonance} is also more transparent in this
		form.}\BibitemShut {Stop}%
	\bibitem [{Note4()}]{Note4}%
	\BibitemOpen
	\bibinfo {note} {As discussed in the introduction, the length scale
		characterizing this crossover in the literature is the doublon-hole
		correlation length $\xi $, rather than the lattice constant $a$. However, for
		our strongly-interacting regime $U/\tau \sim 10$, we have $\xi \approx a$,
		and for simplicity we use $a$ as our typical length scale.}\BibitemShut
	{Stop}%
	\bibitem [{Note5()}]{Note5}%
	\BibitemOpen
	\bibinfo {note} {We remark that resonant features have been numerically found
		in a different context, when analyzing nonequilibrium steady states of
		continuously driven Mott insulators in infinite dimensions \cite
		{driven_steady_state}. However, we expect the extremely high field strengths
		considered in that case to either produce significant gap suppression, which
		would in turn shift down the resonance conditions, or break down the
		insulating state altogether. These consequences of dynamical gap
		renormalization seem to be absent from steady-state double occupancy plots in
		that study.}\BibitemShut {Stop}%
	\bibitem [{Note6()}]{Note6}%
	\BibitemOpen
	\bibinfo {note} {At first, it may seem surprising that the regionsAt first,
		it may seem surprising that the regions in Fig. \ref
		{fig:quasiparticles_and_multiphoton}b where $\mu $ matches the multiphoton
		order $n$ are very narrow, compared to the area marked `Multiphoton
		excitation' in Figure \ref {fig:regime_sketch_2d}. Part of the cause is the
		strongly-peaked density of states $\rho _{\protect \text {dh}}$ in our model,
		whereas in a real material it would be broader. On the other hand, gap
		renormalization effects also alter electric field dependence, and therefore
		reduce its utility for identifying the multiphoton mechanism. We will find
		that momentum distribution of photocarriers offers a better tool for this
		purpose.}\BibitemShut {Stop}%
	\bibitem [{\citenamefont {Ito}\ \emph {et~al.}(2023)\citenamefont {Ito},
		\citenamefont {Sch{\"u}ler}, \citenamefont {Meierhofer}, \citenamefont
		{Schlauderer}, \citenamefont {Freudenstein}, \citenamefont {Reimann},
		\citenamefont {Afanasiev}, \citenamefont {Kokh}, \citenamefont
		{Tereshchenko}, \citenamefont {G{\"u}dde}, \citenamefont {Sentef},
		\citenamefont {H{\"o}fer},\ and\ \citenamefont
		{Huber}}]{band_build_up_original}%
	\BibitemOpen
	\bibfield  {author} {\bibinfo {author} {\bibfnamefont {S.}~\bibnamefont
			{Ito}}, \bibinfo {author} {\bibfnamefont {M.}~\bibnamefont {Sch{\"u}ler}},
		\bibinfo {author} {\bibfnamefont {M.}~\bibnamefont {Meierhofer}}, \bibinfo
		{author} {\bibfnamefont {S.}~\bibnamefont {Schlauderer}}, \bibinfo {author}
		{\bibfnamefont {J.}~\bibnamefont {Freudenstein}}, \bibinfo {author}
		{\bibfnamefont {J.}~\bibnamefont {Reimann}}, \bibinfo {author} {\bibfnamefont
			{D.}~\bibnamefont {Afanasiev}}, \bibinfo {author} {\bibfnamefont {K.~A.}\
			\bibnamefont {Kokh}}, \bibinfo {author} {\bibfnamefont {O.~E.}\ \bibnamefont
			{Tereshchenko}}, \bibinfo {author} {\bibfnamefont {J.}~\bibnamefont
			{G{\"u}dde}}, \bibinfo {author} {\bibfnamefont {M.~A.}\ \bibnamefont
			{Sentef}}, \bibinfo {author} {\bibfnamefont {U.}~\bibnamefont {H{\"o}fer}},\
		and\ \bibinfo {author} {\bibfnamefont {R.}~\bibnamefont {Huber}},\ }\bibfield
	{title} {\bibinfo {title} {Build-up and dephasing of floquet--bloch bands on
			subcycle timescales},\ }\href {https://doi.org/10.1038/s41586-023-05850-x}
	{\bibfield  {journal} {\bibinfo  {journal} {Nature}\ }\textbf {\bibinfo
			{volume} {616}},\ \bibinfo {pages} {696} (\bibinfo {year}
		{2023})}\BibitemShut {NoStop}%
	\bibitem [{\citenamefont {Abergel}(2023)}]{band_build_up_nw}%
	\BibitemOpen
	\bibfield  {author} {\bibinfo {author} {\bibfnamefont {D.}~\bibnamefont
			{Abergel}},\ }\bibfield  {title} {\bibinfo {title} {Band build-up},\ }\href
	{https://doi.org/10.1038/s41567-023-02068-6} {\bibfield  {journal} {\bibinfo
			{journal} {Nature Physics}\ }\textbf {\bibinfo {volume} {19}},\ \bibinfo
		{pages} {621} (\bibinfo {year} {2023})}\BibitemShut {NoStop}%
	\bibitem [{\citenamefont {Rösch}\ and\ \citenamefont
		{Gunnarsson}(2004)}]{rosch_electron_phonon_2004}%
	\BibitemOpen
	\bibfield  {author} {\bibinfo {author} {\bibfnamefont {O.}~\bibnamefont
			{Rösch}}\ and\ \bibinfo {author} {\bibfnamefont {O.}~\bibnamefont
			{Gunnarsson}},\ }\bibfield  {title} {\bibinfo {title} {Electron-{Phonon}
			{Interaction} in the
			\$t{\textbackslash}mathrm\{{\textbackslash}text\{{\textbackslash}ensuremath\{-\}\}\}{J}\$
			{Model}},\ }\href {https://doi.org/10.1103/PhysRevLett.92.146403} {\bibfield
		{journal} {\bibinfo  {journal} {Physical Review Letters}\ }\textbf {\bibinfo
			{volume} {92}},\ \bibinfo {pages} {146403} (\bibinfo {year} {2004})},\
	\bibinfo {note} {publisher: American Physical Society}\BibitemShut {NoStop}%
	\bibitem [{\citenamefont {Devereaux}\ \emph {et~al.}(1999)\citenamefont
		{Devereaux}, \citenamefont {Virosztek},\ and\ \citenamefont
		{Zawadowski}}]{devereaux_neutron_1999}%
	\BibitemOpen
	\bibfield  {author} {\bibinfo {author} {\bibfnamefont {T.~P.}\ \bibnamefont
			{Devereaux}}, \bibinfo {author} {\bibfnamefont {A.}~\bibnamefont
			{Virosztek}},\ and\ \bibinfo {author} {\bibfnamefont {A.}~\bibnamefont
			{Zawadowski}},\ }\bibfield  {title} {\bibinfo {title} {Neutron scattering and
			the \$\{{B}\}\_\{1g\}\$ phonon in the cuprates},\ }\href
	{https://doi.org/10.1103/PhysRevB.59.14618} {\bibfield  {journal} {\bibinfo
			{journal} {Physical Review B}\ }\textbf {\bibinfo {volume} {59}},\ \bibinfo
		{pages} {14618} (\bibinfo {year} {1999})},\ \bibinfo {note} {publisher:
		American Physical Society}\BibitemShut {NoStop}%
	\bibitem [{\citenamefont {Strohmaier}\ \emph {et~al.}(2010)\citenamefont
		{Strohmaier}, \citenamefont {Greif}, \citenamefont {J\"ordens}, \citenamefont
		{Tarruell}, \citenamefont {Moritz}, \citenamefont {Esslinger}, \citenamefont
		{Sensarma}, \citenamefont {Pekker}, \citenamefont {Altman},\ and\
		\citenamefont {Demler}}]{long_lived_doublons_exp}%
	\BibitemOpen
	\bibfield  {author} {\bibinfo {author} {\bibfnamefont {N.}~\bibnamefont
			{Strohmaier}}, \bibinfo {author} {\bibfnamefont {D.}~\bibnamefont {Greif}},
		\bibinfo {author} {\bibfnamefont {R.}~\bibnamefont {J\"ordens}}, \bibinfo
		{author} {\bibfnamefont {L.}~\bibnamefont {Tarruell}}, \bibinfo {author}
		{\bibfnamefont {H.}~\bibnamefont {Moritz}}, \bibinfo {author} {\bibfnamefont
			{T.}~\bibnamefont {Esslinger}}, \bibinfo {author} {\bibfnamefont
			{R.}~\bibnamefont {Sensarma}}, \bibinfo {author} {\bibfnamefont
			{D.}~\bibnamefont {Pekker}}, \bibinfo {author} {\bibfnamefont
			{E.}~\bibnamefont {Altman}},\ and\ \bibinfo {author} {\bibfnamefont
			{E.}~\bibnamefont {Demler}},\ }\bibfield  {title} {\bibinfo {title}
		{Observation of elastic doublon decay in the fermi-hubbard model},\ }\href
	{https://doi.org/10.1103/PhysRevLett.104.080401} {\bibfield  {journal}
		{\bibinfo  {journal} {Phys. Rev. Lett.}\ }\textbf {\bibinfo {volume} {104}},\
		\bibinfo {pages} {080401} (\bibinfo {year} {2010})}\BibitemShut {NoStop}%
	\bibitem [{\citenamefont {Sensarma}\ \emph {et~al.}(2010)\citenamefont
		{Sensarma}, \citenamefont {Pekker}, \citenamefont {Altman}, \citenamefont
		{Demler}, \citenamefont {Strohmaier}, \citenamefont {Greif}, \citenamefont
		{J\"ordens}, \citenamefont {Tarruell}, \citenamefont {Moritz},\ and\
		\citenamefont {Esslinger}}]{long_lived_doublons_th}%
	\BibitemOpen
	\bibfield  {author} {\bibinfo {author} {\bibfnamefont {R.}~\bibnamefont
			{Sensarma}}, \bibinfo {author} {\bibfnamefont {D.}~\bibnamefont {Pekker}},
		\bibinfo {author} {\bibfnamefont {E.}~\bibnamefont {Altman}}, \bibinfo
		{author} {\bibfnamefont {E.}~\bibnamefont {Demler}}, \bibinfo {author}
		{\bibfnamefont {N.}~\bibnamefont {Strohmaier}}, \bibinfo {author}
		{\bibfnamefont {D.}~\bibnamefont {Greif}}, \bibinfo {author} {\bibfnamefont
			{R.}~\bibnamefont {J\"ordens}}, \bibinfo {author} {\bibfnamefont
			{L.}~\bibnamefont {Tarruell}}, \bibinfo {author} {\bibfnamefont
			{H.}~\bibnamefont {Moritz}},\ and\ \bibinfo {author} {\bibfnamefont
			{T.}~\bibnamefont {Esslinger}},\ }\bibfield  {title} {\bibinfo {title}
		{Lifetime of double occupancies in the fermi-hubbard model},\ }\href
	{https://doi.org/10.1103/PhysRevB.82.224302} {\bibfield  {journal} {\bibinfo
			{journal} {Phys. Rev. B}\ }\textbf {\bibinfo {volume} {82}},\ \bibinfo
		{pages} {224302} (\bibinfo {year} {2010})}\BibitemShut {NoStop}%
	\bibitem [{\citenamefont {Werner}\ and\ \citenamefont
		{Eckstein}(2013)}]{phonon_enhanced_doublon_recombination}%
	\BibitemOpen
	\bibfield  {author} {\bibinfo {author} {\bibfnamefont {P.}~\bibnamefont
			{Werner}}\ and\ \bibinfo {author} {\bibfnamefont {M.}~\bibnamefont
			{Eckstein}},\ }\bibfield  {title} {\bibinfo {title} {Phonon-enhanced
			relaxation and excitation in the holstein-hubbard model},\ }\href
	{https://doi.org/10.1103/PhysRevB.88.165108} {\bibfield  {journal} {\bibinfo
			{journal} {Phys. Rev. B}\ }\textbf {\bibinfo {volume} {88}},\ \bibinfo
		{pages} {165108} (\bibinfo {year} {2013})}\BibitemShut {NoStop}%
	\bibitem [{Note7()}]{Note7}%
	\BibitemOpen
	\bibinfo {note} {Electron-phonon coupling has not been explicitly considered
		in previous sections because acoustic phonon energies are too low to
		significantly alter the high-energy photoexcitaton process. Optical phonons
		may contribute to $\Gamma $, an argument which is sketched in Section \ref
		{ssec:phonon_scattering}.}\BibitemShut {Stop}%
	\bibitem [{\citenamefont {Mehio}\ \emph {et~al.}(2023)\citenamefont {Mehio},
		\citenamefont {Li}, \citenamefont {Ning}, \citenamefont
		{Lenar{\v{c}}i{\v{c}}}, \citenamefont {Han}, \citenamefont {Buchhold},
		\citenamefont {Porter}, \citenamefont {Laurita}, \citenamefont {Wilson},\
		and\ \citenamefont {Hsieh}}]{hubbard_excitons}%
	\BibitemOpen
	\bibfield  {author} {\bibinfo {author} {\bibfnamefont {O.}~\bibnamefont
			{Mehio}}, \bibinfo {author} {\bibfnamefont {X.}~\bibnamefont {Li}}, \bibinfo
		{author} {\bibfnamefont {H.}~\bibnamefont {Ning}}, \bibinfo {author}
		{\bibfnamefont {Z.}~\bibnamefont {Lenar{\v{c}}i{\v{c}}}}, \bibinfo {author}
		{\bibfnamefont {Y.}~\bibnamefont {Han}}, \bibinfo {author} {\bibfnamefont
			{M.}~\bibnamefont {Buchhold}}, \bibinfo {author} {\bibfnamefont
			{Z.}~\bibnamefont {Porter}}, \bibinfo {author} {\bibfnamefont {N.~J.}\
			\bibnamefont {Laurita}}, \bibinfo {author} {\bibfnamefont {S.~D.}\
			\bibnamefont {Wilson}},\ and\ \bibinfo {author} {\bibfnamefont
			{D.}~\bibnamefont {Hsieh}},\ }\bibfield  {title} {\bibinfo {title} {A hubbard
			exciton fluid in a photo-doped antiferromagnetic mott insulator},\ }\href
	{https://doi.org/10.1038/s41567-023-02204-2} {\bibfield  {journal} {\bibinfo
			{journal} {Nature Physics}\ }\textbf {\bibinfo {volume} {19}},\ \bibinfo
		{pages} {1876} (\bibinfo {year} {2023})}\BibitemShut {NoStop}%
	\bibitem [{\citenamefont {Stafford}\ and\ \citenamefont
		{Millis}(1993)}]{millis_correlation_length}%
	\BibitemOpen
	\bibfield  {author} {\bibinfo {author} {\bibfnamefont {C.~A.}\ \bibnamefont
			{Stafford}}\ and\ \bibinfo {author} {\bibfnamefont {A.~J.}\ \bibnamefont
			{Millis}},\ }\bibfield  {title} {\bibinfo {title} {Scaling theory of the
			mott-hubbard metal-insulator transition in one dimension},\ }\href
	{https://doi.org/10.1103/PhysRevB.48.1409} {\bibfield  {journal} {\bibinfo
			{journal} {Phys. Rev. B}\ }\textbf {\bibinfo {volume} {48}},\ \bibinfo
		{pages} {1409} (\bibinfo {year} {1993})}\BibitemShut {NoStop}%
	\bibitem [{Note8()}]{Note8}%
	\BibitemOpen
	\bibinfo {note} {At low temperatures, we expect an AFM phase, where the Néel
		order assumption is justified. Above $T_N$, we retain strong short-range spin
		correlations even in the absence of long-range order. As the typical
		doublon-hole correlation length is short - on the order of a few lattice
		constants - they are still evolving in an effective AFM background.
		Therefore, present results are expected to qualitatively hold well above
		$T_N$.}\BibitemShut {Stop}%
	\bibitem [{Note9()}]{Note9}%
	\BibitemOpen
	\bibinfo {note} {During the drive, Peierls phases should also be added to the
		hopping terms in \protect \eqref {eq:ssh_hamiltonian_definition}; however, we
		will only consider pulses of much shorter duration compared to the
		characteristic timescale of the low-energy acoustic phonons. This justifies
		ignoring the direct effect of the electric field on the SSH part of the
		Hamiltonian.}\BibitemShut {Stop}%
	\bibitem [{Note10()}]{Note10}%
	\BibitemOpen
	\bibinfo {note} {In real space, this reads $S^z_{\protect \mathbf Q} = {1
			\over N} \DOTSB \sum@ \slimits@ _{\protect \mathbf j} e^{i \protect \mathbf Q
			\cdot \protect \mathbf j} \protect \tmspace +\thickmuskip {.2777em}
		(n_{\protect \mathbf j \uparrow } - n_{\protect \mathbf j \downarrow
		})$.}\BibitemShut {Stop}%
	\bibitem [{Note11()}]{Note11}%
	\BibitemOpen
	\bibinfo {note} {However, since $S(t)$ is time-dependent, this term should be
		considered towards conservation of energy.}\BibitemShut {Stop}%
	\bibitem [{Note12()}]{Note12}%
	\BibitemOpen
	\bibinfo {note} {We are ignoring $\Gamma ^\protect \text {relax}$ in the
		equation for $\rho _z$, but keeping it in those for $\rho _{x/y}$. The reason
		is that, in the former case, it only leads to a (slow) redistribution of the
		photoexcited carriers between momentum states; in the latter, it strongly
		impacts excitation rates under weak drive. See Section \ref
		{sec:analytic_approximate_solutions} for further discussion.}\BibitemShut
	{Stop}%
\end{thebibliography}
\end{document}